\documentstyle[12pt]{article}
\newcommand{\be}{\begin{equation}}
\newcommand{\ee}{\end{equation}}

\newcommand{\eea}{\end{array}}

\title{REMARKS ON THE WHITHAM EQUATIONS}
\author{Robert Carroll\\Mathematics Department\\University of Illinois\\
Urbana, IL 61801\thanks{email: rcarroll@symcom.math.uiuc.edu}}
\date{November, 1995}

\begin{document}

\bibliographystyle{plain}
\maketitle

\begin{abstract}
We survey some topics involving the Whitham equations, concentrating
on the role of $\psi\psi^*$ (or square eigenfunctions) in averaging
and in producing Cauchy kernels and differentials on Riemann surfaces.
\end{abstract}

\tableofcontents

\section{INTRODUCTION}

\renewcommand{\theequation}{1.\arabic{equation}}\setcounter{equation}{0}
The Whitham equations and averaging methods have become a very important
part of modern mathematical physics (see e.g. \cite{aa,ba,
cb,da,db,dd,de,df,ea,fa,fd,fb,fc,
gb,gc,gj,hc,ka,kb,kc,kd,ke,kf,kg,kh,ki,mc,mf,mg,me,nc,na,nb,sc,va,wa}).
Most of the necessary background and details
are to be found somewhere in these references but there are also a
number of formulas and assertions which need clarification
or amendment.  We will look at some of this here and attempt to organize
a subset of information in a coherent and more or less rigorous
manner with the aim of ``understanding" the basic ideas.
Connections to dispersionless theory are indicated and
it also seems desirable to indicate explicitly the main
features that emerge which are relevant to work in topological
field theory (TFT) etc., so we will display a lot of this.
More recently there has also been a surge of activity connected to
$N=2$ supersymmetric gauge theories and some of this is
sketched briefly as well
(but certainly not explained).  Thus
the presentation was designed to be mainly expository but the
development also leads to some new connections and results.
In particular we indicate various formulas for some Cauchy type kernels,
give some connections of classical inverse scattering to averaging of
square eigenfunctions for KdV, and provide some relations between
averaged quantities for $\psi^*\psi$ and various differentials in
general situations.
In this spirit the classical Whitham theory can be phrased via $\psi\psi^*$
which generates Cauchy type kernels as well and leads, in
a sense indicated, to the
dispersionless limit situation in genus zero.
Relations between averaging and moduli spaces
are developed following Dubrovin and Krichever and some Landau-
Ginzburg (LG) type examples are displayed.  Many basic formulas
involving theta functions, differentials, etc. are also exhibited
in order to make the text useful as a launching pad for further
investigation.  We would like to acknowledge valuable discussions
with P. Grinevich, Y. Kodama, and A. Yaremchuk on related topics.

\section{SOME BACKGROUND}
\renewcommand{\theequation}{2.\arabic{equation}}\setcounter{equation}{0}

We start from the viewpoint of Riemann surfaces.  Thus (cf. \cite
{bb,ca,dc,dd,fe,fh,ga,ia,ma} take an arbitrary Riemann surface $\Sigma$
of genus $g$, pick a point $Q$ and a local variable $1/k$ near $Q$
such that $k(Q) = \infty$.  Let $D = P_1 + \cdots + P_g$ be a positive
divisor of degree $g$ and write $\psi$ for the (unique up to a constant
multiplier) Baker-Akhiezer (BA) function characterized by the properties
({\bf A}) $\psi$ is meromorphic on $\Sigma$ except for $Q$ where $\psi
(P)exp(-q(k))$ is analytic ($q(k) = kx + k^2y + k^3t$ will do for
illustration and (*) $\psi\sim exp(q(k))[1 + \sum_1^{\infty}(\xi_j/
k^j)]$ near $Q$) ({\bf B}) On $\Sigma/Q,\,\,\psi$ has only a finite
number of poles.  In fact $\psi$ can be taken in the form ($P\in\Sigma,\,\,
P_0\not= Q$)
\be
\psi(x,y,t,P) = \alpha exp[\int^P_{P_0}(x\Omega^1 + y\Omega^2 + t\Omega^3)]
\cdot\frac{\Theta({\cal A}(P) + xU + yV + tW + z_0)}{\Theta({\cal A}
(P) + z_0)}
\label{AA}
\ee
where $\Omega^1 = dk + \cdots,\,\,\Omega^2 = d(k^2) + \cdots,\,\,
\Omega^3 = d(k^3) + \cdots, U_j = \int_{b_j}\Omega^1,\,\,V_j = \int_
{b_j}\Omega^2,\,\,W_j = \int_{b_j}\Omega^3\,\,(j = 1,\cdots,g),\,\,z_0
= -{\cal A}(D) - K$, and $\alpha$ is a normalizing factor
(see below).  The symbol $\sim$ will be used generally to mean ``corresponds
to" or ``is associated with"; occasionally it also denotes asymptotic
behavior and this should be clear from the context.
Here the
$\Omega_j$ are meromorphic differentials of second kind normalized via
$\int_{a_k}\Omega_j = 0\,\,(a_j,\,b_j$ are canonical homology cycles)
and we note that $\Omega = x\Omega^1 + u\Omega^2 + t\Omega^3\sim
dq(k)$ normalized;
${\cal A}$ is the Abel-Jacobi map (${\cal A}(P) = \int^P\omega_k$ where
the $\omega_k$ are normalized holomorphic differentials, $k = 1,\cdots,g,
\,\,\int_{a_j}\omega_k = \delta_{jk}$), and $K = (K_j)\sim$ Riemann
constants ($2K = -{\cal A}(K_{\Sigma})$ where $K_{\sigma}$ is the
canonical class of $\Sigma\sim$ equivalence class of meromorphic
differentials) so $\Theta({\cal A}(P) + z_0)$ has exactly $g$ zeros
(or vanishes identically).  The paths of integration are to be the
same in computing $\int_{P_0}^P\Omega^i$ or ${\cal A}(P)$ and it is
shown in \cite{ca,dc} that $\psi$ is well defined (i.e. path independent).
Then the $\xi_j$ in (*) can be computed
formally and one determines Lax operators $L$ and $A$ such that
$\partial_y\psi = L\psi$ with $\partial_t\psi = A\psi$.  Indeed, given
the $\xi_j$ write $u = -2\partial_x\xi_1$ with $w = 3\xi_1\partial_x\xi_1
-3\partial^2_x\xi_1 - 3\partial_x\xi_2$.  Then formally, near $Q$, one
has $(-\partial_y + \partial_x^2 + u)\psi = O(1/k)exp(q)$ and
$(-\partial_t + \partial^3_x + (3/2)u\partial_x + w)\psi = O(1/k)exp(q)$
(i.e. this choice of $u,\,w$ makes the coefficients of $k^nexp(q)$ vanish
for $n = 0,1,2,3$ - note the similarity here to D-bar dressing techniques
as in \cite{cc}).  Now define $L = \partial_x^2 + u$ and $A = \partial^3_x
+ (3/2)u\partial_x + w$ so $\partial_y\psi = L\psi$ and $\partial_t\psi
= A\psi$.  This follows from the uniqueness of BA functions with the same
essential singularity and pole divisors.  Then we have, via compatibility
$L_t - A_y = [A,L]$, a KP-1 equation $(3/4)u_{yy} = \partial_x[u_t
-(1/4)(6uu_x + u_{xxx})]$ and therefore KP equations are parametrized
by nonspecial divisors or equivalently by points in general position
on the Jacobian variety $J(\Sigma)$.  For more in this spirit we refer
to work on the Schottky problem and the Novikov conjecture in \cite
{ca,dg,mb,sa}.  The flow variables $x,y,t$ arise in ({\bf A}) via
$q(k)$ and then miraculously reappear in the theta function via
$xU+yV+tW$; thus the Riemann surface itself contributes to establish
these as linear flow variables on the Jacobian and in a certain sense
defines the flow variables.
The pole positions
$P_i$ do not vary with $x,y,t$ and $(\dagger)\,\,
u = 2\partial^2_x log\Theta(xU + yV  + tW + z_0) + c$
exhibits $\theta$ as a tau function (see \cite{dg} for $c$).
\\[3mm]\indent
Now a divisor $D^{*}$ of degree $g$ is dual to $D$ (relative to $Q$) if
$D + D^{*}$ is the null divisor of a meromorphic differential $\Omega
= dk + (\beta/k^2)dk + \cdots$ with a double pole at $Q$ (look at
$\zeta = 1/k$ to recognize the double pole).  Thus
$D + D^{*} -2Q\sim K_{\Sigma}$ so ${\cal A}(D^{*}) - {\cal A}(Q) + K =
-[{\cal A}(D) - {\cal A}(Q) + K]$.  One can define then a function
$\psi^{*}(x,y,t,P) = exp(-kx-k^2y-k^3t)[1 + \xi_1^{*}/k) + \cdots]$
based on $D^{*}$ (dual BA function, $\not= \psi^{\dagger}$
- see Section 6.8),
and a differential $\Omega$ with zero divisor $D+D^*$, such that
$\phi = \psi\psi^{*}\Omega$ is
meromorphic, having for poles only
a double pole at $Q$ (the zeros of $\Omega$ cancel
the poles of $\psi\psi^{*}$).  Thus $\psi\psi^*\Omega\sim \psi\psi^*(1+
(\beta/k^2+\cdots)dk$ is meromorphic with a second order pole at $\infty$,
and no other poles.
For $L^{*} = L$ and $A^{*} = -A + 2w
-(3/2)u_x$ one has then $(\partial_y + L^{*})\psi^{*} = 0$ and
$(\partial_t + A^{*})\psi^{*} = 0$.  Now
the prescription above seems to specify for $\psi^*$
($\vec{U}=xU+yV+tW,\,\,z_0^* =
-{\cal A}(D^*)-K$)
\be
\psi^*\sim e^{-\int^P_{P_o}(x\Omega_1+y\Omega_2+t\Omega_3)}
\cdot\frac{\theta({\cal A}(P)-\vec{U}+z_0^*)}
{\theta({\cal A}(P)+z_0^*)}
\label{star}
\ee
\indent
Various further hypotheses on $\Sigma$ lead to the traditional KP situations
(KP-1 or KP-2) or to various reductions (e.g. nKdV).  We mention again
especially \cite{da,db,dc,dd,de,ka,kc,ke,na} and also some recent work
on 2-D periodic problems as in \cite{kb,kf,kg,kh}.  One is generally
interested in real smooth solutions and in this respect KP-1 ($\sim
x_{2j}$ imaginary, $x_2 = y$) presents problems - it is not even formally
integrable in general (cf. \cite{kb,kc} for details).  We will not discuss
general theory here but always assume we have some integrable (finite
zone) situation related to an appropriate Riemann surface (and perhaps
periodic potentials).  The analysis relative to
asymptotics and averaging will be cursory and formal at times, in keeping
with "tradition" here, but we have tried to be somewhat more precise
(there are of course some precise theorems in the
literature but a discussion of these would often involve
more hypotheses and detail than appropriate for this
exposition).  In particular we try to clarify some of Krichever's work
which can be difficult to understand at times.
Our interest
is basically to sketch a framework, with formulas and results relative to
Riemann surfaces, which is sufficiently detailed to be believable, and
in which some meaningful subset of a full theory can be
displayed.

\section{THE WHITHAM EQUATIONS AND AVERAGING}
\renewcommand{\theequation}{3.\arabic{equation}}\setcounter{equation}{0}

Averaging can be rather mysterious at first due to some hasty treatments
and bad choices of notation - plus many inherent difficulties.  Some
of the clearest exposition seems to be in \cite{aa,ba,cb,fa,fd,mc,wa}
whereas the more extensive developments in e.g.
\cite{da,db,dd,de,df,gb,ka,kb,kd,ke,na,nb} may be confusing at first,
until one realizes what is going on.  We will try to be careful and distinguish
situations with periodic potentials from more general averaging
situations involving ergodic ideas.

\subsection{Remark on PDE connections}
One choice of background situation
involves an examination of dispersionless limits with no concern for
periodicity or Riemann surfaces (cf. here
\cite{ae,ch,ci,cl,ta}).  Thus given
e.g. a KdV equation
$u_t + 6uu_x + u_{xxx} = 0$ set $\epsilon x = X$ and $\epsilon t = T$
leading to $u^{\epsilon}_T + 6u^{\epsilon}u^{\epsilon}_X +
\epsilon^2 u^{\epsilon}_{XXX} = 0$.  The study of $u^{\epsilon}\to
\hat{u}$ where $\hat{u}$ satisfies the Euler equation $\hat{u}_T + 6
\hat{u}\hat{u}_X = 0$ is a very delicate matter, of great interest in
applications and in PDE but we do not discuss this here (cf. \cite{la}).
On the other hand the purely algebraic
passage of the background mathematics of KdV (involving Lax operators,
the KdV hierarchy, tau functions, vertex operators, etc.) to the
corresponding background mathematics of the dispersionless theory, is
relatively easy and will be indicated below.
Moreover it is of great importance in an entirely
different direction, namely in the study of
topological field theory, strings, and 2-D gravity (see e.g.
\cite{ae,ch,ci,cl,dl,di,dj,dk,kn,km,kc,kj,kl,ta,vb}).
We will insert such material later as appropriate.

\subsection{Preliminary ideas}
Let us follow \cite{kc} (cf. also \cite{kb}) in order to have a
suitably complicated example leading directly to matters of interest here,
so as background we consider the KP equation
\be
\frac{3}{4}\sigma^2u_{yy} = \partial_x[u_t -\frac{3}{2}uu_x +
\frac{1}{4}u_{xxx}]
\label{AC}
\ee
($\sigma^2 = 1\sim$ KP-2; $\sigma^2 = -1\sim$ KP-1).  Here $\sigma L
\sim \partial^2_x - u$ with $\sigma\partial_y\psi =
(\partial^2_x - u)\psi$ and $A\sim
\partial^3_x - (3/2)u\partial_x + w$, which is slightly different than
before (to connect notations let $u\to -u$ and insert $\sigma$).  Then
one has the compatibility equations $[\partial_y - L,\partial_t - A] = 0$
as before and it should be noted that for any function $g(t),\,\,t\sim
x,y,t,t_4,\cdots$, operators $\tilde{L} = gLg^{-1} + \partial_yg\,g^{-1}$
and $\tilde{A} = gAg^{-1} + \partial_t g\,g^{-1}$ could be used,
corresponding to a new wave function $\tilde{\psi} = g\psi$.  The notation
of \cite{kc} also involves differentials $\Omega_i\sim\Omega^i$ where
$\Omega_i = dk^i(1 + O(1/k)]$ and $\int_{a_k}\Omega_j = 0$ with
$U^k_j = (1/2\pi i)\int_{b_k}\Omega_j$ (so $U^k_1\sim U_k,\,\,U^k_2\sim
V_k$, and $U^k_3\sim W_k$, up to factors of $2\pi i$).  There will be solutions
$\psi$ as before in (\ref{AA}) with potentials $u$ given via $(\dagger)$.
We recall that Riemann theta functions are more precisely written as
$\Theta(z|B)$ where $B\sim$ a certain matrix (cf. below),
and $z\sim (z_1,\cdots,z_g)
\sim xU+yV+tW+z_0$ for example.  One knows via \cite{ca,dg,mb,sa} that
$u$ of the form $(\dagger)\,\,u = 2\partial^2_xlog\Theta(z|B) + c$ is a
solution of KP if and only if the matrix $B$ defining $\Theta$ is the
b-period matrix of a Riemann surface determined via $\int_{b_k}d\omega_j$
(with $z$ as defined).
Now consider the spectral theory of 2-D periodic operators (**) $(\sigma
\partial_y - \partial_x^2 + u(x,y))\psi = 0$ where $u(x,y) =
u(x+a_1,y) = u(x,y+a_2)$.  Bloch solutions are defined via
\be
\psi(x+a_1,y,w_1,w_2) =  w_1\psi(x,y,w_1,w_2);
\label{AD}
\ee
$$\psi(x,y+a_2,w_1,w_2) = w_2\psi(x,y,w_1,w_2)$$
and we assume $\psi(0,0,w_1,w_2) = 1$.  The pairs $Q=(w_1,w_2)$ for which
there exists such solutions is called the Floquet set $\Gamma$ and the
multivalued functions $p(Q)$ and $E(Q)$ such that $w_1 = exp(ipa_1)$ with
$w_2 = exp(iEa_2)$ are called quasi-momentum and quasi-energy respectively.
By a gauge transformation $\psi\to exp(h(y))\psi$, with $\partial_y
h(y)$ periodic one obtains solutions of (**) with a new potential
$\tilde{u} = u - \sigma\partial_y h$ so we can assume $\int_0^{a_1}
u(x,y)dx = 0$.

\subsection{Floquet theory}
For $M_0 = \sigma\partial_y
-\partial_x^2$ (with $u=0$) the Floquet set is parametrized by $k\in
{\bf C}$ such that $w_1^0 = exp(ika_1)$ and $w_2^0 = exp(-k^2a_2/\sigma)$
and the Bloch solutions are $\psi(x,y,k) = exp(ikx-k^2y/\sigma)$.
The adjoint Bloch solutions are $\psi^{+}(x,y,k) = exp(-ikx+k^2y/\sigma)$
satisfying $(\sigma\partial_y + \partial_x^2)\psi^{+} = 0$.  The image
of the map $k\to (w_1^0,w_2^0)\in{\bf C}^2$ is the Floquet set for $M_0$
corresponding to the Riemann surface with intersections corresponding
to pairs $k\not= k'$ such that $w_i^0(k) = w_i^0(k'),\,\,i = 1,2$.
This means $k-k' = (2\pi N/a_1)$ and $k^2-(k')^2 = (2\pi i\sigma M/a_2)$
where $N,\,M$ are integers, creating "resonant" points $k = k_{N,M} =
(\pi N/a_1) - (i\sigma Ma_1/Na_2),\,\,N\not= 0,\,\,k' = k_{-N,-M}$.
Then for $k_0\not=k_{N,M}$ and $u$ sufficiently small
one can construct a formal Bloch solution
of (**) in the form of a convergent perturbation series (for any
$\sigma$).  Thus outside of some neighborhoods of the resonant points
one can obtain a Bloch solution $\tilde{\psi}(x,y,k_0)$ which is
analytic in $k_0$, but the extension of $\tilde{\psi}$ to a resonant
domain can be tricky.  For $\Re\sigma = 0$ (as in KP-1)
the resonant points are dense
on the real axis whereas for $\Re\sigma\not= 0$ (as in KP-2)
there are only a finite
number of resonant points in any finite domain of {\bf C}.  In the latter
case one can glue handles between domains around resonant points
and create a Riemann surface $\Gamma$
of Bloch solutions $\psi(x,y,Q),\,\,Q\in\Gamma$.
Moreover, if the potential
$u$ can be analytically extended into a domain $|\Im x|<\tau_1,\,\,
|\Im y|<\tau_2$, then the technique can also
be adapted even when $u$ is not small.
Such Bloch solutions, normalized by $\psi(0,0,Q) = 1$, are meromorphic
on $\Gamma$ and in the case of a finite number of handles a one point
compactification of $\Gamma$ is obtained so that $\psi$ is in fact the
BA function for $\Gamma$.  Generally speaking finite zone situations
as in (\ref{AA}) with potentials given by Riemann
theta functions are quasi-periodic in nature.  To single out
conditions for periodicity one asks for meromorphic differentials
$dp$ and $dE$ on $\Gamma$ having their only singularities at $Q\sim$
point at $\infty$ of the form $dp = dk(1+O(k^{-2}))$ and $dE = i\sigma^{-1}
dk^2(1 + O(k^{-3}))$, normalized so that all periods are real, and
satisfying, for any cycle $C$ on $\Gamma$, $\oint_Cdp = (2\pi n_C/a_1)$
with $\oint_CdE = (2\pi m_C/a_2)$ where $n_C,\,m_C$ are integers.
Then the corresponding potentials $u(x,y)$ will have periods $a_1$ and
$a_2$ in $x$ and $y$, with multipliers $w_1(P) = exp(ia_1\int^P dp)$
and $w_2(P) = exp(ia_2\int^P dE)$.
We go now to finite zone (or quasiperiodic) situations as in (\ref{AA})
with potentials as in $(\dagger)$, where $z\sim xU + yV + tW
+ z_0$ can be written as $(z_k) = [x\int_{b_k}\Omega^1 + y\int_{b_k}
\Omega^2 + t\int_{b_k}\Omega^3 + z^0_k] = (\zeta_k + z^0_k),\,\,
k = 1,\cdots,g$.  Since $\Theta(z+2\pi iN) = \Theta(z)$ we could set
$\zeta = i\theta$ so that as a function of $\theta = (\theta_k),\,\,u$ is
periodic of period $2\pi$ in each variable $\theta_k$.  Now one wants to
consider modulated finite zone situations where solutions are of the form
$u = u_0(xU+yV+tW|I) = u_0(\theta_1,\cdots,\theta_g|I_1,\cdots,I_n)$ where
$u_0$ is periodic in the $\theta_j$ with $U,V,W = U,V,W(I)$.  One assumes
there will be slow variables $X = \epsilon x,\,\,Y = \epsilon y,$ and
$T = \epsilon t$ with fast variables $x,y,t$ so that "asymptotic" solutions
\be
u = u_0(\frac{1}{\epsilon}S(X,Y,T)|I(X,Y,T)) + \epsilon u_1(x,y,t) +
\epsilon^2u_2(x,y,t) + \cdots
\label{AE}
\ee
can be envisioned.  In practice the parameters $I_k$ depend on the
moduli of our Riemann surface, which are then allowed to change
smoothly with the slow variables $X,Y,T$ (so $U,V,W$ also can change),
and one looks for solutions (\ref{AE}) with uniformly bounded $u_1$.
We do not look for convergence of the series in (\ref{AE}) nor check
any other features of ``asymptotic solution" or ``asymptotic series".
Such procedures are standard in the study
of what are called weakly deformed soliton
lattices.

\subsection{KdV averaging}
We will motivate this study with
examples from KdV, where one will see explicitly the nature of things.
First from \cite{mc}, in a slightly different notation,
write $q_t = 6qq_x - q_{xxx}$ with Lax pair
$L = -\partial^2_x + q,\,\,B = -4\partial^3_x + 3(q\partial_x + \partial_x q),
\,\,L_t = [B,L],\,\,L\psi = \lambda\psi,$ and $\psi_t = B\psi$.
Let $\psi$ and $\phi$ be two solutions of the Lax pair equations and
set $\Psi = \psi\phi$; these are the very important ``square eigenfunctions"
which arise in many ways with interesting and varied meanings.  Evidently
$\Psi$ satisfies
\be
[-\partial^3_x + 2(q\partial_x+\partial_x q)]\Psi = 4\lambda\partial_x\Psi;
\,\,\partial_t\Psi = -2q_x\Psi + 2(q+2\lambda)\partial_x\Psi
\label{AF}
\ee
{}From (\ref{AF}) one finds immediately the conservation law ({\bf C}):
$\partial_t[\Psi] + \partial_x[6(q-2\lambda)\Psi - 2\partial^2_x\Psi] = 0$.
If one looks for solutions of (\ref{AF}) of the form $\Psi(x,t,\lambda) =
1 + \sum_1^{\infty}[\Psi_j(x,t)]\lambda^{-j}$ as $\lambda\to\infty$ then
one obtains a recursion relation for polynomial densities
\be
\partial_x\Psi_{j+1} = [-\frac{1}{2}\partial^3_x + (q\partial_x +
\partial_x q)]\Psi_j\,\,(j = 1,2,\cdots);\,\,\Psi_0 = 1
\label{AG}
\ee
Now consider the operator $L = -\partial^2_x + q$ in $L^2(-\infty,\infty)$
with spectrum consisting of closed intervals separated by exactly $N$ gaps
in the spectrum.  The $2N+1$ endpoints $\lambda_k$
of these spectral bands are
denoted by $-\infty<\lambda_0<\lambda_1<\cdots<\lambda_{2N}<\infty$
and called the
simple spectrum of $L$.  They can be viewed as constants of motion for
KdV when $L$ has this form.  We are dealing here with the hyperelliptic
Riemann surface determined via $R^2(\lambda)=\prod_0^{2N}(\lambda - \lambda_k)$
and one can think of a manifold ${\cal M}$ of $N$-phase waves with
fixed simple spectrum as an $N$-torus based on $\theta_j\in [0,2\pi)$.
Hamiltonians in the KdV hierarchy generate flows on this torus and one
writes $q = q_N(\theta_1,\cdots,\theta_N)$ (the $\theta$ variables
originate as in our previous discussion if we use theta functions
for the integration - cf. also below).
Now there is no $y$ variable so let us write
$\theta_j = x\kappa_j+tw_j$ (we will continue to use $d\omega_j$ for
normalized holomorphic differentials)
For details concerning the Riemann surface we refer to \cite{ca,fa} and
will summarize here as follows.  For any $q_N$ as indicated one can find
functions $\mu_j(x,t)$ via $\Psi(x,t,\lambda) = \prod_1^N(\lambda -
\mu_j(x,t))$ where $\mu_j(x,t)\in [\lambda_{2j-1},\lambda_{2j}]$ and
satisfies
\be
\partial_x\mu_j = -2i(R(\mu_j)/\prod_{i\not= j}(\mu_j-\mu_i));
\label{AH}
\ee
$$\partial_t\mu_j = -2i[2(\sum_0^{2N}\lambda_k - 2\sum_{i\not= j}\mu_i)]
\cdot (R(\mu_j)/\prod_{i\not= j}(\mu_j - \mu_i)$$
In fact the $\mu_j$ live on the Riemann surface of $R(\lambda)$
in the spectral gaps
and as $x$ increases $\mu_j$ travels from $\lambda_{2j-1}$ to $\lambda_{2j}$
on one sheet and then returns to $\lambda_{2j-1}$ on the other sheet; this
path will be called the $j^{th}\,\mu$-cycle ($\sim a_j$).
In the present context we
will write the theta function used for integration purposes as
$\Theta({\bf z},\tau) = \sum_{m\in {\bf Z}^N}exp[\pi i(2({\bf m},{\bf z})
+ ({\bf m},\tau{\bf m}))]$ where ${\bf z}\in {\bf Z}^N$ and $\tau$
denotes the $N\times N$ period matrix ($\tau$ is symmetric with $\Im\tau
> 0$).  We take canonical cuts $a_i,\,b_i\,\,(i = 1,\cdots,N)$ where
$a_j\sim (\lambda_{2j-2},\lambda_{2j-1})$ as in \cite{fa} (where a picture
is drawn).  Let $d\omega_j$ be holomorphic diffentials normalized via
$\int_{a_j} d\omega_k = \delta_{jk}$ (the cycle $a_j$ corresponds to
a loop around the cut $a_j$).  Then $q_N$ can be represented in the form
\be
q_N(x,t) = \Lambda + \Gamma -2\partial^2_xlog\Theta({\bf z}(x,t);\tau);\,\,
\Lambda = \sum_0^{2N}\lambda_j;
\label{AI}
\ee
$$\tau = (\tau_{ij}) = (\oint_{b_i}d\omega_j);\,\,\tau^{*}_{ij} = -\tau_{ij};
\,\,\Gamma = -2\sum_1^N\oint_{a_j}\lambda d\omega_j$$
and ${\bf z}(x,t) = -2i[{\bf c}^N(x-x_0) + 2(\Lambda{\bf c}^N +
2{\bf c}^{N-1})t] + {\bf d}$ where $({\bf c}^N)_i = c_{iN}$ arises
from the representation $d\omega_i =
(\sum_1^N c_{ij}\lambda^{j-1})[d\lambda/R(\lambda)]$ (${\bf d}$ is a
constant whose value is not important here).  Then the wave number
and frequency vectors can be defined via $\vec{\kappa} = -4i\pi
\tau^{-1}{\bf c}^N$ and $\vec{w} = -8i\pi\tau^{-1}[\Lambda
{\bf c}^N + 2{\bf c}^{N-1}]$ with $\theta_j(x,t) = \kappa_j x
+ w_j t + \theta_j^0$ (where the $\theta_j^0$ represent initial phases).
\\[3mm]\indent
To model the modulated wave now one writes
now $q = q_N(\theta_1,\cdots,\theta_N;\vec{\lambda})$ where $\lambda_j
\sim\lambda_j(X,T)$ and $\vec{\lambda}\sim(\lambda_j)$.
Then consider the first $2N+1$ polynomial
conservation laws arising from (\ref{AF}) - (\ref{AG})
and ${\bf C}$ for example (cf. below for KP)
and write these as $\partial_t{\cal T}_j(q)
+ \partial_x{\cal X}_j(q) = 0$ (explicit formulas are given in
\cite{fa} roughly as follows).
We note that the adjoint linear KdV equation (governing
the evolution of conserved densities)
is $\partial_t\gamma_j + \partial^3_x\gamma_j - 6q\partial_x
\gamma_j = 0\,\,(\gamma_j\sim\nabla H_j)$ and (\ref{AG}) has the form
$\partial\gamma_{j+1} = (-(1/2)\partial^3 + q\partial +\partial q)\gamma_j$.
One then rewrites this to show that $6q\partial_x\gamma_j = \partial_x
[6\gamma_{j+1}-6q\gamma_j+3\partial^2\gamma_j]$ so that the adjoint
equation becomes
\be
\partial_t\gamma_j +\partial[-2\partial^2\gamma_j + 6q\gamma_j
-6\gamma_{j+1}]=0
\label{NF}
\ee
which leads to (\ref{AJ}) and (\ref{AN}) below (after simplification
of (\ref{NF})).
Then comes
the crucial averaging step.  Keep the slow variables $X,T$ constant
and average over the fast variable $x$ to obtain
\be
\partial_T<{\cal T}_j(q_N)> + \partial_X<{\cal X}_j(q_N)> = 0
\label{AJ}
\ee
(note e.g. $\partial_t = \epsilon\partial_T$).
The procedure involves averages
\be
<{\cal T}_j(q_N)> = lim_{L\to\infty}\frac{1}{2L}
\int_{-L}^L{\cal T}_j(q_N)dx
\label{AK}
\ee
for example (with a similar expression for $<{\cal X}_j(q_N)>$) and an
argument based on ergodicity is used.  Thus if the wave numbers
$\kappa_j$ are incommensurate the trajectory $\{q_N(x,t);\,\,x\in (-\infty,
\infty)\}$ will densely cover the torus ${\cal M}$.  Hence we can replace
$x$ averages with
\be
<{\cal T}_j(q_N)> = \frac{1}{(2\pi)^N}\int_0^{2\pi}\cdots\int_0^{2\pi}
{\cal T}_j(q_N(\vec{\theta}))\prod_1^Nd\theta_j
\label{AL}
\ee
For computational purposes one can change the $\theta$ integrals to
$\mu$ integrals and obtain simpler calculations. By this procedure
one obtains a system of $2N+1$ first order partial differential equations
for the $2N+1$ points $\lambda_j(X,T)$, or equivalently for the
physical characteristics $(\vec{\kappa}(X,T),\vec{w}(X,T))$ (plus
$<q_N>$).
\\[3mm]\indent
The above argument may or may not
have sounded convincing but it was in any case very loose.
Let us be more precise following \cite{fa}.  One looks at the KdV
Hamiltonians beginning with $H = H(q) = lim_{L\to\infty}(1/2L)\int^L_
{-L}(q^2 + (1/2)q_x^2)dx$ (this form is appropriate for quasi-periodic
situations).  Then $q_t = \{q,H\}$ where $\{f,g\} = lim_{L\to\infty}
(1/2L)\int^L_{-L}(\delta f/\delta q)\partial_x(\delta g/\delta q)dx$
(averaged Gardner bracket).  The other Hamiltonians are found via
\be
\partial\frac{\delta H_{m+1}}{\delta q} = (q\partial + \partial q -
\frac{1}{2}\partial^3)\frac{\delta H_m}{\delta q}\,\,(m\geq 0);
\,\,\frac{\delta H_0}{\delta q} = 1
\label{AM}
\ee
where $\gamma_j\sim\nabla H_j\sim(\delta H_j/\delta q)$
(cf. here \cite{ca,fa}).  It is a general situation in the study of
symmetries and conserved gradients (cf. \cite{cd}) that symmetries
will satisfy the linearized KdV equation $(\partial_t -6\partial_x q +
\partial^3_x)Q = 0$ and conserved gradients will satisfy the adjoint
linearized KdV equation $(\partial_t - 6q\partial_x + \partial^3_x)Q^{\dagger}
=0$; the important thing to notice here is that one is linearizing about
a solution $q$ of KdV.  Thus in our averaging processes the function
$q$, presumed known, is inserted in the integrals.  This leads then to
\be
{\cal T}_j(q) = \frac{\delta H_j}{\delta q};\,\,{\cal X}_j(q) =
-2\partial^2_x\frac{\delta H_j}{\delta q} - 6\frac{\delta H_{j+1}}{\delta q}
+ 6q\frac{\delta H_j}{\delta q}
\label{AN}
\ee
with (\ref{AJ}) holding, where $<\partial^2\phi>=0$ implies
\be
<{\cal X}_j> = lim_{L\to\infty}\frac{1}{2L}\int^L_{-L}(-6\frac
{\delta H_{j+1}}{\delta q_N} + 6q_N\frac{\delta H_j}{\delta q_N})dx
\label{AO}
\ee
Note that $\partial_x$ appears in $\chi_j$ for example but in (\ref{AJ})
$\partial_x\to\epsilon\partial_X$.  This is somewhat confusing but
apparently gives correct first order terms in $\epsilon$, after which
$\epsilon$ is cancelled out and eventually allowed to approach $0$.
In \cite{fa} the integrals are then simplified in terms of $\mu$ integrals
and expressed in terms of abelian differentials.  This is a beautiful
and important procedure linking the averaging process to the Riemann
surface and is summarized in
\cite{mc} as follows (see also below for more details and a KP
version).
One defines differentials
\be
\hat{\Omega}_1 = -\frac{1}{2}[\lambda^N - \sum_1^Nc_j\lambda^{j-1}]\frac
{d\lambda}{R(\lambda)}
\label{AP}
\ee
$$
\hat{\Omega}_2 = [-\frac{1}{2}\lambda^{N+1} + \frac{1}{4}(\sum\lambda_j)
\lambda^N + \sum_1^N E_j\lambda^{j-1}]\frac{d\lambda}{R(\lambda)}
$$
where the $c_j,\,\,E_j$ are determined via
$\oint_{b_i}\hat{\Omega}_1 = 0=\oint_{b_i}\hat{\Omega}_2\,\,(i = 1,
2,\cdots,N)$.  Then it can be shown that
\be
<\Psi>\sim
<{\cal T}>\sim\sum_0^{\infty}\frac{<{\cal T}_j>}{(2\mu)^j};\,\,
<{\cal X}>\sim\sum_0^{\infty}\frac{<{\cal X}_j>}{(2\mu)^j}
\label{NZ}
\ee
with
$\hat{\Omega}_1\sim<{\cal T}>(d\xi/\xi^2)$ and $<{\cal X}>(d\xi/\xi^2)\sim
12[(d\xi/\xi^4)-\hat{\Omega}_2]$ where $\mu=\xi^{-2}\to\infty\,\,(\mu\sim
(1/\sqrt{\xi})^{1/2}$) so $d\mu = -2\xi^{-3}d\xi\Rightarrow(d\xi/\xi^2)
\sim -(\xi/2)d\mu\sim-(d\mu/2\sqrt{\mu})$.  Since $\hat{\Omega}_1 =
O(\mu^N/\mu^{N+(1/2)})d\mu = O(\mu^{-(1/2)}d\mu,\,\,
\hat{\Omega}_2 =
O(\mu^{1/2})d\mu$ (with lead term $-(1/2)$) we obtain
$<\Psi>\sim <{\cal T}>=O(1)$ and $<{\cal X}> = O(1)$.
Thus
(\ref{AF}), (\ref{AG}), ({\bf C}) generate all conservation laws
simultaneously with $<{\cal T}_j>$ (resp. $<{\cal X}_j>$) giving rise to
$\hat{\Omega}_1$ (resp. $\hat{\Omega}_2$).
It is then proved that all of the modulational
equations are determined via the equation
\be
\partial_T\hat{\Omega}_1 = 12\partial_X\hat{\Omega}_2
\label{AR}
\ee
where the Riemann surface is thought of as depending on $X,T$ through
the points $\lambda_j(X,T)$.
In particular if the first $2N+1$ averaged conservation laws are satisfied
then so are all higher averaged conservation laws.  These equations
can also be written directly in terms of the $\lambda_j$ as Riemann
invariants via $\partial_T\lambda_j = S_j\partial_X\lambda_j$ for
$j = 0,1,\cdots,2N$ where $S_j$ is a computable characteristic speed.
Thus we have displayed the prototypical model for the Whitham or
modulational equations.

\subsection{Extension to KP}
Given some knowledge of symmetries as sketched in \cite{cd} for example
one is tempted to rush now to an immediate attempt at generalizing the
preceeding results to finite zone KP via the following facts.  The KP
flows can be written as $\partial_nu = K_n(u)$ where the $K_n$ are
symmetries satisfying (in the notation of \cite{cd})
the linearized KP equation $\partial_3\beta = (1/4)\partial^3\beta
+ 3\partial(u\beta) + (3/4)\partial^{-1}\partial_2^2\beta = K'[\beta]$.
The conserved densities or gradients
$\gamma$ satisfy the adjoint linearized KP
equation $\partial_3\gamma = (1/4)\partial^3\gamma + 3u\partial\gamma
+ (3/4)\partial^{-1}\partial_2^2\gamma$.  Then, replacing the square
eigenfunctions by $\psi\psi^{*}$ one has e.g. $\psi\psi^{*} = \sum_0^
{\infty}s_n\lambda^{-n}$ where $s_n\sim$ a $\gamma_n$.
Further $\partial_nu
= K_{n+1} = \partial s_{n+1} = \partial Res\,L^n = \partial\nabla\hat
{I}_{n+1}$ where $\nabla f\sim\delta f/\delta u$
($Res\,L^n = nH^1_{n-1}$ is generally used in the multipotential
theory).  We are working here in a single potential theory where all
potentials $u_i$ in $L = \partial + \sum_1^{\infty}u_{i+1}\partial^{-i}$
are expressed in terms of $u_2=u$ via operators with $\partial$ and
$\partial^{-1}$.  One uses here the Poisson bracket $\{f,g\} =
\int\int(\delta f/\delta u)\partial(\delta g/\delta u)dxdy$ (Gardner
bracket).   Let us retrace
the argument from \cite{fa} and see what applies for KP.  Thus
one has $s_{n+1}\sim\gamma_{n+1}\sim\nabla\hat{I}_{n+1}$
as conserved gradients satisfying the adjoint linear KP equation
(**) $\partial_t\gamma = (1/4)\partial^3\gamma + 3u\partial\gamma
+(3/4)\partial^{-1}\partial^2_y\gamma$.  The nonlocal term $\partial^{-1}$ here
could conceivably change some of the analysis.
We need first a substitute for (\ref{AG}) or
else a direct way of rewriting the adjoint equation as $\partial_t[A]
+\partial_x[B]=0$ or perhaps we want rather $\partial_t[A]+\partial_x[B]
+\partial_y[C] =0$.  To get such a formula we can
simply differentiate the adjoint KP equation
to get
\be
\partial_t[\partial\gamma] +\partial[\frac{1}{4}\partial^3\gamma +
3u\partial\gamma]+\partial_y[\frac{3}{4}\partial_y\gamma]=0
\label{NG}
\ee
This removes all of the nonlocal terms and one doesn't have to deal with
$\partial[(3/4)\partial^{-1}\partial_y^2\gamma]$ for example.
Thus imagine a finite zone situation and write ($\gamma_j
\sim\nabla\hat{I}_j$)
\be
{\cal D}_j = \partial\gamma_j;\,\,{\cal F}_j = \frac{1}{4}\partial^3
\gamma_j+3u\partial\gamma_j;\,\,{\cal G}_j = \frac{3}{4}\partial_y\gamma_j
\label{NH}
\ee
so (\ref{NG}) $\sim\partial_t[{\cal D}_j]+\partial[{\cal F}_j] +
\partial_y[{\cal G}_j] = 0$.  One linearizes around a fixed finite zone
solution $u$ in the adjoint linear KP equation and puts this $u$ into
the $\hat{I}_j$ etc.  Then average in (\ref{NG}) over the $\theta_i$
variables as before to obtain
\be
\partial_T<{\cal D}_j> + \partial_X<{\cal F}_j> +\partial_Y<{\cal G}_j>=0
\label{NI}
\ee
Then modeled on KdV
one expects the terms in (\ref{NI}) to be expressible in terms of
differentials but there is remarkably little information in this direction.
First we expect $<\partial\gamma_j>=0$ and $<\partial_y\gamma_j>=0$ if e.g.
$\partial_y\gamma_j = \sum Y_i(\partial\gamma_j/\partial\theta_i)$ and
$\partial\gamma_j = \sum X_i(\partial\gamma_j/\partial\theta_i)$
(see the analysis below).  Then
$\partial_X<{\cal F}_j> = 0$ ensues which is not very thrilling
(and this would be the situation indicated by
(\ref{YD}) - (\ref{YE}) below).
Let us try then
\be
\partial_t[\gamma] + \partial[\frac{1}{4}\partial^2\gamma+3\partial^{-1}
u\partial\gamma] +\partial_y[\frac{3}{4}\partial_y\partial^{-1}\gamma]=0
\label{NJ}
\ee
Here we expect $<\partial^2\gamma> = 0 =<\partial_y\partial^{-1}\gamma>$
so set ${\cal L} = 3\partial^{-1}(u\partial\gamma)$ with
${\cal L}_j = 3\partial^{-1}(u\partial\gamma_j)$ and deduce that
\be
\partial_T<\gamma> +\partial_X<{\cal L}> =0;\,\,
\partial_T<\gamma_j> + \partial_X<{\cal L}_j> = 0
\label{NK}
\ee
Then one wants to express $<\gamma>$ (resp.
$<\gamma_j>$) and $<{\cal L}>$ (resp. $<{\cal L}_j>$) in terms of
differentials
(and since $<\gamma>=<\psi^*\psi>$ one has some connection to Section 3.4).
However order of magnitude considerations
(see below) suggest that $<\partial^2\gamma>=0
=<\partial_y\partial^{-1}\gamma>$ must be false - due to growth
or whatever - and we take
\be
<\hat{{\cal L}}> = <\frac{1}{4}\partial^2\gamma+3\partial^{-1}(u_0\partial
\gamma)>
\label{NQ}
\ee
with $\hat{{\cal G}} = (3/4)\partial_y
\partial^{-1}\gamma$.  Then
\be
\partial_T<\gamma> + \partial_X<\hat{{\cal L}}> + \partial_Y<\hat{{\cal G}}>
=0
\label{NS}
\ee
which conceivably might be useful.  These matters will be covered after
we develop another approach.

\subsection{Averaging with $\psi^*\psi$ \`a la Krichever}
Let us therefore look at \cite{ka} but in the spirit of \cite{fb}
(cf. also \cite{fc}) which is more carefully done at times.
We will expand upon this with some modifications in order to obtain
a visibly rigorous procedure.  Thus
consider KP in the form (\ref{AC}): $3\sigma^2u_{yy} + \partial_x(4u_t -
6uu_x + u_{xxx}) = 0$ via compatibility $[\partial_y -L,\partial_t - A]
=0$ where $L = \sigma^{-1}(\partial^2-u)$ and $A = \partial^3 - (3/2)
u\partial + w\,\, (\sigma^2 = 1$ is used in \cite{fb} which we follow for
convenience but the procedure should work in general with minor
modifications - note $\partial$ means $\partial_x$).  We have then
$(\partial_y - L)\psi = 0$ with $(\partial_t -A)\psi = 0$ and for the
adjoint or dual wave function
$\psi^{*}$ one writes in \cite{ka} $\psi^{*}L = -\partial_y\psi^{*}$
with $\psi^{*}A = \partial_t\psi^{*}$ where $\psi^{*}(f\partial^j)\equiv
(-\partial)^j(\psi^{*}f)$.  The explicit formulas used in \cite{ka}
(and \cite{fb}) are
\be
\psi = e^{px+Ey+\Omega t+\sum_1^{2g}\sigma_i\zeta_i}\cdot\phi(Ux+Vy
+Wt+\zeta,P)
\label{YD}
\ee
\be
\psi^* = e^{-px-Ey-\Omega t-\sum_1^{2g}\sigma_i\zeta_i}\cdot\phi^*
(-Ux-Vy-Wt-\zeta,P)
\label{YE}
\ee
where $\sigma_i=\sigma_i(P),\,\,p=p(P),\,\,E = E(P),\,\,\Omega =
\Omega(P),$ etc. (see below for more detail).  We will assume these to
be correct although this is not really obvious since the proof in
\cite{ka} is rather ``heuristic" and the form is not immediately
compatible with (\ref{AA}).
However the arguments to follow are essentially independent of this
choice of notation.
Now one sees immediately that
\be
(\psi^{*}L)\psi = \psi^{*}L\psi + \partial_x(\psi^{*}L^1\psi) +
\partial^2_x(\psi^{*}L^2\psi) + \cdots
\label{AS}
\ee
where e.g. $L^r = (-1)^r/r!)(d^r L/d(\partial)^r$.  In particular
\be
\psi^*_{xx}\psi =\psi^*\psi_{xx} -2\partial(\psi^*\psi_x) +
\partial^2(\psi^*\psi)
\label{YG}
\ee
\be
-\psi^*_{xxx}\psi = \psi^*\psi_{xxx} -\partial^3(\psi^*\psi)
+3[\partial^2(\psi^*\psi_x) - \partial(\psi^*\psi_{xx})]
\label{YH}
\ee
This means ($L^*=L,\,\,\psi^*A\sim A^*\psi^* =-\partial^3\psi^*
+(3/2)\partial(\psi^*u)+w\psi^*$)
\be
(\psi^*L)\psi = [(\partial^2-u)\psi^*]\psi = \psi^*_{xx}\psi -u\psi^*\psi
=\psi^*L\psi -2\partial(\psi^*\psi_x) +\partial^2(\psi^*\psi)
\label{YI}
\ee
which implies $L^1 = -2\partial$ and $L^2 = 1$.  Next
\be
(\psi^*A)\psi = -\psi^*_{xxx}\psi +\frac{3}{2}(\psi^*_xu+\psi^*u_x)\psi
+w\psi^*\psi = \psi^*(A\psi) +
\label{YJ}
\ee
$$+ \partial[\psi^*(\frac{3}{2}u-3\partial^2)\psi] + 3\partial^2(\psi^*\psi_x)
-\partial^3(\psi^*\psi)$$
so that $A^1 = -3\partial^2+(3/2)u,\,\,A^2 = 3\partial,$ and $A^3 =
-1$.
We think of a general Riemann surface $\Sigma_g$.  Here one picks
holomorphic differentials
$d\omega_k$ as before and quasi-momenta, quasi-energies, etc. via
$dp\sim\Omega_1,\,\,dE\sim\Omega_2,\,\,d\Omega\sim\Omega_3,\cdots$
where $\lambda\sim k,\,\,p = \int^P_{P_0}\Omega_1,$ etc.),
and use $U^k_1\sim U_k,\,\,U_2^k\sim V_k,$ and $U^k_3\sim W_k$ (general
$\Omega_k\sim\Omega^k$ are discussed at various places
in the text).  Normalize
the $\Omega_k$ now so that $\Re\int_{a_i}\Omega_k = 0 = \Re\int_{b_j}
\Omega_k$ and take then e.g. $U\sim (U_1^k,-(1/2\pi i)\int_{a_m}\Omega_1)
\,\,(k,m = 1,\cdots,g)$ so that $U,\,V,\,W$ are real $2g$ period vectors
(this is essentially equivalent to previous normalizations, e.g.
$\int_{a_m}\Omega_k = 0$).  Then one has BA functions $\psi(x,y,t,P)$ as
in (\ref{YD}) or (\ref{YE}).  As before we look for approximations
as in (\ref{AE}) based on
$u_0(xU+yV+tW|I) = u_o(\theta_j,I_k)$.  For averaging $\theta_j \sim
xU_j + yV_j + tW_j+ \zeta_j,\,\,1\leq j\leq 2g$,
with period $2\pi$ in the
$\theta_j$ seems natural (but note
$\theta_j,\,\theta_{g+j}\sim U_j,$ etc. - cf. (\ref{KB})).
Then again by ergodicity
$<\phi>_x = lim_{L\to\infty}(1/2L)\int_{-L}^L\phi dx$ becomes
$<\phi> = (1/(2\pi)^{2g}\int\cdots\int\phi d^{2g}\theta$ and one
notes that $<\partial_x\phi> = 0$ automatically
for $\phi$ bounded.  In \cite{fb} one thinks
of $\phi(xU+\cdots)$ with $\phi_x = \sum U_i(\partial\phi/\partial\theta_i)$
and $\int\cdots\int(\partial
\phi/\partial\theta_i)d^{2g}\theta = 0$.
\\[3mm]\indent
Now for averaging we think of $u_0\sim u_0(\frac{1}{\epsilon}S|I)$ as
in (\ref{AE}) with $S,I\sim S,I(X,Y,T),
\newline
\partial_X S = U,\,\,\partial_Y
S = V,$ and $\partial_T S = W$.  We think of expanding
about $u_0$ with $\partial_x\to\partial_x + \epsilon\partial_X$.
This step will cover both $x$ and $X$ dependence for subsequent averaging.
Then look at the compatibility condition $(\ddagger):\,\,\partial_tL
-\partial_y A + [L,A] = 0$.  As before we will want the term of first
order in $\epsilon$ upon writing e.g. $L = L_0 + \epsilon L_1 + \cdots$
and $A = A_0 + \epsilon A_1 + \cdots$ where $L_0,A_0$ are to depend
on the slow variables $X,Y,T$.
As indicated in \cite{ka} the term
$[L,A]\to \{L,A\}$ where $\{L,A\}$ arises upon replacing $\partial_x$
be $\partial_x + \epsilon\partial_X$ in all the differential expressions
and taking the terms of first order in $\epsilon$.  However the formula
in \cite{ka} is unclear so we compute some factors explicitly.
In fact,
according to \cite{fb}, one can
write now, to make the coefficient of $\epsilon$ vanish
\be
\partial_t L_1 - \partial_y A_1 + [L_0,A_1] + [L_1,A_0] + F = 0;\,\,
F=\partial_TL - \partial_YA + (L^1\partial_XA - A^1\partial_XL)
\label{AT}
\ee
Thus $F$ is the first order term involving derivatives in the slow
variables.
To clarify this let us write $(\hat{\partial}\sim\partial/\partial X)\,\,
L_{\epsilon} = (\partial +\epsilon
\hat{\partial})^2 - (u_0+\epsilon u_1+\cdots) = \partial^2-u_0 +\epsilon
(2\partial\hat{\partial}-u_1) +O(\epsilon^2)$ and $A_{\epsilon} =
(\partial+\epsilon\hat{\partial})^3 -(3/2)(u_0+\epsilon u_1+\cdots)\cdot
(\partial+\epsilon\hat{\partial}) + w_0+\epsilon w_1+\cdots = \partial^3
-(3/2)u_0\partial+w_0+\epsilon(3\partial^2\hat{\partial} -
(3/2)u_1\partial -(3/2)u_0\hat{\partial} + w_1) + O(\epsilon^2)$.  Write
then $A_{\epsilon} = A_0+\hat{A}_1\epsilon + O(\epsilon^2)$ and $L_{\epsilon}
= L_0 +\hat{L}_1\epsilon +O(\epsilon^2)$ with $\hat{A}_1 = 3\partial^2
\hat{\partial}-(3/2)u_1\partial-(3/2)u_0\hat{\partial} + w_1$ and
$\hat{L}_1 = 2\partial\hat{\partial}-u_1$.  Note that $L^1 = -2\partial\not=
\hat{L}_1$ and $A^1 = -3\partial^2+(3/2)u_0\not=\hat{A}_1$ but we can write
\be
\hat{A}_1 = -A^1\hat{\partial}-(3/2)u_1\partial+w_1 = -A^1\hat{\partial}
+A_1;
\label{YL}
\ee
$$\hat{L}_1 = -L^1\hat{\partial}-u_1 = -L^1\hat{\partial}+L_1$$
with $L_1$ and $A_1$ as in \cite{fb}.  Then $(\ddagger)$ becomes
\be
\partial_tL_0 -\partial_yA_0 +[L_0,A_0] +\epsilon\{\partial_tL_1
-\partial_yA_1 +\partial_TL-\partial_YA +[L_0,\hat{A}_1] +[\hat{L}_1,A_0]\}
+O(\epsilon^2)
\label{YK}
\ee
But for $u_0,\,w_0$ functions of the slow variables only, the $\partial_t
L_0,\,\,\partial_yA_0,\,\,[L_0,A_0]$ terms vanish and we note that
\be
[L_0,\hat{A}_1]+[\hat{L}_1,A_0] = [L_0,-A^1\hat{\partial}] +
[-L^1\hat{\partial},A_0] + [L_0,A_1]+[L_1,A_0] =
\label{YM}
\ee
$$= [L_0,A_1]+[L_1,A_0] +A^1\hat{\partial}L_0-L^1\hat{\partial}A_0
-L_0A^1\hat{\partial} +A_0L^1\hat{\partial}$$
Then one notes that $\partial_XL_0\sim\partial_XL$ and $\partial_XA_0
\sim\partial_XA\,\,(\partial_X\sim\hat{\partial}$) so dropping the
terms in (\ref{YM}) with an inoperative $\hat{\partial}$ on the right
we obtain (\ref{AT}).
Next one writes, using (\ref{AS})
\be
\partial_t(\psi^*L_1\psi) -\partial_y(\psi^*A_1\psi) = \psi^*\{L_{1t}-A_{1y}
+[L_1,A] +[L,A_1]\}\psi =
\label{AU}
\ee
$$ = \psi^*(\partial_tL_1-\partial_yA_1 + [L_0,A_1] + [L_1,A_0])\psi +
\partial_x(\cdots)$$
and via ergodicity in $x,y$, or $t$ flows, averaging of derivatives in
$x,y$, or $t$ gives zero so from (\ref{AT}) and (\ref{AU}) we obtain
the Whitham equations in the form $<\psi^*F\psi> = 0$
(this represents the first order term in $\epsilon$ - the slow variables
are present in $L_0,\,\,A_0,\,\,\psi,$ and $\psi^*$).
In order to spell
this out in \cite{fb} one imagines $X,Y,T$ as a parameter $\xi$ and
considers $L(\xi),\,\,A(\xi)$, etc. (in their perturbed form) with
\be
\psi(\xi) = e^{p(\xi)x+E(\xi)y+\Omega(\xi)t+\sigma\cdot\zeta(\xi)}\cdot
\phi(U(\xi)x + V(\xi)y
+ W(\xi)t+\zeta(\xi)|I(\xi))
\label{AV}
\ee
where $\sum_1^{2g}\sigma_i\zeta_i\sim\sigma\cdot\zeta$
and $\psi^* = exp(-px-Ey-\Omega t-\sigma\cdot\zeta)\phi^*(-Ux-Vy-Wt-\zeta|I)$
(no $\xi$ variation - i.e. assume $p,E,\Omega,U,V,W,I$ fixed).
We recall that
one expects $\lambda_k = \lambda_k(X,Y,T)$ etc. so the Riemann surface
varies with $\xi$.  Also recall that
$x,y,t$ and $X,Y,T$ can be considered as independent variables.  Now as
above, using (\ref{AS}), we can write
\be
\partial_t(\psi^*\psi(\xi)) = \psi^*(A(\xi)-A)\psi(\xi) - \partial_x
(\psi^*A^1\psi(\xi)) + \partial^2_x(\cdots)
\label{AW}
\ee
Note also from (\ref{AV}) for $P$ and $\zeta(\xi)$ fixed
($\theta\sim xU+yV+yW+\zeta$)
\be
\partial_{\xi}\psi^*\psi(\xi)|_{\xi=0}= (\dot{p}x+\dot{E}y+\dot{\Omega}t)
\psi^*\psi +
\label{AX}
\ee
$$ + (\dot{U}x+\dot{V}y +\dot{W}t)\cdot\psi^*\partial_{\theta}\psi +
\dot{I}\cdot\psi^*\partial_I\psi$$
where $\dot{f}\sim\partial f/\partial\xi$.  In \cite{fb} one assumes
(without discussion) that it is also permitted to vary $\xi$ and hold
e.g. the $I_k$ constant while allowing say the $P$ to vary.
Now differentiate the left side $\partial_t
(\psi^*\psi(\xi))$ of (\ref{AW}) in $\xi$ and use (\ref{AX}) to obtain
\be
\partial_{\xi}[\partial_t(\psi^*\psi(\xi))]|_{\xi=0} = \dot{\Omega}
\psi^*\psi + \dot{W}\cdot\psi^*\psi_{\theta} +
\label{AY}
\ee
$$ + \{(\dot{p}x+\dot{E}y+\dot{\Omega}t)\partial_t(\psi^*\psi) +
(\dot{U}x+\dot{V}y+\dot{W}t)\cdot\partial_t(\phi^*\phi_{\theta}) +
\dot{I}\cdot\partial_t(\phi^*\phi_I)\}$$
Fixing $x,y,t$ and averaging (\ref{AY}) in the $\theta$ variables yields
\be
<\partial_{\xi}[\partial_t(\psi^*\psi(\xi))]|_{\xi=0}> =
\dot{\Omega}<\psi^*\psi> + \dot{W}<\psi^*\psi_{\theta}>
\label{AZ}
\ee
Next one differentiates the right side of (\ref{AW}) in $\xi$, averages,
and equates to (\ref{AZ}) to obtain
\be
(\ref{AZ}) = <\psi^*\dot{A}\psi> - <\partial_{\xi}[\partial_x(\psi^*
A^1\psi(\xi))]|_{\xi=0}> =
\label{BA}
\ee
$$<\psi^*\dot{A}\psi> - \dot{p}<\psi^*A^1\psi>
-\dot{U}\cdot <\psi^*A^1\psi_{\theta}>$$
We note here as in (\ref{AX}) - (\ref{AY}) (the result is unaltered if
one disregards $\xi$ dependence in $A^1$, as well as $A^1$ action on
the isolated $x$ term,
and assumes $I$ fixed, since only more terms $\partial_x(\cdots)$ would
occur whose average vanishes)
\be
\partial_x\partial_{\xi}(\psi^*A^1\psi)|_{\xi=0} = \partial_x[\psi^*
A^1(\dot{p}x+\dot{E}y+\dot{\Omega}t)\psi +
\label{BB}
\ee
$$ + \psi^*A^1(\dot{U}x+\dot{V}y+\dot{W}t)\cdot\psi_{\theta}] =
\dot{p}\psi^*A^1\psi + \dot{U}\cdot\psi^*A^1\psi_{\theta} +$$
$$+ (\dot{p}x+\dot{E}y+\dot{\Omega}y)\partial_x(\psi^*A^1\psi) +
(\dot{U}x+\dot{V}y+\dot{W}t)\cdot\partial_x(\psi^*A^1\psi_{\theta})$$
and (\ref{BA}) follows.  Rewriting (\ref{BA}) with $\xi\sim Y$ we obtain
(note $\psi^*\psi_{\theta} = \phi^*\phi_{\theta}$)
\be
-<\psi^*\partial_YA\psi> = -\partial_Y\Omega<\psi^*\psi> -\partial_YW
<\phi^*\phi_{\theta}> -
\label{BC}
\ee
$$ -\partial_Yp<\psi^*A^1\psi> -\partial_YU\cdot<\psi^*A^1
\psi_{\theta}>$$
Similarly one uses (cf. (\ref{AW}))
\be
\partial_y(\psi^*\psi(\xi)) = \psi^*(L(\xi)-L)\psi(\xi) - \partial_x
(\psi^*L^1\psi(\xi)) + \partial_x^2(\cdots)
\label{BD}
\ee
with $\xi\sim T$ to get
\be
<\psi^*\partial_TL\psi> = \partial_TE<\psi^*\psi> + \partial_TV\cdot
<\phi^*\phi_{\theta}> +
\label{BE}
\ee
$$ + \partial_Tp<\psi^*L^1\psi> +\partial_TU\cdot<\psi^*L^1\psi_{\theta}>$$
Finally note that
\be
\partial_y(\psi^*A^1\psi(\xi)) -\partial_t(\psi^*L^1\psi(\xi)) =
\label{BF}
\ee
$$ = \psi^*[L^1(A(\xi)-A) - A^1(L(\xi)-L)]\psi(\xi) + \partial_x(\cdots)$$
Using this with $\xi\sim X$ one gets then as above
$$<\psi^*(L^1\partial_XA - A^1\partial_XL)\psi> = \partial_X\Omega
<\psi^*L^1\psi> - \partial_XE<\psi^*A^1\psi> +$$
\be
+ \partial_XW\cdot<\psi^*L^1\psi_{\theta}> - \partial_XV\cdot
<\psi^*A^1\psi_{\theta}>
\label{BG}
\ee
We recall that $\partial_XS = U,\,\,\partial_YS = V,$ and $\partial_TS
= W$ so there are compatibility relations
\be
\partial_YU = \partial_XV;\,\,\partial_TU = \partial_XW;\,\,\partial_TV
= \partial_YW
\label{BH}
\ee
Now add up (\ref{BC}) and (\ref{BE}), and subtract (\ref{BG}) to get
\be
<\psi^*[\partial_TL - \partial_YA + L^1\partial_XA - A^1\partial_XL]\psi>
= <\psi^*F\psi> = 0 =
\label{BI}
\ee
$$ = (\partial_TE-\partial_Y\Omega)<\psi^*\psi> + (\partial_Tp -
\partial_X\Omega)<\psi^*L^1\psi> + (\partial_XE - \partial_Yp)
<\psi^*A^1\psi>$$
We observe that if one lets the point $P$ on $\Sigma$ vary with $\xi$,
while holding $\theta_k$ and $I_j$ fixed, then
\be
\partial_{\xi}(\psi^*\psi(\xi))|_{\xi=0} = (xdp + ydE + td\Omega)\phi^*
\phi
\label{BJ}
\ee
which, together with (\ref{AW}) and (\ref{BD}), yields e.g.
\be
d\Omega<\phi^*\phi> = -dp<\psi^*A^1\psi>;\,\,dE<\phi^*\phi> =
-dp<\psi^*L^1\psi>
\label{BK}
\ee
(cf. (\ref{BA}) - (\ref{BB})).  It follows then from (\ref{BI}) and
(\ref{BK}) that
\be
0 = (\Omega_Y-E_T)dp + (p_T-\Omega_X)dE + (E_X-p_Y)d\Omega
\label{BL}
\ee
Thus one arrives at a version of the Whitham equations in the form
\be
p_T = \Omega_X;\,\,p_Y = E_X;\,\,E_T = \Omega_Y
\label{BM}
\ee
where the last equation establishes compatibility of the first two via
$p_{TY} - p_{YT} = \Omega_{XY} - E_{XT} = \partial_X(\Omega_Y  - E_T) = 0$.
We feel that this derivation from \cite{fb} is important since it
again exhibits again the role of square eigenfunctions (now in the form
$\psi^*\psi$) in dealing with averaging processes.  In view of the
geometrical nature of such square eigenfunctions (cf. \cite{cd,cf} for
example) one might look for underlying geometrical objects related to
the results of averaging (cf. here \cite{cg}).  Another (new) direction we will
discuss later involves the Cauchy kernels expressed via $\psi^*\psi$
and their dispersionless limits (cf. \cite{ch,gd,ge,gf,oa}).  Connections
to Section 3.5 are also discussed later.

\subsection{Relations to moduli spaces}
Now following \cite{kc,kj} let ${\cal M}_{gN}$ be the moduli space of smooth
algebraic curves $\Gamma_g$ of genus $g$ with local coordinates $k_{\alpha}^
{-1}(P)$ in the neighborhoods of $N$ punctures $P_{\alpha}\,\,(k_{\alpha}^
{-1}(P_{\alpha}) = 0$) so ${\cal M}_{gN}\sim \{\Gamma_g,\,P_{\alpha},\,
k_{\alpha}^{-1},\,\alpha = 1,\cdots,N\}$ (see Section 7.3 for a
discussion in the context of LG theory and Hurwitz spaces of the
moduli space $M_{gN}\not= {\cal M}_{gN}$).
First look at $g=0\,\,
(\Gamma_0\sim{\bf P}^1$) and for simplicity take $N=1,\,\,(P_1\sim\infty$)
with $k_1(P) = P + \sum_1^{\infty}v_sP^{-s}$, so ${\cal M}_{0,1}\sim
\{v_s,\,s = 1,\cdots\}$.  Set ${\cal A} = \{A_i = (1,i)\}$ and define
meromorphic functions
\be
\Omega_i(P) = \Omega_{1,i}(P) = \sum_1^i w_{1,i,s}P^s = k_1^i(P) +
O(k_1^{-1})
\label{CA}
\ee
($P\sim$ point or local coordinate here in a flagrant abuse of notation -
see below).
If we keep $g=0$ but allow more punctures with $k_{\alpha}(P) =
\sum_{-1}^{\infty}v_{\alpha,s}(P-P_{\alpha})^s$ then one defines
${\cal A} = \{A = (\alpha,i),\,\,\alpha = 1,\cdots,N,\,\,i=1,2,\cdots;
\,\,for\,\,i=0,\,\alpha \not= 1\}$ and ${\cal M}_{0,N} = \{P_{\alpha},
v_{\alpha,s}\}$ with meromorphic functions ($\Omega_{1,i}$ as in (\ref{CA}))
\be
\Omega_{\alpha,i}(P) = \sum_{-1}^iw_{\alpha,i,s}(P-P_{\alpha})^{-s} =
k_{\alpha}^i(P) + O(1);\,\,\Omega_{\alpha,i}(\infty) = 0\,\,(\alpha
\not= 1)
\label{CB}
\ee
One can write then in fact
\be
\Omega_{\alpha,i}(P) = \frac{1}{2\pi i}\oint_{C_{\alpha}}\frac
{k_{\alpha}^i(z_{\alpha})dz_{\alpha}}{P-z_{\alpha}};\,\,\Omega_{\alpha,0}(P) =
-log(P-P_{\alpha})\,\,(\alpha\not= 1)
\label{CC}
\ee
where $C_{\alpha}$ is a small cycle around $P_{\alpha}$.
\\[3mm]\indent
More generally for ${\cal M}_{gN}$
one thinks of meromorphic differentials $d\Omega_A$ on $\Gamma_g$ of
two types:  $({\bf A})\,\,d\Omega_{\alpha,i}$ is holomorphic outside
$P_{\alpha}$ with $d\Omega_{\alpha,i} = d(k_{\alpha}^i + O(k_{\alpha}^{-1}))$
near $P_{\alpha}$ and $({\bf B})\,\,\Omega_{\alpha,0}\,\,(\alpha\not= 1)$ is
a differential with simple poles at $P_1$ and $P_{\alpha}$ having residues
$\pm 1$ respectively so that
\be
d\Omega_{\alpha,0} = dk_{\alpha}(k_{\alpha}^{-1} + O(k_{\alpha}^{-1}))
= -dk_1(k_1^{-1} + O(k_1^{-1}))
\label{CD}
\ee
Here $\Omega_{\alpha,i}(P) = \int^Pd\Omega_{\alpha,i}$ and for simplification
of formulas later one complexifies the Whitham hierarchy leading to the moduli
space ${\cal M}^*_{gN} = \{\Gamma_g,\,P_{\alpha},\,k_{\alpha}^{-1},\,a_i,b_i
\in H_1(\Gamma_g,{\bf Z})\}$ where $a_i,\,b_i$ are a canonical homology
basis.  Then one normalizes the $\Omega_A$ via $\oint_{a_i}d\Omega_A = 0$
and sets $p(P) = \Omega_{1,1}(P) = \int^Pd\Omega_{1,1}$.  If one works
on ${\cal M}_{gN}$ the normalization must be changed to $\Im\oint_cd\Omega_A
=0$ for $c\in H_1(\Gamma_g,{\bf Z})$.  Now the multivalued function
$p(P)$ can be used as a coordinate everywhere on $\Gamma_g$ except at
points $\Pi_s$ where $dp(\Pi_s) = 0$.  One can accordingly change $P$
to $p$ in formulas such as (\ref{CA}) - (\ref{CC}) and we will do
this without additional fuss.  Thus a full system of local
coordinates on ${\cal M}^*_{gN}$ is given by $\{p_{\alpha} = p(P_{\alpha}),\,
v_{\alpha,s},\,\alpha = 1,\cdots,N,\,s = -1,0,1,\cdots;\,\pi_s = p(\Pi_s),\,\,
s = 1,\cdots,2g,\,U_i^p = \oint_{b_i}dp,\,i = 1,\cdots,g\}$.
The associated compatible system of evolution equations will be
\be
\partial_Ak_{\alpha}(p,T) = \{k_{\alpha}(p,T),\Omega_A(p,T)\}
\label{CE}
\ee
\be
\partial_AU^p_i = \partial_XU^A_i\,\,(U^A_i = \oint_{b_i}d\Omega_A);\,\,
\partial_A\pi_s = \partial_Ap(\Pi_s) = \partial_X\Omega_A(\Pi_s)
\label{CF}
\ee
and we will say more about this below.

\subsection{Zero curvature equations}
Now we have seen in Section 3.4 and in Section 3.6 how differentials
of the form $dp\sim\Omega_1,\,\,dE\sim\Omega_2,\,\,d\Omega\sim\Omega_3$
arise in Whitham equations such as (\ref{BM}) (or (\ref{AR})).  Evidently
further variables $T_i$ could be appended with say
$T_i\sim\Omega_i\sim k_1^i$ (and one expects $\Omega_{\alpha,i}$ will be
associated to some $T_{\alpha,i}$ as indicated in (\ref{CE}) -
(\ref{CF}) - all of the relevant times are discussed below
and the context of Hurwitz spaces is spelled out
following \cite{di}).  In any event
one can expect to have a family of differentials $\Omega_A\,\,(A\sim
(\alpha,i)$ for example), such that $\partial_A\Omega_B = \partial_B\Omega_A$
when the $\Omega_A$ are expressed in suitable variables.  In connection
with equations such as (\ref{CE}) - (\ref{CF}) (and (\ref{AR}),
(\ref{BM})) we make some observations on the possible forms of Whitham
equations.  In dealing with dispersionless KP (= dKP) for example there
are two natural algebraic forms which emerge (cf. \cite{ae,ch,kc,kj,ta})
which will be displayed below; further there are various geometrical
developments in connection with Frobenius manifolds etc. \`a la
\cite{di,dj}.  A (zero curvature) form which arises in this way is
\be
\partial_A\Omega_B - \partial_B\Omega_A + \{\Omega_A,\Omega_B\} = 0;\,\,
\{f,g\} = f_Xg_p - f_pg_X
\label{CG}
\ee
where $p\sim\Omega_{A_0}\,\,(T_{A_0}\sim X,\,\,A_0\sim (1,1),\,\,
p = \Omega_{1,1}(P) = \int^P d\Omega_{1,1}$).
We note that $\{f,g\}$ here is $\{g,f\}$ in \cite{ci,ta}.
Such equations (\ref{CG})
can be regarded as compatibility equations for
\be
\partial_AE = \{E,\Omega_A\}
\label{CH}
\ee
where $E$ is an arbitrary function of $(p,T)$ (easy exercise, using the
Jacobi identity).  When $\partial_pE\not= 0$ one can write $p=p(E,T)$
and $\partial_Af(p,T) = \partial_AF(E,T) + (\partial F/\partial E)
\partial_AE\,\,(F(E,T) = f(p,T)$).  Then for such an $E$ (\ref{CG})
becomes
\be
\partial_A\Omega_B(E,T) = \partial_B\Omega_A(E,T)
\label{CI}
\ee
Indeed we note that ($\Omega'_A\sim (\partial\Omega_A/\partial E),\,\,
E'\sim(\partial E/\partial p)$)
\be
(\ref{CG}) = \partial_A\Omega_B + \Omega'_BE_A -\partial_B\Omega_A -
\Omega'_AE_B + (\partial_X\Omega_A +\Omega'_AE_X)\Omega'_BE' -
\label{CJ}
\ee
$$- \Omega'_AE'(\partial_X\Omega_B + \Omega'_BE_X) = \partial_A\Omega_B
-\partial_B\Omega_A + $$
$$ +\Omega'_B(E_A+E'\partial_X\Omega_A) -
\Omega'_A(E_B + E'\partial_X\Omega_B) = 0$$
But (\ref{CH}) implies for example that $E_A = E_X\partial_p\Omega_A(p,T) -
E'\partial_X\Omega_A(p,T) = E_X\Omega'_AE' - E'(\partial_X\Omega_A(E,T)
+ \Omega'_AE_X) = -E'\partial_X\Omega_A\Rightarrow E_A + E'\partial_X
\Omega_A(E,T) = 0$.  Similarly $E_B + E'\partial_X\Omega_B(E,T) = 0$ and
(\ref{CJ}) implies (\ref{CI}).  From this point, in coordinates $E,T$
we can introduce a ``potential" $S(E,T)$ via (***) $\Omega_A(E,T) =
\partial_AS(E,T)\,\,(S\sim \sum\Omega_AT_A$) and write
for $Q = \partial S/\partial_E\,\,(\delta\sim$ full exterior derivative)
\be
\omega = \delta S(E,T) - QdE = \partial_ESdE + \sum\partial_ASdT_A -QdE
= \sum \Omega_AdT_A
\label{CK}
\ee
We note also that $p = \partial S/\partial X = \Omega_{A_0}$.  It follows
that $\delta\omega = \sum\delta\Omega_A\wedge dT_A$ formally and from
(\ref{CK}) one obtains $\delta\omega = dE\wedge\delta Q\sim\delta E
\wedge\delta Q$.  This indicates a very special role for $E$ and $Q=
\partial S/\partial E$ in the theory and this will be developed in a
general way for dKP later following \cite{ci,ta}.  In fact,
by writing out $\delta\omega$ (with $\Omega_{1,1} = p$), one will
now have a ``classical" string equation $\{Q,E\} = 1$ when $Q,E$ are
considered as functions of $p,X$ (also then $\partial_AQ = \{Q,\Omega_A\})$.

\subsection{Times}
It is appropriate here to make a few comments
about times $T_A$ in a general sense (cf. here \cite{ae,ch,cj,di}.
Thus we will often have situations where say
\be
\lambda = p + \sum_1^{\infty}a_np^{-n};\,\,p = \lambda +
\sum_1^{\infty}p_m\lambda^{-m}
\label{CL}
\ee
describe inverse functions.  Often this arises in polynomial Landau-
Ginzburg (LG) situations where e.g. $p^{n+1} + \alpha_1 p^{n-1} +
\cdots + \alpha_n\sim\lambda^{n+1}$ and one thinks of Puiseaux series
etc. but (\ref{CL}) can be more general.  Then one notes that
$dp = d\lambda(1-\sum_1^{\infty}mp_m\lambda^{-m-1}$) so
\be
p_m = -\frac{1}{m}Res_{p=\infty}\lambda^mdp = -\frac{1}{m}
Res_{\lambda=\infty}\lambda^m\frac{dp}{d\lambda}d\lambda
\label{CM}
\ee
These coefficients $p_m$ will turn out to be universal time coordinates
in a sense to be indicated below.  $\bullet$
\\[3mm]\indent
Let us consider now some examples from \cite{kc,kj} before bringing in the
dKP background.  Thus one defines algebraic orbits of the Whitham hierarchy
to be solutions $\Omega_A$ say obtained via a global solution of (\ref{CH}).
For genus zero global means $E$ is a meromorphic solution of (\ref{CH}) such
that $\{E(p,T),k_{\alpha}(p,T)\} = 0$.

\subsection
{Lax reduction, $N=1,\,\,
g=0$}  Let
$P_1\sim\infty$ with local parameter $k_1(p) = k(p) = p + \sum_1^{\infty}
v_sp^{-s}$ (recall $p\sim P$ here).  Suppose some power $k^n(p)\sim
\lambda^n$ is a polynomial
\be
E = k^n(p) = p^n + u_{n-2}p^{n-2} + \cdots + u_0
\label{CN}
\ee
We think here of $k(p,T),\,E(p,T)$, etc. and it is convenient to take
$E = \prod_1^n(p-p_k)$ with distinct real roots $p_k$ leading to an
n-sheeted Riemann surface $\Sigma_E$ of genus $0$ associated with $E$
(cf. Sections 4,5 and \cite{ch,gh}).
This means that the $v_s$ are
(polynomial) functions of the $u_i$.  The dispersionless Lax equations
corresponding to this example will be, as in (\ref{CH})
\be
\partial_iE(p,T) = \{E(p,T),\Omega_i(p,T)\};\,\,\Omega_i(p,T) =
[E^{\frac{i}{n}}(p,T)]_{+}
\label{CO}
\ee
(see also Section 4 for a more thorough development).
We note here that from $\partial_ip(E) = \partial_X\Omega_i(E)$ as in
(\ref{CI}) and (\ref{CO}) there results (using $\partial_X\Omega_i(p) =
\partial_X\Omega_i(E) + \partial_E\Omega_i(E)E_X$ and $\partial_p\Omega_i
(p) = \partial_E\Omega_i(E)E'$)
\be
\partial_iE = E_X\partial_p\Omega_i - E'(\partial_X\Omega_i(E) +
\partial_E\Omega_i(E)E_X) = -E'\partial_ip(E)
\label{CP}
\ee
Such an equation is also used in \cite{kj,kl} but it's origin there is
not clear
(whereas here just its meaning is unclear - a shift $X\to -X$ seems
indicated).  One defines now a generating
function
\be
S = \sum_1^{\infty}T_i\Omega_i = \sum_1^{\infty}T_ik^i + O(k^{-1});\,\,
\Omega_i = k^i_{+}
\label{CQ}
\ee
Let $q_s$ satisfy $(dE/dp)(q_s) = 0$
(i.e. $(dE/dp) = np^{n-1} + (n-2)u_{n-2}p^{n-3} + \cdots + u_1 =
\prod_1^n(p-q_s);\,\,q_n = -\sum_1^{n-1}q_s$) and consider $q_s,\,u_0$ as
a new set of independent coordinates.  Now let the $u_j$ and $T_i$ be
related via $(dS/dp)(q_s) = 0$ (this is a stipulation which determines a
dependence of the $u_j$ and $T_i$, leading to $u_j=u_j(T_i)$).
Then one shows that $\partial_iS(E,T) =
\Omega_i(E,T)$ (the argument in \cite{kj,kl} is however somewhat curious
and matters will be clarified in Section 4).  We note also in passing
from (\ref{CO}), namely $\partial_iE = \partial_XE\partial_p\Omega_i -
\partial_pE\partial_X\Omega_i$, that at $q_s$ one has
\be
\partial_iE(q_s,T) = E_X\partial_p\Omega_i(q_s,T)
\label{CR}
\ee
and this (Riemann invariant) form of (\ref{CO}) will be important later
in topological field theory (cf. \cite{ae}).
Another point of view (related to Section 3.9) involves looking at any
formal series $Q(p) = \sum_1^{\infty}b_jp^j$ and defining ``times" via
\be
\hat{T}_i = \frac{1}{i}Res_{p=\infty}k^{-i}(p)Q(p)dE(p)
\label{CS}
\ee
Thus for $Q\sim (dS/dE)$ one has
\be
\hat{T}_i = \frac{1}{i}Res_{k=\infty}k^{-i}dS = \frac{1}{i}Res_{k=\infty}
\sum_1^{\infty}jT_jk^{j-i-1} = T_i
\label{CT}
\ee
Variations of this approach will appear in Section 4.
It is clear however that one can
use (\ref{CS}) to define times $\hat{T}_i$ and as indicated
in (\ref{CT}) $Q\sim dS/dE$ yields $\hat{T}_i = T_i$.
This leads to the more general picture below.  $\bullet$
\\[3mm]\indent
Take now the general situation with a ``big phase space" as the moduli space
${\cal N}_g = \{\Gamma_g,\,dQ,\,dE,\,P_{\alpha},\,n_{\alpha},\,
k_{\alpha}^{-1},\,
a_i,b_i\in H_1(\Gamma_g)\}$ where $dE$ is a fixed normalized meromorphic
differential having poles of orders $n_{\alpha}+1$ at $P_{\alpha}$ and
$dQ$ is a fixed normalized differential holomorphic outside of the
punctures.  Here $E$ will have the form
\be
E = p^n + u_{n-2}p^{n-2} + \cdots + u_0 +
\sum_{\alpha = 2}^N\sum_1^{n_{\alpha}}
\nu_{\alpha,s}(p-p_{\alpha})^{-s}
\label{CU}
\ee
Generally in the Hurwitz space context $E$ determines
the ramification (cf. Section 5.3).
One also takes $k_{\alpha}^{n_{\alpha}}(p) = E(p)$
and $dS = QdE$.
Then times are defined via
\be
T_{\alpha,i} = \frac{1}{i}Res_{P_{\alpha}}(k_{\alpha}^{-i}(p)Q(p)dE(p)),\,\,
(i>0);\,\,T_{\alpha,0} = Res_{P_{\alpha}}QdE
\label{CV}
\ee
and the formulas ($i = 1,\cdots,g$)
\be
T_{h,i} = \oint_{a_i}dS;\,\,\,T_{Q,i} = -\oint_{b_i}dE;\,\,\,T_{E,i} =
\oint_{b_i}dQ
\label{CW}
\ee
Now one associates differentials to the time variables $T_A$.  One should
have as in (\ref{CA}) - (\ref{CB}), $\Omega_A$ corresponding to $T_A$  with
$\oint_{a_i}d\Omega_A = 0$, to
which one adds holomorphic differentials $d\Omega_{h,k}$ normalized via
\be
\oint_{a_i}d\Omega_{h,k} = \delta_{ik}
\label{CX}
\ee
to give the correct number of $T_A$ (cf. \cite{dj,kj}).
The differentials $d\Omega_{E,i}$ and $d\Omega_{Q,i}$ are holomorphic
on $\Gamma_g$ except for the $a_j$ cycles where they have jumps
\be
d\Omega^{+}_{E,i} - d\Omega^{-}_{E,i} = \delta_{ij}dE;\,\,
d\Omega^{+}_{Q,i} - d\Omega^{-}_{Q,i} = \delta_{ij}dQ
\label{CY}
\ee
The normalization conditions are
\be
\oint_{a_i}d\Omega_{E,j} = \oint_{a_i}d\Omega_{Q,j} = 0
\label{CZ}
\ee
Let then ${\cal D}_g\subset {\cal N}_g$ be the subset of the big phase space
where $dE,\,dQ$ have no common zeros on $\Gamma_g$.  Then the system
of times $T_A$ above defines a system of coordinates on ${\cal D}_g$
and the corresponding dependence of $\Gamma_g(T)$ and $dE(T)$ on $T$
represents a solution of the universal Whitham hierarchy on ${\cal M}^*_{gN}$.
This is equivalent to saying that the Whitham hierarchy can be considered
as a way to define the special coordinate system on the moduli space
of curves with punctures and jets of local coordinates in the neighborhoods
of the punctures.  The $n_{\alpha}$ jets are the equivalence classes of
coordinates where $k'_{\alpha}(P)\equiv k_{\alpha}(P)$ means
$k'_{\alpha}(P) = k_{\alpha}(P) + O(k_{\alpha}(P)^{-n_{\alpha}-1}$).

\subsection{The tau function}
We will make various remarks about the tau function here in the moduli
space context and refer to Section 4 for other points of view.  Generally
the tau function of e.g. KP theory does not itself tend to a limit in
dKP, but writing (in a WKB spirit) $\tau = exp[(1/\epsilon^2)F(T)]$ one
can invoke limiting procedures and $F(T) = log\tau^{dKP}$ is the quantity
of interest in the dispersionless theory (cf. \cite{ae,ca,ch,ci,cl,ta} for
discussion).  We follow here mainly \cite{kc,kj,kl} (cf. also \cite{di,dj}).
For genus zero and general $E$ one can express solutions of (\ref{CH})
implicitly via
\be
\frac{dS}{dp}(q_s,T) = 0;\,\,S(p,T) = \sum_A(T_A-T^0_A)\Omega_A(p,T)
\label{DA}
\ee
where $(dE/dp)(q_s) = 0$ and for $B\sim (\beta,i),\,\,\partial_BS(p,T) =
\Omega_B(p,T)$.  One defines then
\be
F = \frac{1}{2}\sum_1^N (Res_{\alpha}[\sum_{i=1}^{\infty}\tilde{T}_{\alpha,i}
k_{\alpha}^idS(p,T)] + \tilde{T}_{\alpha,0}s_{\alpha}(T))
\label{DB}
\ee
where $\tilde{T}_{\alpha,i}\sim T_{\alpha,i} - T^0_{\alpha,i}$ and
$s_{\alpha}$ is defined via
\be
S(p,T) = \sum_1^N(\sum_{i=1}^{\infty}\tilde{T}_{\alpha,i}k^i_{\alpha}
+ T_{\alpha,0}log\,k_{\alpha} + s_{\alpha}) + O(k^{-1})
\label{DC}
\ee
($T_{1,0} = -\sum_2^NT_{\alpha,0}$).  One makes cuts now connecting
$P_1\sim \infty$ with the $P_{\alpha}$ and chooses a branch of
$S(p,T)$.  Then
\be
s_{\alpha} = \frac{1}{2\pi i}\oint_{\sigma_{\alpha}}log(p-p_{\alpha})dS
\label{DD}
\ee
where $\sigma_{\alpha}$ is a contour surrounding the corresponding
cut (we will be somewhat cavalier about interchanging $p_{\alpha}$ and
$P_{\alpha}$).  The function $S$ has jumps on the cuts and is
holomorphic outside of the cuts and punctures.  One can also write
(\ref{DB}) in the form
\be
F = \int \bar{d}S\wedge dS
\label{DE}
\ee
which will be clarified as we go along (many details are missing in
this first sketch \`a la Krichever).
\\[3mm]\indent
Now note from (\ref{DB}) for $A = (\alpha,i>0)$
\be
2\partial_AF = Res_{\alpha}(k^i_{\alpha}dS) + \sum_1^N(\sum_{j=1}^{\infty}
Res_{\beta}(\tilde{T}_{\beta,i}k^j_{\beta}d\Omega_A) +
\tilde{T}_{\beta,0}\Omega_A(p_{\beta}))
\label{DF}
\ee
where $\partial_As_{\beta} = \Omega_A(p_{\beta})$.  Now $\sum_1^N
Res_{\alpha}(\Omega_Ad\Omega_B) = 0$ implies that $Res_{\beta}(k^j_{\beta}
d\Omega_{\alpha,i}) = Res_{\alpha}(k^i_{\alpha}d\Omega_{\beta,j})$ for
$j>0$.  Further one obtains $\Omega_A(p_{\beta}) = Res_{\beta}
(\Omega_Ad\,log(p-p_{\beta})) = Res_{\alpha}(k^i_{\alpha}d\Omega_{\beta,0})$.
Putting these equations into (\ref{DF}) yields
\be
\partial_{\alpha,i}F(T) = Res(k^i_{\alpha}dS(p,T))\,\,(i>0)
\label{DG}
\ee
and similarly one obtains $(\spadesuit)\,\,
\partial_{\alpha,0}F(T) = s_{\alpha}$.
Hence the expansion of $S(p,T)$ at $P_{\alpha}$ has the form
\be
S = \sum_1^N(\sum_{i=1}^{\infty}\tilde{T}_{\alpha,i}k^i_{\alpha}
+ \tilde{T}_{\alpha,0}log\,k_{\alpha} + \partial_{\alpha,0}F +
\sum_1^{\infty}\frac{1}{j}\partial_{\alpha,j}F\,k_{\alpha}^{-j})
\label{DH}
\ee
which implies
\be
\partial^2_{A,B}F = Res_{\alpha}(k^i_{\alpha}d\Omega_B)\,\,(A = (\alpha,
i>0));
\label{DI}
\ee
$$\partial_{\alpha,0}\partial_{\beta,0}F = log(p_{\alpha}-p_{\beta})$$
Next, from (\ref{DG}) and $(\spadesuit)$, (\ref{DB}) becomes $2F =
\sum\tilde{T}_A\partial_AF$ from which $2F_B = \sum\tilde{T}_AF_{AB}
+ F_B$ and hence we have
\be
2F = \sum\tilde{T}_BF_B = \sum\sum\tilde{T}_A\tilde{T}_BF_{AB}
\label{DJ}
\ee
The quantities $F_{AB}$ will play the role of two point (correlation)
functions in TFT and are in many respects the fundamental objects of the
theory (see especially \cite{ae,ch,ci,cl,di,dj,ta})
\\[3mm]\indent
We look next at genus $g$ and ${\cal N}_g$ etc. as above.  Then the
moduli space ${\cal N}_g$ is foliated via leaves determined by the
periods of $dE$, namely $V_k = \oint_{b_k}dE$ (corresponding to $-T_{Q,k}$
in (\ref{CW})).  The discussion in \cite{kc,kl} is customarily mysterious
here but we write down the following formulas involving $F$ (clarification
is needed).  Thus one takes
\be
F = F_0 + \frac{1}{4\pi i}\sum_1^g\oint_{a^{-}_k}T_{E,k}EdS -
\oint_{b_k}T_{h,k}dS + T_{h,k}T_{E,k}E_k
\label{DK}
\ee
where $F_0$ is given by (\ref{DB}) with $\tilde{T}_{\alpha,i}$ replaced
by $T_{\alpha,i}$.  The first integral in (\ref{DK}) is taken over the
left side of the $a_k$ cycle and $E_k = E(P_k)$ where $P_k$ is the
intersection point of the $a_k$ and $b_k$ cycles; the last term with
$E_k$ apparently makes $F$ dependent only on the homology class of cycles.
A picture can be contrived of course but more detail would be helpful here.
For this F it is claimed that (\ref{DG}) and $(\spadesuit)$ hold along
with
\be
\partial_{h,k}F = \frac{1}{2\pi i}[T_{E,k}E_k - \oint_{b_k}dS];
\label{DL}
\ee
$$\partial_{E,k}F = \frac{1}{2\pi i}\oint_{a_k}EdS;\,\,\partial_{Q,k}F =
\frac{1}{4\pi i}[\oint_{a_k}QdS - 2T_{E,k}T_{h,k}]$$
Additional derivatives are given by
\be
\partial^2_{(h,k),A}F = \frac{1}{2\pi i}[E_k\delta_{(E,k),A} + Q_k
\delta_{(Q,k),A} - \oint_{b_k}d\Omega_A];
\label{DM}
\ee
$$\partial^2_{(E,k),A}F = \frac{1}{2\pi i}\oint_{a_k}Ed\Omega_A;\,\,
\partial^2_{(Q,k),A}F = \frac{1}{2\pi i}[\oint_{a_k}Qd\Omega_A -
\partial_A(T_{E,k}T_{h,k})]$$
Further one obtains from (\ref{DM})
the b-period matrix of normalized holomorphic
differentials on $\Gamma_g$ via
\be
\partial^2_{(h,i),(h,j)}F = -\oint_{b_i}d\Omega_{h,j}
\label{DN}
\ee
(cf. also \cite{dk}).

\subsection{Landau-Ginzburg (LG) theory}
Let us go now to a Landau-Ginzburg (LG) situation with polynomial
$E(p)$ as in (\ref{CN}).
Thus work with $N=1,\,\,g=0,$ and $S_m^{+}(p,T_1,\cdots,T_m) = \sum_1^m
T_i\Omega_i(p),\,\,\Omega_i = E_{+}^{i/n},\,\,E = k^n = p^n +
\cdots + u_0,$ and think of $dS = QdE = QE'dp$ or $dS/dE = Q$ (cf.
(\ref{CT}), etc.).  For $m=n+1$ one will have a situation (cf. \cite{kl})
$\partial_pS^{+}_m = (b_1p+b_0)E'$ or $\partial_ES^{+}_{n+1} = b_1p+b_0$
and one chooses $b_1 = 1/n,\,\,b_0=0$ for the standard $A_{n-1}$ type
LG theory ($B_{n+1} = p/n$ in \cite{kl} - note also $\partial_pS_m^{+}
= 0$ when $E'=0$ as desired in (\ref{CR}) etc. and see here also
Sections 4 and 5).  The LG theory here involves a superpotential $W$ and a
ring of primary chiral fields $\phi_i$ isomorphic to $R = {\bf C}[p]/
\{W'(p)=0\}$ where $nW(p,T_0,\cdots,T_{n-2})\sim E(p,T_0,T_1/2,
\cdots,T_{n-2}/(n-1),1/(n+1))$.  Note the zeros of $W'$ correspond to
zeros $q_s$ of $E'$ so we are in the context of (\ref{CR}).  With a change
of variables $T_i\to T_{i-1}/i$ then one asks for
\be
\phi_i = \frac{1}{i+1}\partial_p(nW)^{\frac{i+1}{n}} = \frac{1}{i+1}
\partial_p\Omega_{i+1} = \pm\partial_i W = \pm\frac{1}{i+1}\partial_{i+1}E
\label{DO}
\ee
This is consistent with (\ref{CR}) again provided $E_X=\pm 1$ (which
is OK at least in certain prototypical situations - cf. \cite{ae,cj,dl})
and indicates how the $q_s$ arise intrinsically, since the ring $R$, as
well as the time variables, are developed around the $q_s$.  Note
however that $E(q_s,T)\sim$ Riemann invariants and $E_X$ is a natural
creature in the Whitham equations (cf. (\ref{CR}) and (\ref{FD})).
In \cite{dl} one obtains a formula (cf. also \cite{cj})
\be
\partial_iF = Res_{\infty}(\frac{(nW)^{\frac{n+i+1}{n}}}{(i+1)(n+i+1)}dp)
\label{DP}
\ee
and from this it can be shown that
\be
F = -\frac{1}{2}Res_{\infty}(S^{-}_{n+1}dS^{+}_{n+1})
\label{DQ}
\ee
where $k^n = nW$ and $S_{n+1} = (k^{n+1}/(n+1)) +
\sum_0^{n-2}(T_j/(j+1))k^{j+1}$.  After the change of variables $T_j
\to (T_{j-1}/j),\,\,S^{+}_{n+1}$ agrees with the $S_{n+1}^{+}$ discussed
at the beginning of Remark 3.7.  Now from $\partial_ES^{+}_{n+1}(k) =
p/n$ one can write (\ref{DP}) as
\be
\partial_iF = Res_{\infty}(\frac{k^{n+i+1}}{(i+1)(n+i+1)}nd\frac{d
S^{+}_{n+1}}{dk^n})=
\label{DR}
\ee
$$= -Res_{\infty}(\frac{k^{i+1}}{i+1}d_ES^{+}_{n+1}) = -Res_{\infty}
(\frac{k^{i+1}_{-}}{i+1}d_ES^{+}_{n+1})$$
(note $d(fg) = (df)g + fdg$ and $Res\,d(fg) = 0$).  Let the right side
of (\ref{DR}) be called $F_1$ and compute $\partial_iF_1$ as follows.
Using $k(p)$ as local parameter one knows that $\partial_iS^{+}_{n+1}(k) =
k^{i+1}_{+}/(i+1) = \Omega_i/(i+1)$ (since $S^{+}_{n+1} =
\sum_1^{n+1}T_i\Omega_i$ and $T_i\to (T_{i-1}/i$, etc.).  Hence
$\partial_iS_{n+1} = k^{i+1}/(i+1)$ is natural with $\partial_i
S^{-}_{n+1} = k^{i+1}_{-}/(i+1)$.  This gives
\be
\partial_iF_1 = -\frac{1}{2(i+1)}Res(k^{i+1}_{-}dS^{+}_{n+1} -
k^{i+1}_{+}dS^{-}_{n+1})
\label{DS}
\ee
But $0=Res(k^{i+1}dS_{n+1})$ since $d_ES_{n+1} = d_{k^n}S_{n+1} =
(1/n)k^{-n+1}d_kS_{n+1}\Rightarrow k^{i+1}d_ES_{n+1} = (1/n)k^{i-n+2}
d_kS_{n+1} = d(\sum\alpha_mk^m)$ and hence $0 = Res(k^{i+1}_{+}
dS^{-}_{n+1} + k^{i+1}_{-}dS^{+}_{n+1})$.  This gives then $\partial_iF_1
=\partial_iF$ from (\ref{DR}) and one will conclude that $F=F_1$.

\section{CLASSICAL KP AND dKP}
\renewcommand{\theequation}{4.\arabic{equation}}\setcounter{equation}{0}

We follow here \cite{ch} (cf. also \cite{dn,kn,ta}) and begin
with two pseudodifferential operators ($\partial =
\partial/\partial x$),
\be
L = \partial + \sum_1^{\infty}u_{n+1}\partial^{-n};\, \ \,W = 1 +
\sum_1^ {\infty}w_n\partial^{-n} \ ,
\label{ZQ}
\ee
called the Lax operator and gauge operator respectively, where the
generalized Leibnitz rule with $\partial^{-1} \partial = \partial
\partial^{-1} = 1$ applies \be
\partial^if = \sum_{j=0}^{\infty}{i \choose j}(\partial^j f)
\partial^ {i-j}
\label{AAB}
\ee
for any $i \in {\bf Z}$, and $L = W\partial\,W^{-1}$. The KP hierarchy
then is determined by
the Lax equations ($\partial_n = \partial/\partial t_n$), \be \partial_n L =
[B_n,L] = B_n L - L B_n \ , \label{AAC} \ee
where $B_n = L^n_{+} $ is the
differential part of $L^n = L^n_{+} + L^n_{-} = \sum_0^{\infty}
\ell_i^n\partial^i + \sum_{-\infty}^ {-1}\ell_i^n\partial^i$. One can also
express this via the Sato equation, \be \partial_n W\,W^{-1} = -L^n_{-}
\label{AAD}
\ee
which is particularly well adapted to the dKP theory. Now define the wave
function
via
\be
\psi = W\,e^{\xi} = w(t,\lambda)e^{\xi};\, \ \,\xi = \sum_1^{\infty}t_n
\lambda^n;
\, \ \,w(t,\lambda) = 1 + \sum_1^{\infty}w_n(t)\lambda^{-n} \ , \label{AAE}
\ee
where $t_1 = x$. There is also an adjoint wave function $\psi^{*} =
W^{*-1} \exp(-\xi) = w^{*}(t,\lambda)\exp(-\xi),\,\,w^{*}(t,\lambda) = 1 +
\sum_1^ {\infty}w_i^{*}(t)\lambda^{-i}$, and one has equations
\be L\psi =
\lambda\psi;\, \ \,\partial_n\psi = B_n\psi;\, \ \,L^{*}\psi^{*} =
\lambda\psi^{*};\, \ \,\partial_n\psi^{*} = -B_n^{*}\psi^{*} \ . \label{AAF}
\ee
Note that the KP hierarchy (\ref{AAC}) is then given by the compatibility
conditions among these equations, treating $\lambda$ as a constant.
Next one
has the fundamental tau function $\tau(t)$ and vertex operators ${\bf X},
\,\,{\bf X}^{*}$ satisfying
\be
\psi(t,\lambda) = \frac{{\bf X}(\lambda)\tau (t)}{\tau (t)} = \frac{e^{\xi}
G_{-}(\lambda)\tau (t)}{\tau (t)} = \frac{e^{\xi}\tau(t-[\lambda^{-1}])}
{\tau (t)};
\label{AAG}
\ee
$$\psi^{*}(t,\lambda) = \frac{{\bf X}^{*}(\lambda)\tau (t)} {\tau (t)} =
\frac{e^{-\xi}
G_{+}(\lambda)\tau (t)}{\tau (t)} = \frac{e^{-\xi}\tau(t+[\lambda^{-1}])}
{\tau (t)} $$
where $G_{\pm}(\lambda) = \exp(\pm\xi(\tilde{\partial},\lambda^{-1}))$ with
$\tilde{\partial} = (\partial_1,(1/2)\partial_2,(1/3)\partial_3, \cdots)$
and $t\pm[\lambda^{-1}]= (t_1\pm \lambda^{-1},t_2 \pm (1/2) \lambda^{-2},
\cdots)$.
One writes also
\be
e^{\xi} = \exp \left({\sum_1^{\infty}t_n\lambda^n}\right) = \sum_0^
{\infty}\chi_j(t_1, t_2, \cdots ,t_j) \lambda^j  \label{AAH} \ee
where
the $\chi_j$ are the elementary Schur polynomials, which arise in many
important formulas (cf. below). \\[3mm]\indent We mention now the famous
bilinear identity which generates the entire KP hierarchy. This has the
form \be
\oint_{\infty}\psi(t,\lambda)\psi^{*}(t',\lambda)d\lambda = 0 \label{AAI}
\ee where $\oint_{\infty}(\cdot)d\lambda$ is the residue integral about
$\infty$, which we also denote $Res_{\lambda}[(\cdot)d\lambda]$. Using
(\ref{AAG}) this can also be written in terms of tau functions as \be
\oint_{\infty}\tau(t-[\lambda^{-1}])\tau(t'+[\lambda^{-1}])
e^{\xi(t,\lambda)-\xi(t',\lambda)}d\lambda = 0  \label{AAJ}
\ee This leads to the characterization of the tau function in
bilinear form expressed via ($t\to t-y,\,\,t'\to t+y$)
\be
\left(\sum_0^{\infty}\chi_n(-2y)\chi_{n+1}(\tilde{\partial})
e^{\sum_1^{\infty}
y_i \partial_i}\right)\tau\,\cdot\,\tau = 0  \label{AAK}
\ee
where $\partial^m_j a\,\cdot\,b = (\partial^m/
\partial s_j^m) a(t_j+s_j)b(t_j-s_j)|_{s=0}$ and $\tilde{\partial} =
(\partial_1,(1/2) \partial_2,(1/3)\partial_3,\cdots)$.
In particular, we have from the
coefficients of $y_n$ in (\ref{AAK}),
\be \label{hirota}
\partial_1\partial_n\tau \cdot \tau = 2 \chi_{n+1} (\tilde{\partial})
\tau \cdot \tau
\ee
which are called the Hirota bilinear equations.  One has also
the Fay identity via (cf.
\cite{ab,ch} - c.p. means cyclic permutations) \be \sum_{c.p.}
(s_0-s_1)(s_2-s_3)\tau(t+[s_0]+[s_1])\tau(t+[s_2]+[s_3]) = 0  \label{ZR}
\ee
which can be derived from the bilinear identity (\ref{AAJ}).
Differentiating this in $s_0$, then setting $s_0 = s_3 = 0$, then
dividing by $s_1 s_2$, and finally shifting $t\to t-[s_2]$, leads to
the differential Fay identity,
\begin{eqnarray}
\nonumber
& &\tau(t)\partial\tau(t+[s_1]-[s_2]) - \tau(t+[s_1] -[s_2])\partial
\tau(t) \\ & &= (s_1^{-1}-s_2^{-1}) \left[\tau(t+[s_1]-[s_2]) \tau(t) -
\tau(t+[s_1])\tau(t-[s_2])\right]  \label{AAL} \end{eqnarray}
The
Hirota equations
(\ref{hirota}) can be also derived from (\ref{AAL}) by taking the
limit $s_1 \to s_2$. The identity (\ref{AAL}) will play an important
role later. \\[3mm]\indent
Now for the dispersionless theory (dKP) one can think of fast and slow
variables, etc., or averaging procedures, but simply one takes
$t_n\to\epsilon t_n = T_n\,\,(t_1 = x\to \epsilon x = X)$ in the
KP equation $u_t = (1/4)u_{xxx} + 3uu_x + (3/4)\partial^{-1}u_{yy},
\,\, (y=t_2,\,\,t=t_3)$, with $\partial_n\to \epsilon\partial/\partial
T_n$ and $u(t_n)\to U(T_n)$ to obtain $\partial_T U = 3UU_X +
(3/4)\partial^ {-1}U_{YY}$ when $\epsilon\to 0\,\,(\partial =
\partial/\partial X$ now). Thus the dispersion term $u_{xxx}$ is
removed. In terms of hierarchies we write \be
L_{\epsilon} = \epsilon\partial + \sum_1^{\infty}u_{n+1}(T/\epsilon)
(\epsilon\partial)^{-n}
\label{AAM}
\ee
and think of $u_n( T/\epsilon)= U_n(T) + O(\epsilon)$, etc. One takes
then a WKB form for the wave function with the action $S$
\be \psi = \exp \left[\frac{1}{\epsilon}S(T,\lambda) \right] \label{AAN}
\ee
Replacing now $\partial_n$ by $\epsilon\partial_n$, where $\partial_n =
\partial/\partial T_n$ now, we define $P = \partial S = S_X$. Then
$\epsilon^i\partial^i\psi\to P^i\psi$ as $\epsilon\to 0$ and the equation
$L\psi = \lambda\psi$ becomes
\be
\lambda = P + \sum_1^{\infty}U_{n+1}P^{-n};\, \ \,P = \lambda -
\sum_1^{\infty}P_{i+1}\lambda^{-i}  \label{AAO} \ee
where the second
equation is simply the inversion of the first. We also note from
$\partial_n\psi =
B_n\psi = \sum_0^nb_{nm}(\epsilon\partial)^m\psi$ that one obtains
$\partial_n S = {\cal B}_n(P) = \lambda^n_{+}$ where the subscript (+)
refers now to powers of $P$ (note $\epsilon\partial_n\psi/\psi \to
\partial_n S$). Thus $B_n = L^n_{+}\to {\cal B}_n(P) = \lambda^n_{+} =
\sum_0^nb_{nm}P^m$ and the KP hierarchy goes to \be \partial_n P =
\partial {\cal B}_n
\label{YC}
\ee
which is the dKP hierarchy
(note $\partial_n S = {\cal B}_n\Rightarrow \partial_n P =
\partial{\cal B}_n$).
The action $S$ in (\ref{AN}) can be computed from (\ref{AG}) in the
limit $\epsilon \to 0$ as
\be
\label{action}
S = \sum_{1}^{\infty} T_n \lambda^n - \sum_{1}^{\infty} {\partial_mF
\over m} \lambda^{-m}
\ee
where the function $F=F(T)$ (free energy) is defined by
\be
\label{tau}
\tau = \exp \left[ {1 \over \epsilon^2} F(T) \right] \ee The formula
(\ref{action}) then solves the dKP hierarchy (\ref{YC}), i.e. $P={\cal B}_1 =
\partial S$ and
\be
\label{B}
{\cal B}_n = \partial_n S =
\lambda^n - \sum_{1}^{\infty} {F_{nm} \over m} \lambda^{-m}  \ee
where $F_{nm} = \partial_n\partial_m F$ which play an important role in
the theory of dKP.
\\[3mm]\indent
Now following \cite{ta} one writes the differential Fay identity (\ref{AAL})
with $\epsilon\partial_n$ replacing $\partial_n$, looks at logarithms,
and passes $\epsilon\to 0$ (using (\ref{tau})).  Then
only the second order derivatives survive, and one
gets the dispersionless differential Fay identity
\be \sum_{m,n=1}^{\infty}\mu^{-m}\lambda^{-n}\frac{F_{mn}}
{mn} = \log \left(1- \sum_1^{\infty}\frac{\mu^{-n}-\lambda^{-n}}{\mu-\lambda}
\frac{F_{1n}}{n} \right)
\label{AAT}
\ee
Although (\ref{AAT}) only uses a subset of the Pl\"ucker relations defining
the KP hierarchy it was shown in \cite{ta} that this subset is sufficient to
determine KP; hence (\ref{AAT}) characterizes the function $F$ for dKP.
Following \cite{ch,cl}, we now derive a dispersionless limit of the Hirota
bilinear equations (\ref{hirota}), which we call the dispersionless
Hirota equations. We first note from (\ref{action}) and (\ref{AAO})
that $F_{1n} = nP_{n+1}$
so
\be
\sum_1^{\infty}\lambda^{-n}\frac{F_{1n}}{n} = \sum_1^{\infty}P_{n+1}
\lambda^{-n} = \lambda - P(\lambda)
\label{AAU}
\ee
Consequently the right side of (\ref{AAT}) becomes $\log[\frac{P(\mu) -
P(\lambda)}{\mu-\lambda}]$ and for $\mu\to \lambda$ with $\dot{P} =
\partial_{\lambda}P$ we have
\be
\log\dot{P}(\lambda) = \sum_{m,n=1}^{\infty}\lambda^{-m-n}\frac{F_{mn}}
{mn} = \sum_{j=1}^{\infty} \left(\sum_{n+m=j} {F_{mn} \over mn} \right)
\lambda^{-j}
\label{AAV}
\ee
Then using the elementary Schur polynomial defined in (\ref{AAH}) and
(\ref{AAO}), we obtain
$$
\dot{P}(\lambda) = \sum_0^{\infty} \chi_j(Z_2, \cdots,Z_j) \lambda^{-j}
= 1 + \sum_1^{\infty}F_{1j}
\lambda^{-j-1};$$
\be
Z_i = \sum_{m+n=i} {F_{mn} \over mn}\,\,\,\,(Z_1 = 0)
\label{AAW}
\ee
Thus we obtain the dispersionless Hirota
equations, \be \label{F}
F_{1j} = \chi_{j+1}(Z_1=0,Z_2, \cdots,Z_{j+1})
\ee These can
be also derived directly from (\ref{hirota}) with (\ref{tau}) in the limit
$\epsilon \to 0$ or by expanding (\ref{AAV}) in powers of $\lambda^{-n}$
as in  \cite{ch,cl}).
The equations (\ref{F}) then characterize dKP.
\\[3mm]\indent
It is also interesting to note that the dispersionless Hirota equations
(\ref{F})  can be regarded as algebraic equations for
``symbols" $F_{mn}$, which are defined via (\ref{B}), i.e. \be
{\cal B}_n := \lambda^n_+= \lambda^n - \sum_1^{\infty}\frac{F_{nm}}{m}
\lambda^{-m}
\label{AAY}
\ee
and in fact
\be
F_{nm} = F_{mn} = Res_P[\lambda^m d \lambda^n_+]  \label{AAZ} \ee
Thus
for $\lambda,\,\,P$ given algebraically as in (\ref{AAO}),
with no a priori connection to dKP, and for ${\cal B}_n$ defined as in
(\ref{AAY}) via a formal collection of symbols with two
indices $F_{mn}$, it follows that the dispersionless Hirota equations
(\ref{F}) are nothing but polynomial identities among
$F_{mn}$.  In particular one has from \cite{ch}
\\[3mm]\indent {\bf THEOREM 4.1}.$\,\,$
(\ref{AAZ}) with (\ref{F}) completely characterizes and solves the
dKP hierarchy.
\\[3mm]\indent
Now one very natural way of developing dKP begins with (\ref{AAO}) and
(\ref{YC}) since
eventually the $P_{j+1}$ can serve as universal coordinates (cf. here
\cite{ae} for a discussion of this in connection with topological field
theory = TFT). This point of view is also natural in terms of developing
a Hamilton-Jacobi theory involving ideas from the hodograph $-$ Riemann
invariant approach (cf. \cite{ci,gh,kn,kq,kr} and in
connecting
NKdV ideas to TFT, strings, and quantum gravity.
It is natural here to work with $Q_n := (1/n){\cal B}_n$ and note that
$\partial_n S = {\cal B}_n$ corresponds to $\partial_n P = \partial
{\cal B}_n = n\partial Q_n$.
In this connection one often uses different time variables, say $T'_n =
nT_n$, so that $\partial'_nP = \partial Q_n$, and $G_{mn} = F_{mn}/mn$ is
used in place of $F_{mn}$. Here however we will retain the $T_n$ notation
with $\partial_n S = nQ_n$ and $\partial_n P = n\partial Q_n$ since one
will be connecting a number of formulas to standard KP notation. Now given
(\ref{AAO}) and (\ref{YC}) the equation
$\partial_n P = n\partial Q_n$ corresponds to Benney's moment equations
and is equivalent to a system of Hamiltonian equations defining the dKP
hierarchy (cf. \cite{ci,kn});
the Hamilton-Jacobi equations are
$\partial_n S = nQ_n$ with Hamiltonians $nQ_n(X, P=\partial S)$).
There is now an important formula involving the functions $Q_n$ from
\cite{kn}, namely
the generating function of $\partial_P Q_n(\lambda)$ is given by
\be \frac{1}{P(\mu) - P(\lambda)} = \sum_1^{\infty}\partial_P Q_n(\lambda)
\mu^{-n}  \label{ABF}
\ee
In particular one notes
\be \oint_{\infty} {\mu^n \over
P(\mu) - P(\lambda)} d\mu = \partial_P Q_{n+1}(\lambda) \ , \label{canonicalT}
\ee
which gives a key formula in the Hamilton-Jacobi method for the dKP \cite{kn}.
Also note here that the function
$P(\lambda)$ alone provides all the information necessary for the dKP theory.
It is proved in \cite{ch} that
\\[3mm]\indent {\bf THEOREM 4.2}.$\,\,$
The kernel formula (\ref{ABF}) is
equivalent to the dispersionless differential Fay identity (\ref{AAT}).
\\[3mm]\indent
The proof uses
\be \label{ShP}
\partial_P Q_n = \chi_{n-1}(Q_1,\cdots,Q_{n-1})
\ee
where $\chi_n(Q_1,\cdots,Q_n)$ can be expressed as a polynomial in $Q_1 = P$
with the coefficients given by polynomials in the $P_{j+1}$.
Indeed
\be
\chi_n = det
\left[
\begin{array}{ccccccc}
P & -1 & 0 & 0 & 0 & \cdots & 0\\
P_2 & P & -1 & 0 & 0 & \cdots & 0\\
P_3 & P_2 & P & -1 & 0 & \cdots & 0\\
\vdots & \vdots & \vdots & \vdots & \vdots & \ddots & \vdots\\ P_n & P_{n-1}
& \cdots & P_4 & P_3 & P_2 & P
\end{array}
\right] = \partial_P Q_{n+1}
\label{ABM}
\ee
and this leads to the observation that the $F_{mn}$ can be
expressed
as polynomials in $P_{j+1} = F_{1j}/j$.
Thus
the dispersionless Hirota
equations can be solved totally algebraically via $F_{mn} =
\Phi_{mn}(P_2,P_3,\cdots,P_{m+n})$ where $\Phi_{mn}$ is a polynomial in
the $P_{j+1}$ so the $F_{1n} = nP_{n+1}$ are generating elements for the
$F_{mn}$, and serve as universal coordinates.  Indeed
formulas such as (\ref{ABM}) and (\ref{ShP})
indicate that in fact dKP theory can be characterized
using only elementary Schur polynomials since these provide all the
information necessary for the kernel (\ref{ABF}) or equivalently for the
dispersionless differential Fay identity. This amounts
also to observing that in the passage from KP to dKP only certain Schur
polynomials survive the limiting process $\epsilon\to 0$. Such terms
involve second derivatives of $F$ and these may be characterized in terms of
Young diagrams with only vertical or horizontal boxes. This is also related
to the explicit form of the hodograph transformation where one needs only
$\partial_P Q_n = \chi_{n-1}(Q_1,\cdots,Q_{n-1})$ and the $P_{j+1}$ in the
expansion of $P$ (cf. \cite{ch}).
Given KP and dKP theory we can now discuss nKdV or dnKdV easily although
many special aspects of nKdV for example are not visible in KP.  In
particular for the $F_{ij}$ one will have $F_{nj} = F_{jn} = 0$
for dnKdV.
We note also (cf. \cite{km}) that from (\ref{ShP}) one has
\be
\frac{1}{P(\mu)-P(\lambda)} = \sum_1^{\infty}\partial_PQ_n\mu^{-n} =
\sum_0^{\infty}\chi_n(Q)\mu^{-n}=exp(\sum_1^{\infty}Q_m\mu^{-m})
\label{ABN}
\ee

\section{HURWITZ SPACES}
\renewcommand{\theequation}{5.\arabic{equation}}\setcounter{equation}{0}

The preceeding development leads one to work in the framework of Hurwitz
spaces and Frobenius manifolds developed by Dubrovin (cf. \cite{di}).
In fact we could have started here by hindsight but felt it desirable
to pursue the more tortuous path traced in order to illustrate various
points of view, derivations, and techniques (in particular the nature
of averaging).  We will extract here freely from \cite{di}.
\\[3mm]\indent
Hurwitz spaces are defined to be moduli spaces of pairs $(\Sigma_g,\lambda)$
where $\Sigma_g$ is a smooth algebraic curve of genus $g$ and $n+1$
sheets; $\lambda$
is a meromorphic function on $\Sigma_g$ of degree $n+1$ (which can be
used to realize $\Sigma_g$ as an n-sheeted covering
over ${\bf C}P^1$).  One will assume here that the
ramification over $\infty$ is fixed as indicated below.  Thus let
$M = M_{g;n_0,\cdots,n_m}$ be a moduli space of dimension $n=2g+n_0+\cdots
+n_m+2m$ of sets $(\Sigma_g;\infty_0,\cdots,\infty_m;\lambda)$ where
$\Sigma_g$ is a Riemann surface with marked points $\infty_0,\cdots,
\infty_m$ and $\lambda$ is a
meromorphic function $\lambda:\,\,\Sigma_g\to CP^1$ with
$\lambda^{-1}(\infty) = \infty_0\cup\cdots\cup\infty_m$
and having
degree $n_i+1$ near $\infty_i$.  The critical values of $\lambda$,
defined via
\be
u^j = \lambda(P_j);\,\,d\lambda|_{P_j} = 0\,\,(j=1,\cdots,n)
\label{IA}
\ee
are to be local coordinates in open domains $\hat{M}$ where $u^j\not=
u^i$ for $i\not= j$ (here $\hat{M}$ is to denote a covering of our
eventual Hurwitz space in which $\lambda$ is a variable).
These are the ramification points of the surface
$\lambda:\,\,\Sigma\to CP^1$ and the $P_j$ are branch points of
$\Sigma_g$.  One asks also that the one dimensional affine group
acts on $\hat{M}$ via
\be
(\Sigma_g;\infty_0,\cdots,\infty_m;\lambda)\to (\Sigma_g;\infty_0,
\cdots,\infty_m;a\lambda+b);\,\,u^i\mapsto au^i+b
\label{IB}
\ee
We note that $\lambda(P_j) = u^j$ is a point in a moduli space while
$P_j$ refers to a particular Riemann surface.  There is a strong
interaction between $\Sigma_g$ and the function $\lambda$ as indicated
in the examples to follow.

\subsection{Examples}
({\bf A})  For $g=0,\,m=0,\,n_0=n$ the
Hurwitz space consists of all polynomials of the form
$\lambda(p) = p^{n+1} + a_np^{n-1} +\cdots+ a_1\,\,(a_i\in {\bf C})$
(here $p$ denotes a point on $\Sigma_g$ and strictly one should
use $P$ - cf. below).
The affine transformations $\lambda\mapsto a\lambda + b$ act via
$p\mapsto a^{1/(n+1)}p,\,\,a_i\mapsto a_ia^{(n-i+1)/(n+2)}\,\,(i> 1),\,\,
a_1\mapsto aa_1+b$. This corresponds to $A_n$ as in Section 5.3 and the
extension to $M_{gn}$ is displayed in Section 7.3.
({\bf B})  For $g=0,\,m=n,\,n_0=\cdots =n_m = 0$
the Hurwitz space is all rational functions of the form $\lambda(p)
= p +\sum_1^n[q_i/(p-p_i)]$ with affine group action $p\mapsto ap + b,\,\,
p_i\mapsto p_i+(b/a),\,\,q_i\mapsto aq_i$.
({\bf C})  For $g>0,\,\,m=0,\,\,n_0=1$ the Hurwitz space consists of all
hyperelliptic curves $\mu^2 = \prod_1^{2g+1}(\lambda-u^j)$.  The critical
values $u^1,\cdots,u^{2g+1}$ of the projection $(\lambda,\mu)\mapsto
\lambda$ are the local coordinates on the moduli space
(note $\mu^2 = Poly(\lambda,u^j)=P(\lambda,u^j)\Rightarrow 2\mu=
\partial_{\lambda}P(d\lambda/d\mu\Rightarrow d\lambda/d\mu=0$ for
$\mu=0$ or $\lambda=u^j$).  Other examples are indicated below.
$\bullet$
\\[3mm]\indent  Evidently example ({\bf C})
is defined somewhat differently and we will try to clarify this
later.  One notes here from \cite{ja} for example that if $\Sigma_g$
is the Riemann surface of an irreducible algebraic equation $P(z,w)
=0$ of degree $n$ in $w$ and if the branch points have orders $n_i$ then
the genus is $g=1-n+(1/2)\sum_1^r n_i$ (Riemann-Hurwitz formula).  Thus
e.g. for $\lambda(p)$ in the first example ({\bf A}) (with $p\sim w$)
if one assumes $\lambda'(p) = \prod_1^{n}(p-p_s)$ and distinct $p_s$
then $n_s = 1$ and there is a branch point of order $n$ at $\infty$.
Hence $g=1-(n+1)+(1/2)\sum_1^n 1 + (1/2)n = 0$ as required here.  This
shows also how an n-sheeted surface can have genus $0$.  For
hyperelliptic situations we recall the following facts (cf. \cite{za}).
Consider $R=\mu^2 = \prod_0^{2m}(\lambda-\lambda_k)$ as in the beginning
of Section 5.1 with branch points $\lambda_k$ and $\infty$.  One makes
cuts $(\lambda_1,\lambda_2),\,(\lambda_3,\lambda_4),\cdots,(\lambda_{2m-1},
\lambda_{2m})$ as spectral gaps with $a_i$ cycles corresponding to
$(\lambda_{2i-1},\lambda_{2i})$ around these cuts vertically and $b_j$
cycles horizontal around $(\lambda_0,\lambda_1),\cdots,(\lambda_{2m},
\infty)$ for example (see \cite{ca,za}).  This produces a surface with
$g=m$ visible holes.  Note the Riemann-Hurwitz formula gives also
($w\sim \mu$) $g=1-2+(1/2)\sum_0^{2m} 1 + (1/2) = m$.  The
hyperelliptic situation will be developed further below.
\\[3mm]\indent
Now one constructs a covering $\hat{M}$ of $M = M_{g;n_0,\cdots,n_m}$ in order
to describe multivalued quadratic differentials for which the one forms
$\Omega_Q = \sum_1^ndu^iRes_{P_i}(Q/d\lambda)$ will define metrics on the
Hurwitz space via $\Omega(a\cdot b) = <a,b>_{\Omega}$.  Thus take
$\hat{M}$ to consist of sets $(\Sigma_g;\infty_0,\cdots,\infty_m;\lambda;
k_0,\cdots,k_m;a_1,\cdots,a_g,b_1,\cdots,b_g)\in M$ with the same
$(\Sigma_g,\infty_0,\cdots,\infty_m,\lambda)$ as before and with a marked
symplectic basis $a_i,\,b_i\in H_1(\Sigma_g,{\bf Z})$ and marked branches
of roots near $\infty_i$ of orders $n_i+1\,\,(i = 0,\cdots,m)$, namely
$k_i^{n_i+1}(P) = \lambda(P)$ near $\infty_i$.  The admissable quadratic
differentials will be constructed as $Q = \phi^2$ for certain
primary differentials
$\phi$ on $\Sigma_g$ or on a covering.
The primary differentials have the following forms.  ({\bf D})  Normalized
Abelian differentials of the second kind on $\Sigma_g$ with poles only
at $\infty_0,\cdots,\infty_m$ of orders less than the corresponding
orders of the differential $d\lambda$.  Explicitly $\phi = \phi_
{t^{i;\alpha}}\,\,(i = 0,\cdots,m,\,\,\alpha = 1,\cdots,n_i)$ is the
normalized Abelian differential of second kind with a pole at $\infty_i$
of the form
\be
\phi_{t^{i;\alpha}} = -\frac{1}{\alpha}dk_i^{\alpha} + regular\,\,terms
\,\,near\,\,\infty_i;\,\,\oint_{a_j}\phi_{t^{i;\alpha}} = 0
\label{IC}
\ee
({\bf E})  Next one considers
\be
\phi = \sum_1^m\delta_i\phi_{v^i}\,\,(i = 1,\cdots,m)
\label{ID}
\ee
with the $\delta_i$ independent of the point in $\hat{M}$.  Here
$\phi_{v^i}$ is one of the normalized Abelian differentials of second kind
on $\Sigma_g$ with a pole only at $\infty_i$ with principal part of the
form
\be
\phi_{v^i} = -d\lambda + regular\,\,terms\,\,near\,\,\infty_i;\,\,
\oint_{a_j}\phi_{v^i} = 0
\label{IE}
\ee
({\bf F}) Next take
\be
\phi = \sum_1^m\alpha_i\phi_{w^i};\,\,\oint_{a_j}\phi = 0
\label{IF}
\ee
with $\alpha_1,\cdots,\alpha_m$ independent of the point on $\hat{M}$ and
$\phi_{w^i}$ is the normalized Abelian differential of the third
kind with simple poles at $\infty_0$ and $\infty_i$ with residues $-1$ and
$+1$ respectively.
({\bf G})  Next look at $\phi = \sum_1^g\beta_i\phi_{r^i}$ with
$\beta_i$ independent
of the point in $\hat{M}$ and $\phi_{r^i}$ is the normalized multivalued
differential with increments along the cycles $b_j$ of the form
\be
\phi_{r^i}(P+b_j) - \phi_{r^i}(P) = -\delta_{ij}d\lambda;\,\,
\oint_{a_j}\phi_{r^i} = 0
\label{IG}
\ee
and without other singularities.
({\bf H}) Finally $\phi = \sum_1^g\gamma_i\phi_{s^i}$ with $\gamma_i$
independent of the point on $\hat{M}$ and $\phi_{s^i}$ denotes
holomorphic differentials normalized by $\oint_{a_j}\phi_{s^i} = \delta_{ij}$.
Then one will pick a primary differential $Q=\phi^2$ to
develop a Frobenius structure on
$\hat{M}$.  In constructing the superpotential one introduces a
multivalued function $p$ on $\Sigma_g$ via $p(P) = v.p.\,\int_{\infty_0}^P
\phi$ where divergent parts have been subtracted in the principal
value integral.  Then $\phi = dp$ and $\lambda = \lambda(p)$ can be used.

\subsection{WDVV and Frobenius manifolds}
We prepare the way by extracting
some material on the WDVV equations and Frobenius manifolds from \cite{di,dj}.
We will sketch (without proofs) only
a brief selection of this
extensively developed material.  The main creature
is a quasihomogeneous function $F(t),\,\,t\sim (t_1,...,t_n)$
such that the third derivatives $\partial_{\alpha}\partial_{\beta}
\partial_{\gamma}F = c_{\alpha\beta\gamma}(t)$ correspond to
correlation functions $<\phi_{\alpha}\phi_{\beta}\phi_{\gamma}>_0$
and in the LG models $exp(F)$ is the dispersionless tau
function (we use $t_k$ and $T_k$ interchangably here - $T_k$ is correct
in our notation but $t_k$ is used in \cite{di,dj}).
These functions $c_{\alpha\beta\gamma}$ satisfy a system
of partial differential equations (PDE) called the WDVV (Witten-
Dijkgraaf-Verlinde-Verlinde) equations based on algebraic properties of
TFT.
Thus (cf. \cite{di,dj}) one takes a 2-D TFT with say n
primary fields $\phi_i$ and $<\,\,,\,\,>_0\,\sim$ genus zero correlation
function.  The two point functions $<\phi_{\alpha}\phi_{\beta}>_0 =
\eta_{\alpha\beta} = \eta_{\beta\alpha}$ determine a nondegenerate
scalar product (see below) on the space of primaries and the $c_{\alpha
\beta\gamma}$ determine the structure of the primary chiral algebra.
More axiomatically one looks for $F(t)$ with $c_{\alpha\beta\gamma} =
\partial_{\alpha}\partial_{\beta}\partial_{\gamma}F$ such that
$\eta_{\alpha\beta} = c_{1\alpha\beta}$ is a constant nondegenerate
matrix with $\eta^{\alpha\beta} = (\eta_{\alpha\beta})^{-1}$.  Then one
specifies $c^{\gamma}_{\alpha\beta} = \eta^{\gamma\epsilon}
c_{\epsilon\alpha\beta}$ and such $c_{\alpha\beta\gamma}(t)$ must
define in the n-dimensional space with basis $e_i\,\,(1\leq i\leq n)$
an associative algebra $A_t$ with
\be
e_{\alpha}\cdot e_{\beta} = c^{\gamma}_{\alpha\beta}(t)e_{\gamma};\,\,
c^{\beta}_{1\alpha}(t) = \delta^{\beta}_{\alpha}
\label{IH}
\ee
Here $F$ should be quasihomogeneous in the sense $F(c^{d_1}t_1,...,
c^{d_n}t_n) = c^{d_F}F(t_1,...,t_n)$ (we also
use $t_i$ and $t^i$ interchangably)
and this can be expressed infinitesimally via $E = E^{\alpha}(t)\partial_
{\alpha}$ and (${\cal L}\sim$ Lie derivative)
\be
{\cal L}_E F = E^{\alpha}(t)\partial_{\alpha}F(t) = d_F\cdot F(t);\,\,
E^{\alpha} = d_{\alpha}t^{\alpha}
\label{II}
\ee
Note for $e = \partial_1$ ($\sim$ unity vector field) ${\cal L}_E e
=-d_1e$.  One can modify quasihomogeneity by requiring
\be
{\cal L}_E F = d_F F + A_{\alpha\beta}t^{\alpha}t^{\beta} +
B_{\alpha}t^{\alpha} + C
\label{IJ}
\ee
since the third derivatives are not affected.  Further if
$d_F\not= 0,\,\,d_F - d_{\alpha}\not= 0,\,\,d_F - d_{\alpha} - d_{\beta}
\not= 0$ for any $\alpha,\,\beta$ then the extra terms in (\ref{IJ})
could be killed by adding a quadratic function to $F$.
The degrees $d_1,...,d_n,\,d_F$
are well defined up to a nonzero factor and we consider here only the case
$d_1\not= 0$ with normalization so that $d_1 = 1$.  Often one writes
$d_{\alpha} = 1-q_{\alpha}\,\,(q_1 = 0),\,\,d_F = 3-d,$ so $q_n = d$
and $q_{\alpha} + q_{n-\alpha +1} = d$.  Then associativity implies the
WDVV equations
\be
\partial_{\alpha}\partial_{\beta}\partial_{\lambda}F\eta^{\lambda\mu}
\partial_{\gamma}\partial_{\delta}\partial_{\mu}F = \partial_{\gamma}
\partial_{\beta}\partial_{\lambda}F\eta^{\lambda\mu}\partial_{\alpha}
\partial_{\delta}\partial_{\mu}F
\label{IK}
\ee
for $1\leq \alpha,\,\beta,\,\gamma,\,\delta\leq n$.  A solution $F$ of
(\ref{IK}) is called a primary free energy.
\\[3mm]\indent
Now one says $A$ is a (commutative) Frobenius algebra if (1) $A$ is a
commutative associative algebra with unity $e$ (2) $(a,b)\mapsto
<a,b>:A\times A\to {\bf C}$ is ${\bf C}$ bilinear symmetric nondegenerate
with $<ab,c> = <a,bc>$.  Then $<a,b> = <e,ab>\,\,(e\sim e_1$) will be a
suitable scalar product and for $e_i$ a basis in $A$ one can specify
$<e_i,e_j> = \eta_{ij},\,\,e_ie_j = c^k_{ij}e_k$ to obtain a structure
of Frobenius algebra on $A$ ($c_{ijk} = \eta_{is}c^s_{jk} =
c_{jik} = c_{ikj}$).  A theory of Frobenius manifolds can now be
developed as follows.  $M$ is a Frobenius manifold if a structure of
Frobenius algebra is specified on every tangent space $T_t M$ depending
smoothly on $t$ such that (1) $<\,,\,>$ is a flat metric (2) $\nabla e = 0$
for the Levi-Civita connection $\nabla$ of $<\,,\,>$ (3) For $c(u,v,w)
= <u\cdot v,w>$ one requires $(\nabla_z c)(u,v,w)$ to be symmetric in
$u,v,w,z$ and (4) $\nabla(\nabla E) = 0$ for a vector field $E$ =
Euler vector field (cf. (\ref{IJ})) such that the corresponding one
parameter diffeomorphism group acts via conformal transformations on
the metric $<\,,\,>$ and by rescalings on the F-algebra $T_t M$.
The infinitesimal form of (4) is
\be
\nabla_{\gamma}(\nabla_{\beta} E^{\alpha}) = 0;\,\,{\cal L}_E c^{\gamma}_
{\alpha\beta} = c^{\gamma}_{\alpha\beta};\,\,{\cal L}_E e = -e;\,\,
{\cal L}_E\eta_{\alpha\beta} = D\eta_{\alpha\beta}\,\,(D=2-d)
\label{IL}
\ee
One proves that any solution of WDVV with $d_1\not= 0$ defined in a
domain of $t$ determines a structure of F-manifold via
\be
\partial_{\alpha}\cdot\partial_{\beta} = c^{\gamma}_{\alpha\beta}
\partial_{\gamma};\,\,<\partial_{\alpha},\partial_{\beta}> = \eta_
{\alpha\beta};\,\,\partial_{\alpha} = \frac{\partial}{\partial t_{\alpha}};
\,\,e=\partial_1
\label{IM}
\ee
with $E$ of (\ref{IJ}).  An example of particular interest arises from
$M = \{\lambda(p) = p^{n+1} + a_np^{n-1} + \cdots + a_1\}$ with
$T_{\lambda}M\sim$ polynomials of degree $<n$.  Then
the Frobenius algebra $A_{\lambda}$ on
$T_{\lambda}M$ is ${\bf C}[p]/(\lambda'(p))$ with $<f,g>_{\lambda} =
Res_{p=\infty}[f(p)g(p)/\lambda'(p)]$.  Here $e = \partial/\partial a_1$ and
$E = (1/n+1)\sum(n-i+1)a_i\partial_i$.

\subsection{$A_{n-1}$}
We will write out first the
situation for the $A_{n-1}$ topological
minimal model following \cite{dj} since it is more detailed than
\cite{di} and will give a more complete picture.  There is a lot of
duplication with previous formulas but we will stay here with the
notation of \cite{dj}.  The dispersionless theory will also be presented
again in this notation.  Thus one considers the coupling space $M$ of
all polynomials of degree n of the form $M = \{\lambda(p) = p^n +
a_{n-2}p^{n-2} + \cdots + a_0\}$ and the associated F-algebra is
$A_{\lambda} = {\bf C}[p]/(\lambda'(p)=0)$ with scalar product
$<f(p),g(p)> = -(1/n)Res_{p=\infty}[f(p)g(p)/\lambda'(p)]$
($p\sim P$).  There is an
affine structure on $M$ defined as follows.  Let $\phi_{\alpha}(p,
\lambda),\,\,\alpha = 1,...,n-1$, be an orthogonal basis of $A_{\lambda}$
satisfying
\be
<\phi_{\alpha},\phi_{\beta}> = \eta_{\alpha\beta} = \delta_{\alpha +
\beta,n};\,\,deg\,\phi_{\alpha} = \alpha -1
\label{IN}
\ee
Generally fields can be defined via
\be
\phi_{\alpha}(p,\lambda) = \frac{n}{\alpha}\partial_p\lambda_{+}^
{\frac{\alpha}{n}}
\label{IO}
\ee
The dependence of $\lambda (p)$ (i.e. of its coefficients) on the
flat coordinates $t_{\alpha}$
(better $T_{\alpha}$) is determined via
\be
\partial_{\alpha}\lambda(p) = -\phi_{\alpha}(p,\lambda);\,\,\,\alpha =
1,\cdots,n-1
\label{IP}
\ee
We recall the Gelfand-Dickey hierarchies now via $\partial_q L =
[L,L_{+}^{\frac{q}{n}}]$ for $L = \partial^n + a_{n-2}\partial^{n-2} +
\cdots + a_0$.  The dispersionless hierarchy is formed as before
via $X = \epsilon x,\,\,T_q = \epsilon t_q,\,\,\epsilon\to 0$ and one
obtains in the spirit of \cite{fa} ($\partial_q\sim\partial/\partial
T_q$ now and $\partial\sim\partial/\partial X$)
\be
\partial_{\alpha} dp|_{\lambda = c} = \partial_X d\Phi_{\alpha}|_{\lambda = c};
\,\,\Phi_{\alpha} = \lambda_{+}^{\frac{\alpha}{n}}
\label{IQ}
\ee
This can be rewritten via Wronskians as
\be
\partial_{\alpha}\lambda(p) = \partial_p\lambda(\partial_X\Phi_{\alpha})|_{p=k}
-\partial_p\Phi_{\alpha}(\partial_X\lambda)|_{p=k}
\label{IR}
\ee
(these are all standard calculations in the
dispersionless theory (cf. \cite{ta}).
Now one has diagonal coordinates (Riemann invariants) for our
dispersionless hierarchy as the critical values $u_1,\cdots,u_{n-1}$
of $\lambda(p)$, i.e.
\be
u_i = \lambda(p_i);\,\,\lambda'(p_i) = 0
\label{IS}
\ee
and the characteristic speeds are (cf. also \cite{dj}, equation (5.58),
where $\partial_p\Phi_q\sim -(\Omega_q/\Omega_1),\,\,\Omega_1\sim dp$
in the small phase space)
\be
v_{q,i}(u) = -\partial_p\Phi_q|_{p=p_i};\,\,\partial_q u^i =
v_{q,i}\partial_X u^i\,\,(i=1,\cdots,N=n-1)
\label{IT}
\ee
We assume here that $\lambda'(p)=0$ has simple zeros so the algebra
$A_{\lambda}$ will be decomposable by definition and the corresponding
metric $ds^2$ in these coordinates is
\be
ds^2 = \sum_1^N\frac{(du_i)^2}{\lambda''(p_i)}
\label{IU}
\ee
Now one knows that the functions on $M$
\be
h_{\alpha,q} = -\frac{1}{(\alpha/n)_q}Res_{p=\infty}\lambda^
{(\alpha/n)+q}dp;\,\,\alpha = 1,\cdots,n-1,\,\,q\geq 0
\label{IV}
\ee
where $(a)_q = a(a+1)\cdots(a+q-1)$, are basic conservation
densities and
\be
\partial h_{\alpha,q} = h_{\alpha,q-1}\,\,(q\geq 1);\,\,\partial
h_{\alpha,0} = c\,\,(\partial = \sum_1^N\partial_i,\,\,\partial_i
=\frac{\partial}{\partial u_i})
\label{IW}
\ee
(recall $s_n\sim Res\,L^n\sim nH^1_{n-1}$ and $Res\,L^n\to Res_P\lambda^n$
so (\ref{IV}) is natural enough).
This follows from noting that a translation $\lambda\mapsto \lambda +
\epsilon,\,\,p\mapsto p,\,\,a_i\mapsto a_i\,\,(i\not= 0),\,\,a_0
\mapsto a_0+\epsilon$ is equivalent to translation along $u_1 +\cdots +
u_N$ so that
\be
\partial h_{\alpha,q} = D_{\epsilon}h_{\alpha,q}(u_i+\epsilon)|_
{\epsilon = 0} =
\label{IX}
\ee
$$\frac{1}{(\alpha/n)_q}Res_p D_{\epsilon}[(\lambda + \epsilon)^{(\alpha/n)
+q}dp(\lambda + \epsilon)]_{\epsilon = 0} =\frac{1}{(\alpha/n)_{q-1}}
Res_p\lambda^{(\alpha/n)+q-1}dp$$
This gives (\ref{IW}) for $q\geq 1$ and $\partial h_{\alpha,0} =
Res_p\lambda^{(\alpha/n)-1}dp = -\delta_{\alpha,n-1}$.  In particular
for the flat coordinates we have
\be
T_{\alpha} = -nRes_p\frac{\lambda^{(n-\alpha)/n}}{n-\alpha}dp;\,\,
\alpha = 1,\cdots,N=n-1
\label{IY}
\ee
(cf. Remark 3.5 and Example 3.6).
One has in fact a generating function for the $h_{\alpha,q}$
in the form
\be
x_{\alpha}(t,z) = -\frac{n}{\alpha}Res_p\,{}_1F_1(1;1+\frac{\alpha}{n};
z\lambda)dp;\,\,\alpha = 1,\cdots,N
\label{IZ}
\ee
where ${}_1F_1(a;c;z) = \sum_0^{\infty}\frac{(a)_m z^m}{(c)_m m!}$ is
the Kummer hypergeometric function.
\\[3mm]\indent
Now returning to \cite{di} let $\phi$ be one of our primary differentials
and let $\hat{M}_{\phi}$ be the open domain in $\hat{M}$ specified by
the condition $\phi(P_i)\not= 0\,\,(i=1,\cdots,n)$.
The main result says that for any primary differential $\phi$ of the types
({\bf D}) to ({\bf H}) the multiplication $\partial_i\cdot\partial_j
=\delta_{ij}\partial_i$ for $\partial_i = \partial/\partial u^i$, the
unity and Euler vector field $e = \sum_1^n\partial_i$ and $E = \sum_1^n
u^i\partial_i$, and the one form $\Omega_{\phi^2}$ determine on
$\hat{M}_{\phi}$
a structure of Frobenius manifold.  The flat coordinates $T^A,\,\,A = 1,\cdots,
n$, consist of the five parts
\be
T^A = (T^{i;\alpha},\,0\leq i\leq m,\,1\leq \alpha\leq n;\,p^j,\,q^j,\,
1\leq j\leq m;\,r^j,\,s^j,\,1\leq j\leq g)
\label{JA}
\ee
where
\be
T^{i;\alpha} = Res_{\infty_i}k_i^{-\alpha}pd\lambda,\,\,i=0,\cdots,m,\,
\alpha = 1,\cdots,n_i;
\label{JB}
\ee
$$p^i = v.p.\int_{\infty_0}^{\infty_i}dp,\,\,q^i = -Res_{\infty_i}
\lambda dp,\,\,i=1,\cdots,m;$$
$$r^i = \oint_{b_i}dp,\,\,s^i = -\frac{1}{2\pi i}\oint_{a_i}\lambda dp,\,\,
i=1,\cdots,g$$
The metric will have the form
\be
\eta_{T^{i;\alpha}T^{j,\beta}} = \frac{1}{n_i+1}\delta_{ij}\delta_
{\alpha+\beta,n_i+1};
\label{JC}
\ee
$$\eta_{v^iw^j} = \frac{1}{n_i+1}\delta_{ij};\,\,\eta_{r^ks^j} =
\frac{1}{2\pi i}\delta_{kj}$$
The function $\lambda = \lambda(p)$ will be the superpotential of this
Frobenius manifold in the sense that $u^i(T) = \lambda(q^i(T),T)$
where $(d\lambda/dp)(q^i(T)) = 0\,\,(i=1,\cdots,n$), plus other stipulations
which will be indicated as needed.  Finally for any other primary
differential $\phi$ the one form $\Omega_{\phi}$ is an admissable one
form on the Frobenius manifold in a sense to be indicated.
\\[3mm]\indent
We check some of this following \cite{di}.  First one declares that the
$u^i\,\,(1\leq i\leq n)$ are the canonical coordinates for a multiplication
$\partial_i\cdot\partial_j = \delta_{ij}\partial_i\,\,(\partial_i
=\partial/\partial u^i)$ and defines the metric corresponding to
$\Omega_{\phi^2}$ via $<\partial',\partial''>_{\phi} = \Omega_{\phi}
(\partial'\cdot\partial'')$ for any two tangent vector fields $\partial',
\partial''$ on $\hat{M}$.  This metric will also be diagonal in these
canonical coordinates and one writes
\be
ds^2_{\phi} = \sum_1^n\eta_{ii}(du^i)^2;\,\,\eta_{ii} = Res_{P_i}\frac
{\phi^2}{d\lambda}
\label{JD}
\ee
One can show now that, for any primary differential,
this is a flat Darboux-Egoroff metric in an open domain $\hat{M}_{\phi}$
leading to a Frobenius structure on $\hat{M}_{\phi}$.  This holds for
any $\phi$ and $\cup\hat{M}_{\phi}$ provides a Frobenius structure on
$\hat{M}$.  The proof is nontrivial and we refer to \cite{di}.  Next
one shows
\be
\partial_{T^{i;\alpha}}\lambda(p)dp = -\phi_{T^{i;\alpha}};
\label{JE}
\ee
$$\partial_{v^i}\lambda(p)dp = -\phi_{v^i};\,\,\partial_{w^i}\lambda(p)
dp = -\phi_{w^i};$$
$$\partial_{r^i}\lambda(p)dp = -\phi_{r^i};\,\,\partial_{s^i}\lambda(p)
dp = -\phi_{s^i}$$
This is achieved via the thermodynamic identity
\be
\partial_{\alpha}(\lambda dp)|_{p=c} = -\partial_{\alpha}(pd\lambda)|_
{\lambda = c'}
\label{JF}
\ee
For (\ref{JF}) note that $\lambda = \lambda(p(\lambda,T),T)$ implies
$0=\partial_{\alpha}\lambda|_{p=c} + (dp/d\lambda)\partial_{\alpha}p|_
{\lambda=c'}$ which says $\partial_{\alpha}\lambda|_{p=c}dp =
-\partial_{\alpha}p|_{\lambda=c'}d\lambda$ which is what (\ref{JF}) means.
Now for example derivatives $(\partial p(\lambda)/\partial T^A)|_{\lambda=c'}$
are holomorphic on $\Sigma_g/\infty$ away from the $P_j$ where
they have simple poles, and $d\lambda$ vanishes precisely at such $P_j$
so (\ref{JF}) is holomorphic on $\Sigma_g/\infty$.  A calculation of
singularities gives (\ref{JE}).  Finally let $T^A$ be one of the
coordinates (\ref{JB}).  Then to prove (\ref{JC}) look at $\phi_A =
-\partial_{T^A}\lambda dp$.  Some calculation (not shown here) shows that
\be
<\partial_{T^A},\partial_{T^B}>_{\phi} = \sum_{|\lambda|<\infty}
Res_{d\lambda=0}\frac{\phi_A\phi_B}{d\lambda} = \partial_e
<\phi_A\phi_B>
\label{JG}
\ee
where $<\,\,>$ is a complicated explicit bilinear pairing of suitable
differentials.  further nontrivial calculation yields (\ref{JC}).  The
structure constants for the Frobenius structure will now be
\be
c_{ABC}= \sum_1^nRes_{P_i}\frac{\phi_{T^A}\phi_{T^B}\phi_{T^C}}
{d\lambda dp}
\label{JH}
\ee
and the Frobenius structure can be extended to the entire moduli space
$\hat{M}$ via the condition that
$(\phi_{T^A}\phi_{T^B} - c_{AB}^C\phi_{T^C}dp)/d\lambda$ be holomorphic
for $|w|<\infty$.  It turns out also that
\be
F = -\frac{1}{2}<pd\lambda\,\, pd\lambda>;\,\,\partial_{T^A}\partial_{T^B}F =
-<\phi_A\phi_B>
\label{JI}
\ee
In particular
\be
\partial_{s^{\alpha}}\partial_{s^{\beta}}F = -\tau_{\alpha\beta} =
\oint_{b_{\beta}}\phi_{s^{\alpha}}
\label{JJ}
\ee
This means that the Jacobians $J(\Sigma_g) = {\bf C}^g/{m+\tau n},\,\,
m,n\in {\bf Z}^g$ are Lagrangian manifolds for the symplectic structure
$\sum_1^g ds^{\alpha}\wedge dz_{\alpha}$ where $z_1,\cdots,z_g$ are
natural coordinates on $J(\Sigma_g)$ coming from the linear coordinates
in ${\bf C}^g$.  Thus the Jacobians are complex Liouville tori and the
coordinates $z_{\alpha},s^{\alpha}$ are complex action-angle variables
on the tori.

\section{CAUCHY TYPE KERNELS}
\renewcommand{\theequation}{6.\arabic{equation}}\setcounter{equation}{0}

\subsection{Review of KdV}
For convenience we begin with a hyperelliptic Riemann surface (corresponding
to the KdV situation) in the form $R^2(\lambda) = \prod_0^{2g}(\lambda -
\lambda_k)$ as in Remark 3.3.  Then there are normalized holomorphic
differentials
\be
\omega_i\sim\frac{\sum_1^g\alpha_{ij}\lambda^{j-1}d\lambda}{R(\lambda)};\,\,
\oint_{a_k}\omega_i = \delta_{ik}
\label{FA}
\ee
(again $\omega_i$ is used instead of $d\omega_i$).
We refer here to \cite{ae,ba,bb,ca,cj,fa,gf,gg,km,kn} for hyperelliptic
surfaces.  There is often some difference in notation or normalization in
the literature and we will try to specify explicitly our selection here
(to follow \cite{za}).
Thus with $R^2$ as above one has branch points $\lambda_0,\cdots,\lambda_{2g},
\infty$
(whereas for $R^2 = \prod_0^{2g-1}(\lambda-\lambda_k)$ there are branch
points $\lambda_0,\cdots,\lambda_{2g-1}$).  Again $a_i\sim
(\lambda_{2i-1},\lambda_{2i}),\,\,i = 1,\cdots,g$, and we will typically
take $\lambda_i$ real and finite
with $\lambda_0<\lambda_1<\cdots<\lambda_{2g}$.
The period matrix is
$B_{ij} = \oint_{b_j}\omega_i$.  We note that in \cite{gf,gg} for
example one uses cuts $(-\infty,\lambda_0),\cdots,(\lambda_{2g-1},
\lambda_{2g})$ instead of $(\lambda_0,\lambda_1),\cdots,(\lambda_{2g},
\infty)$ as is indicated here; such choices are obviously equivalent.
The $b_i$ cycles can be drawn e.g. from a common vertex $P_0$ passing
through ($\lambda_{2j-1},\lambda_{2j}$).  We recall that hyperelliptic
Riemann surfaces have a number of very special properties (cf. \cite
{ca,fh,ga,gi,me}) so some statements to follow have correspondingly limited
applicability.  We keep our standard local coordinate $k(P)$ or
$k^{-1}(P)$ at the single point $P_1\sim\infty$.
We use differentials $\Omega_k\sim \Omega_{1,k}$ as in Remark 3.4
with $d\Omega_k\sim d(k^i + O(k^{-1}))$ near $\infty,\,\,\oint_{a_j}
d\Omega_k = 0$, and $p(P) = \int^Pd\Omega_1$.  Set then $U^k_j = \oint_
{b_k}d\Omega_j$ as at the beginning of Section 3.1 (mod $2\pi i$ - $U^k_1
\sim U_k,\,\,U^k_2\sim V_k,\,\,U^k_3\sim W_k$).  Recall also $U^A_i =
\oint_{b_i}d\Omega_A$ and $A\sim (1,k)$ here with $dp = \Omega_1,\,\,
dE = \Omega_2,\,\,d\Omega = \Omega_3$.
\\[3mm]\indent
Now in the hyperelliptic case the $\omega_k$ have the form (\ref{FA}) as
indicated earlier.  We observe that the Riemann surface is generated e.g.
via the spectral bands of Bloch eigenfunctions as in Remark 3.3 and has
nothing to do a priori with a putative
LG potential as in
(\ref{CN}) (cf. however ({\bf C}) in Section 5.1).
For $g\not= 0$ one expects $\omega_i$ as in
(\ref{FA}) with $\alpha_{ij}$ determined by the normalization conditions
$\oint_{a_k}\omega_i = \delta_{ik}$ ($g$ equations in $g$ unknowns for
each $i$).  Note also as $\lambda\to \infty,\,\,\omega_i\sim
(\lambda^{g-1}/\lambda^{g+\frac{1}{2}})d\lambda\sim\lambda^{-3/2}d\lambda$.
For the $\Omega_s$ we write
\be
\Omega_{2n+1} = \frac{\lambda^{g+n} + \sum_1^g\beta_{nj}\lambda^{j-1}}
{R(\lambda)}d\lambda
\label{FB}
\ee
with the $\beta_{ij}$ determined via $\oint_{a_k}\Omega_{2n+1} = 0$
(again $g$ equations in $g$ unknowns for each $n$).  Note here as
$\lambda\to\infty,\,\,\Omega_{2n+1}\sim\lambda^{n-\frac{1}{2}}d\lambda$
so for $s=2n+1$ and $k^2\sim\lambda,\,\,\Omega_s\sim\lambda^{n-\frac{1}{2}}
d\lambda\sim d\lambda^{n+\frac{1}{2}}\sim dk^{2n+1}\sim dk^s$.  Also note
that the lower order terms in (\ref{FB}) are led by $(\lambda^{g-1}/
\lambda^{g+\frac{1}{2}})d\lambda\sim\lambda^{-\frac{3}{2}}\sim
d(1/k)$ so the asymptotic behavior is correct.
Then one determines $\Omega_s$ by normalization instead of using a
pseudodifferential operator $(k_1^s)_{+} = k_1^s + O(1/k_1)$ to produce
correct negative powers when $g=0$.
In particular $p=\int^P\Omega_1$ with $dp/dP\sim \Omega_1$ formally in
some appropriate sense and a choice $Q_s \sim Q_{2n+1} = \int^P\Omega_{2n+1}$
will yield for $s$ odd, $\partial Q_s/\partial P\sim\Omega_s$, with
\be
\frac{\partial Q_s}{\partial p} =\partial_p Q_s = \frac{\frac{\partial Q_s}
{\partial P}}{\frac{\partial p}{\partial P}} = \frac{\Omega_s}{\Omega_1}
\label{FC}
\ee
formally (which corresponds to the result suggested in \cite{gh,km} where
$Q_s\sim k^s_{+}$).
We note also that the branch points $\lambda_k$
correspond to integrals of motion (cf. \cite{za}) and
one has Riemann invariants $\lambda_k$ satisfying
(cf. \cite{da,fa,gh,gf,kd}, Sections 5.1 and 5.3, and Section 6.2
to follow)
\be
\frac{\partial\lambda_k}{\partial T_s} = -\frac{\Omega_s}{\Omega_1}
(\lambda_k)\frac{\partial\lambda_k}{\partial X}
\label{FD}
\ee
The Hurwitz space here is indicated in the Examples 5.1 (cf. also
Section 7.3).

\subsection{Another look at KdV averaging}
In order to better understand these matters regarding equations such as
(\ref{FD}) we will go over some averaging procedures from \cite{da}
with attention to \cite{dk,kd,ka,ki,kj}.  We think first of KdV with
$R(\lambda) = \mu^2 = \prod_1^{2m+1}(\lambda-\lambda_k)$ and recall
(cf. Section 3.1) that for periodic situations with wave functions
satisfying (\ref{AD}) (i.e. $L\psi = \lambda\psi;\,\,L = -\partial^2 +
\phi;\,\,\partial_t\psi = A\psi;\,\,A = 4\partial^3 -6\phi\partial -3
\phi_x$) one defines the quasi-momentum and quasi-energy via
\be
p(\lambda) = -i\overline{(log\psi)_x};\,\,E(\lambda) =
-i\overline{(log\psi)_t}
\label{FE}
\ee
Recall that the notation $<\,,\,>_x$ simply
means $x$-averaging (or ergodic averaging)
and $\overline{(\log\psi)_x}\sim <(log\psi)_x>_x\not= 0$ here since e.g.
$(log\psi)_x$ is not bounded.
Observe that
(\ref{FE}) applies to any finite zone quasi-periodic situation.
Now in the notation of \cite{da} one takes
$\lambda_1>\cdots>\lambda_{2m+1}$ with spectral bands $[\lambda_{2m+1},
\lambda_{2m}],\cdots,[\lambda_1,\infty)$ on the real axis or $m$ gaps
$(\lambda_{2m},\lambda_{2m-1}),\cdots,(\lambda_2,\lambda_1)$ in the
spectrum (see e.g. \cite{ca,fa} for pictures).  Thus (cf. (\ref{FB}))
\be
p(\lambda) = \int dp(\lambda) = \int \frac{{\cal P}(\lambda)d\lambda}
{2\sqrt{R(\lambda)}};\,\,{\cal P} = \lambda^m+\sum_1^ma_j\lambda^{m-j};
\label{FF}
\ee
$$E(\lambda) = \int dE(\lambda) = \int\frac{6\lambda^{m+1}
+{\cal E}(\lambda)}{\sqrt{R(\lambda}}d\lambda;\,\,{\cal E} =
\sum_0^mb_j\lambda^{m-j};\,\,b_0 = -3\sum_1^{2m+1}\lambda_i$$
and the normalizations are
\be
\int^{\lambda_{2i-1}}_{\lambda_{2i}}dp(\lambda) = \int^{\lambda_{2i-1}}_
{\lambda_{2i}}dE(\lambda) = 0;\,\,i=1,\cdots,m
\label{FG}
\ee
Note as indicated in Remark 3.4 that one will often require normalizations
$\Im\int_c d\Omega_A = 0$ for $c\in H_1(\Gamma_g,{\bf Z})$.  Here
the $a_j$ cycles can be drawn above or around
the gaps $(\lambda_{2m},\lambda_{2m-1}),
\cdots,(\lambda_2,\lambda_1)$ and for $\omega_j$ a basis of holomorphic
differentials ($j = 1,\cdots,m=g$) one has (cf. (\ref{FA}))
\be
\oint_{a_k}\omega_j = 2\pi\delta_{jk}\,\,(j,k=1,\cdots,m);\,\,
\omega_j = \sum_1^m\frac{c_{jq}\lambda^{q-1}d\lambda}{\sqrt{R(\lambda)}}
\label{FH}
\ee
with $iB_{jk} = \oint_{b_k}\omega_j\,\,(j,k=1,\cdots,m)$ (the notation
changes slightly from time to time in this paper but subsections are
consistent).  The matrix of periods $(B_{jk})$ is symmetric, real,
and positive definite and the Riemann theta function is defined by
\be
\theta(\tau|B)=\sum_{-\infty<n_1<\cdots<n_m<\infty}exp(-\frac{1}{2}
\sum_{j,k}B_{jk}n_jn_k + i\sum_jn_j\tau_j)
\label{FI}
\ee
and finite zone solutions $\phi(x,t)$ of the KdV equation $\phi_t =
6\phi\phi_x - \phi_{xxx}$ determined by $\lambda_1,\cdots,\lambda_{2m+1}$
have the form
\be
\phi(x,t) = -2\partial_x^2log\theta(kx+\omega t + \tau|B) + c
\label{FJ}
\ee
where $\tau = (\tau_1,\cdots,\tau_m)$ and
\be
k_j = \oint_{b_j}dp;\,\,\omega_j = \oint_{b_j}dE\,\,(j=1,\cdots,m);\,\,
c = \sum\lambda_i-2\sum_1^m\oint_{a_q}\lambda\omega_q
\label{FK}
\ee
so the branch points $\lambda_i$ of $\Gamma_m$ parametrize the invariant
tori (\ref{FJ}) of KdV (note $k_ix +\omega_i t+\tau_i\sim\theta_i$ in
Remark 3.3 and note that $\omega_j$ is bad notation here).
\\[3mm]\indent
Another set of parameters is provided by the Kruskal integrals
$I_0,\cdots,I_{2m}$ which arise via a generating function
\be
p(\lambda) = -i<(log\psi)_x>_x = \sqrt{\lambda} + \sum_0^{\infty}
\frac{I_s}{(2\sqrt{\lambda})^{2s+1}}
\label{FL}
\ee
where $I_s = <P_s>_x = \overline{P_s}
\,\,(s = 0,1,\cdots$) with $-i(log\psi)_x =
\sqrt{\lambda} + \sum_0^{\infty}[P_s/(2\sqrt{\lambda})^{2s+1}]$.
Similarly
\be
-i\overline{(log\psi)_t} = -i\frac{A\psi}{\psi} = 4(\sqrt{\lambda})^3 +
\sum_0^{\infty}\frac{E_s}{(2\sqrt{\lambda})^{2s+1}}
\label{FM}
\ee
and one knows $\partial_tP_s = \partial_xE_s$ since
$(\clubsuit)\,\, [(log\psi)_x]_t =
[(log\psi)_t]_x$.  The expansions are standard (cf. \cite{ca,ce,ch,cl}).
For completeness we recall from \cite{da} that (modulo total derivatives)
\be
P_0 = \phi;\,\,P_1 =\frac{\phi^2}{2};\,\,P_2 = \frac{\phi^2_x}{2} - \phi^2;
\cdots;\,\,E_0 = 3\phi^2;
\label{FN}
\ee
$$E_1 = 2\phi^3+\frac{3}{2}\phi_x^2;\,\,E_2 = \frac{9}{2}\phi^4 -
\phi_x\phi_{xxx} + \frac{1}{2}\phi^2_{xx}-3\phi^2\phi_{xx} + 6\phi\phi^2_x;
\cdots$$
\indent
Now consider a ``weakly deformed" soliton lattice of the form (\ref{FI})
with the $\lambda_i\,\,(i=1,\cdots,2m+1)$ (or equivalently the parameters
$u^i = I_i;\,\,i=0,\cdots,2m$) slowly varying functions of $x,t$ (e.g.
$u^i = u^i(X,T),\,\,X = \epsilon x,\,T=\epsilon t,\,\,i = 0,1,\cdots,2m$).
Now one wants to obtain a version of (\ref{AR}) directly via
$(\clubsuit)$.  Thus insert the slow variables in $(\clubsuit)$ and average,
using $\epsilon\partial_X$ or $\epsilon\partial_T$ in the external derivatives,
to obtain
\be
\partial_T\overline{(log\psi)_x} = \partial_X\overline{(log\psi)_t}
\label{FO}
\ee
or $\partial_Tp(\lambda) = \partial_XE(\lambda)$.  Then from (\ref{FF})
differentiating in $\lambda$ one gets
\be
\partial_Tdp = \partial_XdE
\label{FP}
\ee
which is equivalent to (\ref{AR}).  Now (recall $\partial_tP_s = \partial_x
E_s$) expanding (\ref{FP}) in powers of $(\sqrt{\lambda})^{-1}$ one
obtains the slow modulation equations in the form
\be
\partial_T u^s = \partial_X\overline{E_s}\,\;\,(s=0,\cdots,2m)
\label{FQ}
\ee
where $\overline{E_s}$ is a function of the $u^i\,\,(0\leq i\leq 2m)$.
This leads to equations
\be
\partial_T\lambda_i = v_i(\lambda_1,\cdots,\lambda_{2m+1})\partial_X
\lambda_i\,\;\,(i=1,\cdots,2m+1)
\label{FR}
\ee
for the branch points $\lambda_k$ as Riemann invariants.  The characteristic
``velocities" have the form
\be
v_i = \frac{dE}{dp}\Big{|}_{\lambda=\lambda_i} = 2\frac{6\lambda_i^{m+1} +
E(\lambda_i)}{P(\lambda_i)}\,\;\,(1\leq i\leq 2m+1)
\label{FS}
\ee
To see this simply multiply (\ref{FP}) by $(\lambda-\lambda_i)^{3/2}$ and
pass to limits as $\lambda\to\lambda_i$.  These equations (\ref{FR}) -
(\ref{FS}) correspond to (\ref{FD}).

\subsection{Formulas for kernels}
We will begin by listing various formulas involving Cauchy type kernels on a
Riemann surface.  Properties and origins will be indicated along with
a sketch of proofs and references.  Then we want to examine relations between
some of these kernels.  One motivation for this investigation is the result
of \cite{ch} which shows that the dispersionless differential Fay identity
is equivalent to the kernel expansion in dKP theory
\be
{\cal K}(\lambda,\mu) =
\frac{1}{P(\mu)-P(\lambda)} = \sum_1^{\infty}\partial_PQ_n(\lambda)\mu^{-n}
\label{GA}
\ee
(cf. Section 4).  To recapitulate one can begin with a Lax operator
$L= \partial + \sum_1^{\infty}u_{n+1}\partial^{-n}$, and pass to
dispersionless limits via
formulas $\psi = exp[(1/\epsilon)S(T,\lambda)],\,\,
T = (T_n),\,\,T_n=\epsilon t_n$) to obtain
\be
\lambda = P + \sum_1^{\infty}U_{n+1}P^{-n};\,\,P = \lambda -
\sum_1^{\infty}P_{i+1}\lambda^{-i}
\label{GB}
\ee
The KP hierarchy then becomes the dKP hierarchy in the form $\partial_n
P = \partial{\cal B}_n$ where ${\cal B}_n = \lambda^n_{+}$ and $P =
\partial_XS$.  Setting $Q_n = (1/n){\cal B}_n$ one arrives at (\ref{GA}),
whose importance was first indicated in \cite{kn} (see Section 4 and cf.
also \cite{ch,ci}).  In the reduction to NKdV the dispersionless
situation involves $\lambda^N_{+} = \lambda^N\sim E$ in (\ref{CU})
(with $P$ replacing $p$).  In particular the action of ${\cal K}(\lambda,
\mu)$ at $\mu=\infty$ is expressed via
\be
\frac{1}{2\pi i}\oint_{\infty}\frac{\mu^n}{P(\mu)-P(\lambda)}d\mu
=\partial_PQ_{n+1}(\lambda)
\label{GC}
\ee
It was stated in \cite{ch} that ${\cal K}$ represents a dispersionless
limit of the Fay prime form which we want to clarify and make
more precise
here.
\\[3mm]\indent
First we refer to \cite{gd,ge,gf} where some interesting kernel results
appear.  To begin with we
consider mainly hyperelliptic situations for illustration
and think of $\mu^2 = \prod_1^{2g+1}(\lambda-\lambda_i)$ with branch
points $\lambda_1,\cdots,\lambda_{2g+1},\infty$.  Thus, following
\cite{gf}, take $\lambda_1<\cdots<\lambda_{2g+1}<\infty$ with spectral
bands $[\lambda_1,\lambda_2],\cdots,[\lambda_{2g+1},\infty)$ and gaps
$(\lambda_2,\lambda_3),\cdots,(\lambda_{2g},\lambda_{2g+1})$.  We will
write $-\lambda= k^2$ here (cf. (\ref{FB}) where $k^2\sim\lambda$) with
\be
\Omega_s = d(k^s) + \sum_{m\geq 1}Q_{sm}k^{-m-1}dk
\label{GD}
\ee
where $\Omega_{2s} = d((-\lambda)^s)$ for $\Gamma_g$ hyperelliptic.
Further set
\be
\oint_{b_j}\Omega_s = (U_s)_j;\,\,\omega_j = \sum_{m\geq 1}q^H_{jm}
k^{-m-1}dk
\label{GE}
\ee
(cf. (\ref{FA})) and the Riemann bilinear relations imply
\be
Q_{ij} = Q_{ji};\,\,(U_j)_k=-2\pi iq^H_{kj}
\label{GF}
\ee
Also $Q_{kl} = 0$ if either $k$ of $l$ is even and in the present
situation $\Im(B_{ij})$ is positive definite ($B_{ij} = \oint_{b_j}\omega_i$).
Note here $\lambda$ belongs to a two sheeted covering of the $k$ plane.
\\[3mm]\indent $\bullet$  The above constructions are the
same as before modulo notation but it is approptiate now to write the
quasi-momentum $dp$ via
\be
dp \sim -idk;\,\,\Im\oint_{a_j}dp = \Im\oint_{b_j}dp = 0
\label{GG}
\ee
This corresponds to the real Whitham normalization discussed in Remark
3.4.  Thus $\Omega_1\sim dk$ and $dp\sim -idk$ here.  $\bullet$
\\[3mm]\indent
Now we recall the (Fay or Klein)
prime form (cf. \cite{hb,fe,ma,ob}), which as used in \cite{gd,ge,gf},
has the properties:  $E(P,Q),\,\,P,Q\in\Gamma_g$, is a holomorphic
$(-1/2)$-form in $P,Q$ with the properties:  ({\bf G})  $E(P,Q) = 0$
if and only if $P=Q$ ({\bf H})  Given $t$ a local coordinate on
$\Gamma_g$ near $Q$, if $P\to Q$ one has (cf. \cite{ca,dm,ha,ma} for
half differentials)
\be
E(P,Q) = \frac{(t(P)-t(Q))[1+O((t(P)-t(Q))^2)]}{\sqrt{dt(P)dt(Q)}}
\label{GK}
\ee
({\bf I})  $E(P,Q)$ is multivalued with periodicity conditions:
$E(P+a_k,Q) = E(P,Q),\,\,E(P+b_k,Q) = \pm E(P,Q)exp(-iB_{kk}+2\pi i
\int_P^Q\omega_k)$ where $a_k,\,b_k$ are the basic cycles ({\bf J})  Using
local coordinates $k^{-1}$ at $\infty$ one takes $\Omega_s$ as usual with
$
\oint_{a_i}\Omega_s = 0$.
Further (shifting to local coordinates)
\be
d_{\zeta}d_z log\,E(\zeta,z) = -\sum_1^{\infty}\Omega_s(\zeta)z^{s-1}dz
\label{GM}
\ee
which implies
\be
d_{\zeta}log\,E(\zeta,z) = -\sum_1^{\infty}\Omega_s(\zeta)\frac{z^s}{s}
+ f(\zeta)
\label{logE}
\ee
({\bf K})  In the neighborhood of $\infty$ one has ($z\sim 1/k$ etc.)
\be
log\frac{E(z,z')}{z-z'} = \sum_{m\geq 2}(z^m+z'^m)\frac{Q_{m0}}{m} +
\sum_{m,n\geq 1}z^mz'^n\frac{Q_{mn}}{mn}\,\,\,(Q_{mn}=Q_{nm})
\label{GN}
\ee
\be
\Omega_s = d(\frac{1}{z^s}) - \sum_{m\geq 1}Q_{sm}z^{m-1}dz\,\,(s\geq 1);
\label{GO}
\ee
$$\vec{\omega} = -\sum_{s\geq 1}\vec{U}_sz^{s-1}dz;\,\,(\vec{U}_s)_m =
\frac{1}{2\pi i}\oint_{b_m}\Omega_s$$
where the terms with $Q_{m0}$ correspond to some normalization and
$\vec{\omega}\sim (\omega_1,\cdots,\omega_g)$ represents the
standard holomorphic differentials (cf. \cite{ge}).
Note that the first equation in
(\ref{GO}) is equivalent to (\ref{GD}) since $d(1/k) = -(1/k^2)dk$
but in the second equation the $(U_s)_m$ differ from (\ref{GE}) by a
factor of $2\pi i$.  The $Q_{m0}$ terms are sometimes omitted in
(\ref{GN}) (cf. below and \cite{ag}, p. 181).
Equation (\ref{GN}) implies now
\be
d_zlog\,E(z,z') - \frac{dz}{z-z'} = \sum_{m\geq 2}Q_{m0}z^{m-1}dz +
\label{newker}
\ee
$$+ \sum_{m,n=1}^{\infty}\frac{z'^n}{n}Q_{mn}z^{m-1}dz=\sum_{m\geq 2}
Q_{m0}z^{m-1}dz +\sum_1^{\infty}\frac{z'^n}{n}[d(\frac{1}{z^n})
-\Omega_n(z)]$$
and consequently
\be
d_zlog\,E(z,z') = \sum_{m\geq 2}Q_{m0}z^{m-1}dz + \frac{dz}{z} -
\sum_1^{\infty}\frac{z'^n}{n}\Omega_n(z)
\label{newkernel}
\ee
Then with such terms one has $f(\zeta) = (d\zeta/\zeta) +\sum_{m\geq 2}
Q_{m0}\zeta^{m-1}d\zeta$ in (\ref{logE}).  We refer here also to (\ref{LK})
where it is shown that ${\cal Z}_p(x) = d_xlog\,E(x,p)$ has a pole
of residue one at $x=p$ and $f(p) = (1/2\pi i)\int_{\partial U}f(x)
{\cal Z}_p(x)$ for $f$ holomorphic in a neighborhood $U$ of $p$
(i.e. ${\cal Z}_p(x)$ is a local Cauchy kernel).
\\[3mm]\indent
Now we recall that (\ref{GO}) $\equiv$ (\ref{GD}) (up to a factor of
$2\pi i$ in the $(U_s)_m)$ and in (\ref{FG}) one sees that $Q_j =
\int^P\Omega_j$ (with $p = \int^P\Omega_1$) leads formally to $\partial_P
Q_j = (\Omega_j/\Omega_1)$ with (\ref{FD}) in the form $\partial_j\lambda_k
= -(\Omega_j/\Omega_1)\partial_X\lambda_k=-\partial_PQ_j\partial_X
\lambda_k$.  On the other hand in the LG
theory for genus $0$ as in Section 3.10 one has (using $\tilde{\Omega}_j$ now
for $\Omega_j$ in Section 3.10) $\tilde{\Omega}_j\sim k^j_{+}\sim\partial_jS$
and via (\ref{CR}), $\partial_jE = -E_X\partial_p\tilde{\Omega}_j$ at
$p=q_s\,\,(E(q_s,T)=E_s$ corresponds to Riemann invariants as do the branch
points $\lambda_j$ in (\ref{FD}) and we shift back $X\to-X$ now)
Now $E\sim k^n$ so $\partial_jE
= nk^{n-1}\partial_j k$ and $\partial_XE = nk^{n-1}\partial_Xk$ implying also
\be
\partial_jk(q_s) = -\partial_p\tilde{\Omega}_j\partial_Xk(q_s)
\label{HA}
\ee
(with complex Riemann invariants $k_s = k(q_s)$)
and $\tilde{\Omega}_j\sim j\tilde{Q}_j$ where $\tilde{Q}_j$ corresponds to the
$Q_j$ of dKP theory in Section 4 (cf. also (\ref{GA}) and (\ref{GC})).
We recall the time variable changes in Section 3.12 and in Section 4, so one
can simply ignore or adjust the factor of $j$
arising in $\partial_j$ for the moment.
Then (\ref{HA}) or (\ref{CR})
compares with (\ref{FD}) using the the analogy $\Omega_j/
\Omega_1\sim \partial_p\tilde{Q}_j\sim \partial_PQ_j$ in
a suitable time scale.  This says that the finite zone (hyperelliptic)
Riemann surface theory with averaging yields branch point equations
(\ref{FD}) for $\lambda_k$ whereas the LG theory for $E$ of Section 3.10
in genus zero yields corresponding equations (\ref{CR}) for Riemann
invariants $E_s$ or $k_s$ relative to the n-sheeted Riemann surface
$\Sigma_E$ of $E$ (the number of branch points may differ of course).
We refer to Section 5 for clarification of roles and will discuss
this all later at some length.
The formula analogous to the kernel expansion
(\ref{GA}) is then (cf. \cite{km} where such formulas were first
suggested)
\be
\hat{{\cal K}}(\lambda,\mu)\sim\sum_1^{\infty}\frac{\Omega_j}
{\Omega_1}(\lambda)\mu^{-j}\,\,or\,\,\tilde{{\cal K}}(\lambda,\mu)
\sim\sum_1^{\infty}\frac{\Omega_j}{\Omega_1}(\lambda)\frac{\mu^{-j}}{j}
\label{HB}
\ee
depending on how we scale the $T_j$.
Now from (\ref{GM}) we have
$d_{\zeta}d_z log\,E(\zeta,z) = \Xi = -\sum_1^{\infty}\Omega_j(\zeta)
z^{j-1}dz$ and we could also take $d_{\zeta}log\,E(\zeta,z)\sim
-\sum_1^{\infty}\Omega_s(\zeta)z^s/s + (d\zeta/\zeta)$.
Note here $\zeta\sim 1/\lambda$ and $z\sim 1/\mu$ with say
$\lambda,\,\mu\to\infty$ so from (\ref{GO}), with abuse of notation,
$\Omega_s(\zeta) = d(1/\zeta^s)-\sum_1^{\infty}Q_{sm}\zeta^{m-1}d\zeta
\sim d(\lambda^s)+\sum_1^{\infty}Q_{sm}\lambda^{-m-1}d\lambda = \Omega_s
(\lambda)$ (the latter form for $\Omega_s$ corresponds to $\lambda\to
\infty$).  Hence via $d_{\zeta}log\,E(\zeta,z) = d_{\lambda}log\,E
(1/\lambda,1/\mu)$, etc. we get
\be
d_{\zeta}d_zlog\,E(\zeta,z)\sim d_{\lambda}d_{\mu}log\,E(\frac{1}{\lambda},
\frac{1}{\mu})=\sum_1^{\infty}\Omega_j(\lambda)\mu^{-j-1}d\mu
\label{more}
\ee
\be
d_{\zeta}log\,E(\zeta,z)\sim d_{\lambda}log\,E(\frac{1}{\lambda},
\frac{1}{\mu})=-\sum_1^{\infty}\Omega_s(\lambda)\frac{\mu^{-s}}
{s} - \frac{d\lambda}{\lambda}
\label{again}
\ee
Now (\ref{HB}) gives two analogues of (\ref{GA}) corresponding to
(\ref{more}) and (\ref{again}).  Since $d_{\zeta}d_zlog\,E(\zeta,z)$ has
a second order pole at $\zeta=z$ (or $\lambda=\mu$) we are led
reject it and to suggest
as an analogue of ${\cal K}$ the formula ($\lambda\sim 1/\zeta,\,\mu\sim
1/z$)
\be
{\cal K}(\lambda,\mu) = \frac{1}{P(\mu)-P(\lambda)}\sim\tilde{{\cal K}}
(\lambda,\mu)\sim\sum_1^{\infty}\frac{\Omega_j}{\Omega_1}(\lambda)
\frac{\mu^{-j}}{j}\sim
\label{anal}
\ee
$$\sim\sum_1^{\infty}\frac{\Omega_j}{\Omega_1}(\zeta)\frac{z^j}{j}
=\frac{\frac{d\zeta}{\zeta}-d_{\zeta}log\,E(\zeta,z)}{\Omega_1(\zeta)}
=\frac{\frac{d\zeta}{\zeta}-{\cal Z}_z(\zeta)}{\Omega_1(\zeta)}$$
As mentioned above in \cite{ag} one writes
\be
log\frac{E(\zeta,z)}{\zeta-z} = \sum_{m,n=1}^{\infty}\zeta^m z^n\frac
{Q_{mn}}{mn}
\label{Eagain}
\ee
\be
d_{\zeta}log\,E(\zeta,z)\sim \frac{d\zeta}{\zeta-z}+\sum_{m,n=1}^{\infty}
\zeta^{m-1}\frac{z^n}{n}Q_{mn}d\zeta\sim {\cal Z}_z(\zeta)
\label{EEE}
\ee
so ${\cal Z}_z(\zeta)$ has the appropriate singularity at $z=\zeta$
(cf. also (\ref{LK}))
Let us summarize heuristically in
\\[3mm]
\indent {\bf THEOREM 6.1}.$\,\,$  For $\lambda\sim (1/\zeta),\,\,
\mu\sim (1/z)$ one has an argument as in
(\ref{FD}) or (\ref{HA}) relating ${\cal K}=1/(P(\mu)-P(\lambda))$ to
a Riemann surface object
$\tilde{{\cal K}}$ in (\ref{HB}) for general Riemann surfaces
(with suitable scaling of time variables), which in turn can be expressed
via ${\cal Z}_z(\zeta)$ as indicated.  The
$\Omega_j$ will be differentials normalized via $\oint_{a_k}\Omega_j=0$
when $g\not= 0$ (depending on slow variables $T_j$),
or represented by suitable powers of a LG polynomial
for the dispersionless situation where $g=0$ (or in other LG situations
with $g=0$ by suitable powers of a LG function).
Thus formally $(d\zeta/\zeta)-{\cal Z}_z(\zeta)\sim \Omega_1(\lambda)
\hat{{\cal K}}(\lambda,\mu)\sim [dP(\lambda)/(P(\mu)-P(\lambda))]$ and
we note that $(d\zeta/\zeta)-{\cal Z}_z(\zeta)\sim (d\zeta/\zeta) -
(d\zeta/(\zeta-z))\sim-(d\lambda/(\lambda-\mu))$.
Below we will also indicate a somewhat different kind of connection
of ${\cal K}$ to the kernel
$\omega$ based on $\psi\psi^*$.
$\bullet$
\\[3mm]\indent
Generally one appears to have a theory here involving Riemann surfaces
alone.  Averaging procedures have nothing to do \`a priori
with the construction
of differentials (but see here Theorem 7.3).
Thus start with a Riemann surface $\Sigma_g$ and form a KP equation
as at the beginning of Section 2 via a BA function, which defines flow
variables $x,\,y,\,t,\cdots\sim t_n$.  Then average as in Section 3.6 to
get equations (\ref{BM}), and proceed to general $\Omega_A$ as in
Section 3.7 for say ${\cal M}_{g,1}$.
No LG function is needed a priori (branch points are determined for
${\cal M}_{g,1}$ via $p$ and additional times arise via $T_A,\,\,
A\sim (\alpha,i)$ for ${\cal M}_{gN}$).
Suppose first, for illustration, that $\Sigma_g$ is
hyperelliptic as in Section 6.1, leading to (\ref{FR}) - (\ref{FS}) or
equivalently to (\ref{FD}).  This means that the modulation equations
determine deformations of the Riemann surface via branch point equations
using natural flow parameters determined from the Riemann surface.  However
KP or KdV has only been used as an intermediate step and disappears
at the end except for the slow variables.  The modulation equations
only involve differentials based on $k_1$
and slow variables (other intrinsic
time variables and differentials arise for additional punctures).
Thus (for one puncture)
one creates a ``generic" point $P_1\sim\infty$ and
builds the $\Omega_A$ and $T_A$ leading to (\ref{FD}).  If one starts
with a LG potential $E$ as in (\ref{CN}) to generate an n-sheeted
Riemann surface of genus $0$ the matter seems somewhat contrived
but in fact here, and especially
in the case of general $E$ as in (\ref{CU}), many
interesting things occur.  A more detailed
study of this situation (\ref{CU}) is
indicated, going beyond the description of Hurwitz spaces and Frobenius
manifolds in Section 5.  One desideratum here is to make
more explicit the relations between the LG potential and the Riemann
surface in the Hurwitz space approach.   Other questions also
arise.  For example let $\Sigma_g$ be given, say hyperelliptic.
If $\Sigma_g$ arises from the Bloch spectrum of a KdV situation then
one has developments as in Section 6.2; the wave function for KdV will
apparently be the BA function for $\Sigma_g$.  However if we choose
a BA function $\psi$ for $\Sigma_g$ as indicated in Section 2, it will
correspond in general to a KP situation.  It seems clear that
for some divisor $D,\,\,\psi$
will be a KdV wave function corresponding to a potential $u$ whose
Bloch spectrum gives rise to $\Sigma_g$.  It might be worthwhile to
make such matters precise and explicit.

\subsection{The kernel $\omega$}
Now go back to the BA function for KP as in (\ref{AA}) so that
$\psi(\vec{t},k)$ is meromorphic in $\Gamma_g/\infty$ with simple
poles at $P_1,\cdots,P_g$ and no other singularities in $\Gamma_g/\infty$
(here $\vec{t} = (t_i)$ with $t_1=x,\,\,t_2=y,\,\,t_3=t,$ etc.).  We use
$k^{-1}$ as the local coordinate at $\infty$ so $\psi\sim exp[\xi
(\vec{t},k)]\cdot (1+\sum_1^{\infty}\chi_i k^{-i})$ for $|k|$ large
where $\xi(\vec{t},k) = \sum_1^{\infty}t_ik^i$.  For $t_{2i} = 0$
this is a KdV situation.
The BA conjugate differential $\psi^{\dagger}(\vec{t},\mu)$
is apparently defined in \cite{gd,ge,gf} to have
simple zeros at $P_1,\cdots,P_g$, to be a holomorphic 1-differential in
$\Gamma_g/\infty$, and to have an essential singularity at $\infty$ of the form
$\psi^{\dagger}\sim  exp[-\xi(\vec{t},k)](1+\sum_1^{\infty}
\chi^{\dagger}_ik^{-i})dk$ (this definition is discussed later in
more detail).  Then it is ``known" that
\be
\oint_C\psi(k,\vec{t})\psi^{\dagger}(k,\vec{t'})dk = 0
\label{GH}
\ee
for $C$ a small contour around $\infty$ (Hirota bilinear identity) and
\be
\int_{-\infty}^{\infty}\psi(k,\vec{t})\psi^{\dagger}(k',\vec{t})dx
=2\pi i\delta(k-k')
\label{GI}
\ee
for $\Im p(k) = \Im p(k')$.  In (\ref{GI}) the contour could apparently
be a closed curve through $\infty$ for example (which would become a straight
line in the - degenerate - scattering situation).  This point is however
consistently overlooked and should be further examined (cf. also
\cite{cc} for KP-1 and the scattering situation).
Next the Cauchy-Baker-Akhiezer
(CBA) kernel $\omega(k,k',\vec{t})$ is defined via
(we give an expanded form in (\ref{ker}))
\be
\omega(k,k',x,y,t,\cdots) = \frac{1}{2\pi i}\int_{\pm\infty}^x
\psi(k,x',y,t,\cdots)\psi^{\dagger}(k',x',y,t,\cdots)dx'
\label{GJ}
\ee
According to \cite{gd,ge,gf}
this kernel is to have the following
properties:  ({\bf A})  $\omega(k,k',\vec{t})$
is a function in $k$ and a one form in $k'$  ({\bf B})  $\omega(k,k',
\vec{t})$ is meromorphic in $k$ in $\Gamma_g/\infty$ with simple poles
at $P_1,\cdots,P_g,k'$ ({\bf C})  As a function of $k',\,\,\omega$ is
meromorphic in $\Gamma/\infty$ with one pole $k$ and zeros at $P_1,\cdots,
P_g$ ({\bf D})  $\omega(k,k',\vec{t}) = O(exp[\xi(k,\vec{t})])$ as $k\to
\infty$ ({\bf E})  $\omega(k,k',\vec{t}) = O(exp[-\xi(k',\vec{t})])$ as
$k'\to\infty$ ({\bf F})  $\omega(k,k',\vec{t})\sim(dk'/2\pi i(k'-k))$ as
$k\to k'$.  We will express $\psi^{\dagger}$ below as a suitable
multiple of $\psi^*$
\\[3mm]\indent
The material on the prime form is well discussed in \cite{fe,ma}
for example (cf. also \cite{hb} and Section 6.5 below)
but the proofs in \cite{gd,ge,gf} of (\ref{GI}), (\ref{GJ}),
and properties ({\bf A}) - ({\bf F}) of the CBA kernel are somewhat
unclear so we will give some discussion in various contexts.
Formula (\ref{GH}) with $\psi^*$ in place of $\psi^{\dagger}$ (Hirota
bilinear identity) can be proved in various traditional manners (cf.
\cite{ca,cc,dn}) so we omit comment.  Now look at the integrands in
(\ref{GI}) and (\ref{GJ}) for large $k,k'$ and $\Im p(k) = \Im p(k')$,
namely, $\psi(k,\vec{t})\psi^{\dagger}(k',\vec{t})\sim exp[\sum
t_n(k^n-k'^n)](1+\sum\chi_ik^{-i})(1+\sum\chi_j^{\dagger}k'^{-j})dk'$.
As $x = t_1$ varies one has a multiplier $exp[x(k-k')]$ with
$|exp[x(k-k')]| = exp[x\Re(k-k')]$ (note $dp\sim -idk$ so for large
$k,\,k', \Im p(k') = \Im p(k)\sim \Re k = \Re k'$).  Thus $exp[x
(k-k')]\sim exp[ix(\Im k - \Im k')]$ and (\ref{GI}) has a Fourier
intergal flavor.  On the other hand from $\partial_n\psi = B_n\psi$ and
$\partial_n\psi^{\dagger} = -B_n^*\psi^{\dagger}$
(true at least for $\psi^{\dagger}\sim\psi^*$) one obtains
($\Im p(k') = \Im p(k)$)
\be
\partial_n\int_{-\infty}^{\infty}\psi\psi^{\dagger}dx = \int_{-\infty}^{\infty}
[(B_n\psi)\psi^{\dagger} - \psi(B_n^*\psi^{\dagger})]dx = 0
\label{GP}
\ee
provided integration by parts is permitted.  This is not obviously
true but one can assume it under reasonable circumstances.
Further for large $t_n$ and $k'\not= k,\,\,|\psi\psi^*|\sim O(|exp
[t_n(k^n-k'^n)]|) = exp[t_n\Re(k^n-k'^n)]\to 0$ for some $n$ when
$t_n\to \infty$ or $-\infty$ (with $\Im p(k') = \Im p(k)$ or not).  This
(with (\ref{GP})) implies that the integral in (\ref{GI}) is $0$ for
$k'\not= k$ and $\Im p(k') = \Im p(k)$.
To show that (\ref{GI}) actually gives a delta function
for $\Im p(k') = \Im p(k)$ one is referred
in \cite{ge} to \cite{kb} for a discrete
version whose proof is said to be extendible.  This is not clear but see also
\cite{ko,kp} which don't seem to explain much either.
Finally it is clear that
the $\pm \infty$ limit of the integral in (\ref{GJ}) may change when
$k$ crosses the path $\Im p(k) = \Im p(k')$.  In this regard recall
$|exp[x(k'-k)]| = exp[x(\Re k' - \Re k)]$ and $\Im p(k)\sim -\Re k$ for
$k$ large.  Hence $\omega$ in (\ref{GJ}) can have a jump discontinuity
across the curve $C:\,\,\Im p(k) = \Im p(k')$.  To see this and to show
that $\omega(k,k')$ is nevertheless continuous for $k\not= k'$ let $I
\subset C$ be a small arc seqment and $D\supset I$ a small open set with
boundary $\partial D$.  Then $\int_D\bar{\partial}\omega dkd\bar{k}
=\int_{\partial D}\omega dk$ by Stoke's theorem.  Think of $I$ as a small
straight line segment with $D$ shrunken down around $I$ to be lines
above and below with little end curves.  One obtains in an obvious notation
\be
\int_D\bar{\partial}\omega dkd\bar{k} = \int_I(\omega_{+} - \omega_{-})dk =
\label{GQ}
\ee
$$=\frac{1}{2\pi}\int_I\int_{-\infty}^{\infty}\psi(k)\psi^{\dagger}(k')dxdk =
\left\{
\begin{array}{cc}
0 & for\,\,k'\not\in I\\
1 & for \,\,k'\in I
\end{array}
\right.
$$
via (\ref{GI}) and (\ref{GJ}) (using the change in integration limit in
(\ref{GJ})).  The other properties in ({\bf A}) - ({\bf E}) are more or
less natural for suitable $\psi^{\dagger}$ as indicated below.
Property ({\bf F})  is suggested by
e.g. $(1/2\pi i)\int_{-\infty}^x\psi\psi^{\dagger}dx'dk'\sim
[1/2\pi i(k-k')]\cdot\int^x_{-\infty}exp[x'(k-k')]\cdot O(1)dx'dk'$  when
say $\Re k > \Re k',\,\,k'\to k$ and $x>0$ with the
expectation here that $1/(k-k')$ will emerge from
the integration (cf. also Theorem 6.3).
We will suggest formulas for $\omega$ and $\psi^{\dagger}$ below
which accord with (\ref{GH}), (\ref{GI}), and
({\bf A}) - ({\bf F}), but a further infusion of rigor would be
welcome.

\subsection{Remarks on general kernels}
We refer here to \cite{zb} for some general information on Cauchy
type kernels on a Riemann surface $\Sigma_g$ (cf. also \cite{ra}).
Take a Riemann surface $\Sigma_g$ with a canonical homology basis
$(a_i,b_i)\,\,1\leq i\leq g$, and abelian differentials $\omega_k,\,\,
1\leq k\leq g$, with $\oint_{a_k}\omega_i=\delta_{ik}$.  The B periods
are $B_{kj} = \oint_{b_k}\omega_j$ as usual.  Let $\omega_{qs}(p)$
(or $d\omega_qs)(p)$) be a normalized (i.e. $\oint_{a_k}\omega_{qs}=0$)
differential with simple poles at $q$ and $s$ with residues $\pm 1$
respectively.  One knows that $\omega_{qs}(p)-\omega_{qs}(r)=\omega_{pr}(q)
-\omega_{pr}(s)$ so $\omega_{qs}(p)$ is a many valued function of $q$
with a single pole at $q=p$.  Further
\be
\oint_{b_k}\omega_{qs}(p)=2\pi i\int_q^s\omega_k
\label{alpha}
\ee
(cf. (\ref{HK})).  Consequently
\be
\oint_{a_k}d_q\omega_{qs}(p)=0;\,\,\oint_{b_k}d_q\omega_{qs}(p)
=2\pi i\omega_k
\label{beta}
\ee
Choose a brance $\hat{\omega}_{qs}(p)$ now which is a single valued
function on $\Sigma_g$ and satisfies the condition $d\hat{\omega}_{ss}
(p)\equiv 0$.  This branch can be defined via
\be
\hat{\omega}_{qs}(p)=\int_s^qd_t\omega_{ts}(p)
\label{gamma}
\ee
where the path of integration does not intersect $a_1,\cdots,a_g$.
One uses $+$ (resp. $-$) to destinguish objects associated with the
left (resp. right) bank of the oriented curves $a_k$.  Thus $a_k^{+}$
(resp. $a_k^{-}$) denotes $a_k$ described in the $+$ (resp. $-$)
sense.  The surface $\Sigma_g$ is on the left of the curves
$a_1^{+}b_1^{+}a_1^{-}b_1^{-}\cdots a_g^{+}b_g^{+}a_g^{-}b_g^{-}
\sim \partial\Sigma_g$ in the standard cutting.
Then from (\ref{beta}) - (\ref{gamma}) one has for $t\in a_k$
\be
\hat{\omega}^{+}_{ts}(p) - \hat{\omega}^{-}_{ts}(p) =
-2\pi i\omega_k(p);\,\,\hat{\omega}^{+}_{qs}(t)-\hat{\omega}^{-}_{qs}
(t)=-2\pi i\int_s^q\omega_k
\label{delta}
\ee
($\pm$ over $t$ would be slightly better notation).
\\[3mm]\indent
Consider now a Cauchy kernel $K=(d\tau/(\tau-z))$ in the complex plane,
which has the properties:  (1) As a function of $z,\,\,K$ is a
meromorphic function with a single pole at $z=\tau$ and a single
zero at $z=\infty$.  (2)  As a function of $\tau,\,\,K$ is a meromorphic
differential with two simple poles at $\tau=z$ and $\tau=\infty$, with
residues $\pm 1$ respectively.  We denote analogues of Cauchy kernels
on $\Sigma_g$ by $A(p,q)dp$ where (3)  $A(p,q)dp=(dp/(p-q))$ +
regular terms as $p\to q$.  Generally for $g>0$ a meromorphic function
cannot have one simple pole so there is no analogue of (1).  Hence,
abandoning (1), there exist infinitely many analogues of a Cauchy
kernel and two especially are constructed in \cite{zb}.  The first is
a discontinuous analogue which is fulfilled by $\hat{\omega}$ above
where $\hat{\omega}$ is discontinuous in $q$ along the $a_k$.  Thus
\be
A(p,q)dp=\hat{\omega}_{qs}(p)
\label{epsilon}
\ee
($s$ can be fixed arbitrarily and $\tau,\,z,\,\infty\sim p,\,q,\,s$).
Except for the discontinuity (1) and (2) are satisfied.
\\[3mm]\indent
The second type is a meromorphic analogue.  For this one says first
that a divisor $D$ is minimal if $ord(D) = g-1$ and $r(D^{-1})=
i(D)=0$ where the meaning of $r,i$ is visible from the Riemann-Roch
theorem $r(D^{-1})=ord(D)+i(D)-g+1$.  However note here that this is
in multiplicative notation for divisors, $D=p_1^{n_1}\cdots p_k^{n_k}$
instead of $\sum n_ip_i$ so one knows (cf. \cite{fg}) $R(1)=1,\,\,
i(1)=g,$ and $ord(D)=0\sim D=1$.  We recall also for $n_i\geq 0$ with
$D\geq 1$ one has ${\bf C}\subset L(D^{-1})\,\,({\bf C}=L(1)$) where
$L$ is standard notation.  Thus if $ord(D)=g-1$ with $r(D^{-1})=0$
then $D$ must have poles.  Now (given a minimal divisor $D$)
one constructs a Cauchy kernel $A(p,q)dp$
having property (3) which as a function of $p$ is a differential with
divisor a multiple of $q^{-1}D\,\,(\sim D-q$ additively)
and as a function of $Q$ the kernel is meromorphic with divisor a
multiple of $p^{-1}D^{-1}$.  Thus pick e.g.
\be
D=\prod_1^kp_j^{n_j}\cdot\prod_0^iq_l^{-m_l-1}\,\,\,(i>0,\,\,k>0,\,\,
n_j\geq 0,\,\,m_l\geq -1)
\label{iota}
\ee
with $ord(D) = \sum_1^kn_j-\sum_0^im_l-i-1=g-1$.  By remarks above there
is at least one $q$, say $q_0$, with a positive multiplicity $m_0+1
>0$.  Then one can write down a formula in determinants for $A(p,q)dp$
(cf. \cite{zb}) with entries
\be
\omega^{\lambda\mu}_{q_1q_2}(p)=\frac{\partial^{\lambda+\mu}
\hat{\omega}_{q_1q_2}(p)}{\partial p^{\lambda}\partial q_1^{\mu}}
dp^{\lambda}dq_1^{\mu};\,\,
\omega^{1,0}_{q_0q}(p)=-d\hat{\omega}_{qq_0}(p)
\label{kappa}
\ee
This kernel will be unique via minimality of its characteristic divisor.
However there exist other meromorphic analogues having different properties
(cf. below).  Examples of various sorts are indicated in \cite{zb}.
Discontinuous Cauchy type kernels are connected to Plemlj-Sokhotskij type
formulas (with eventual application to Riemann-Hilbert problems possible)
and meromorphic Cauchy kernel analogues are related to kernels of
Chibrikova, Gusman-Rodin, Koppelman, Tietze, Vaccaro, and Weierstrass
for example.
\\[3mm]
\indent
Some other analogues of Cauchy kernels arise from the differential
\be
\Xi=\frac{w+\zeta}{2\zeta}\cdot\frac{d\tau}{\tau-z}
\label{mu}
\ee
where $(z,w)\in\Sigma_g,\,\,(\tau,\zeta)\in \Sigma_g$ when $\Sigma_g$
is the hyperelliptic Riemann surface of $w^2-f(z)=0$ for $f(z) = \prod_1^
{2g+1}(z-z_k)$.  One thinks here of $a_k$ cycles surrounding $r_{2k-1},
r_{2k})$ with $(z,\sqrt{f})$ the upper sheet.  One can take as usual
$(dz/w),\cdots,(z^{g-1}dz/w)$ for the $\omega_k$ (unnormalized).  Then
the discontinuous analogue of the Cauchy kernel is
\be
d\hat{\omega}_{(z,w),(z_0,w_0)}(\tau,\zeta) = d\hat{\omega}_{(z,w)}
(\tau,\zeta)-d\hat{\omega}_{(z_0,w_0)}(\tau,\zeta)
\label{nu}
\ee
where
\be
d\hat{\omega}_(z,w)(\tau,\zeta)=\frac{Num}{Den};\,\,
Den=
\left|
\begin{array}{ccc}
\oint_{a_1}\frac{d\tau}{\zeta} & \cdots & \oint_{a_1}\frac{\tau^{g-1}d\tau}
{\zeta}\\
\cdots & \cdots & \cdots\\
\oint_{a_g}\frac{d\tau}{\zeta} & \cdots & \oint_{a_g}\frac{\tau^{g-1}d\tau}
{\zeta}
\end{array}
\right|
\label{zeta}
\ee
$$ Num=
\left|
\begin{array}{cccc}
\frac{w+\zeta}{2\zeta}\cdot\frac{d\tau}{\tau-z} & \frac{d\tau}{\zeta}
& \cdots & \frac{\tau^{g-1}d\tau}{\zeta}\\
\oint_{a_1}\frac{w+\zeta}{2\zeta}\cdot\frac{d\tau}{\tau-z} & \oint_{a_1}
\frac{d\tau}{\zeta} & \cdots & \oint_{a_1}\frac{\tau^{g-1}d\tau}{\zeta}\\
\cdots & \cdots & \cdots & \cdots\\
\oint_{a_g}\frac{w+\zeta}{2\zeta}\cdot\frac{d\tau}{\tau-z} & \oint_{a_g}
\frac{d\tau}{\zeta} & \cdots & \oint_{a_g}\frac{\tau^{g-1}d\tau}{\zeta}
\end{array}
\right|
$$
One could also phrase this in terms of normalized $\omega_j$.  On
the other hand a meromorphic analogue (elementary Weierstrass
function) with characteristic divisor $(z_0,w_0)^{-1}(\tau_1,\zeta_1)
\cdots (\tau_g,\zeta_g)$ has the form
\be
A[(z,w),(\tau,\zeta)]d\tau = A_0[(z,w),(\tau,\zeta)]d\tau-A_0[(z_0,w_0),
(\tau,\zeta)]d\tau
\label{omicron}
\ee
where $A_0[(z,w),(\tau,\zeta)]d\tau= (Num/Den)(d\tau/2\zeta)$ with
\be
Den=
\left|
\begin{array}{cccc}
1 & \tau_1 & \cdots & \tau_1^{g-1}\\
\cdots & \cdots & \cdots & \cdots\\
1 & \tau_g & \cdots & \tau_g^{g-1}
\end{array}
\right|
\label{theta}
\ee
$$ Num=
\left|
\begin{array}{ccccc}
\frac{w+\zeta}{\tau-z} & 1 & \tau & \cdots & \tau^{g-1}\\
\frac{w+\zeta_1}{\tau_1-z} & 1 & \tau_1 & \cdots & \tau_1^{g-1}\\
\cdots & \cdots & \cdots & \cdots & \cdots\\
\frac{w+\zeta_g}{\tau_g-z} & 1 & \tau_g & \cdots & \tau_g^{g-1}
\end{array}
\right|
$$

\subsection{The prime form}
We will collect first further
information and some calculations involving the prime form (cf.
\cite{af,ag,ac,dc,do,hb,ma,md,vb}).
In particular we want to confirm (\ref{GK}) - (\ref{GO}).  Thus work
with a lattice $L_B = {\bf Z}^g + B{\bf Z}^g$ and a torus ${\bf C}^g/
L_B$ (we think here of $B$ with positive imaginary part as period
matrix, so some factors of $i$ may need adjustment relative to
previous notations).
The theta function with characteristic $\alpha,\beta$ is defined via
\be
\theta[\alpha,\beta](z|B) = \sum e^{i\pi(n+\alpha)B(n+\alpha) +
2\pi i(z+\beta)}
\label{HE}
\ee
where the periodicity condition $\theta[\alpha,\beta](z+Bn+m|B) =
exp(-i\pi nBn -2\pi n(z+\beta) + 2\pi i\alpha\cdot m)\theta(z|B)$ determines
$\theta$ up to a constant.  For some arbitrary base point $P_0$ define now
the Abel-Jacobi map $I(D) = \sum\eta_i\int^{P_{i}}_{P_0}\vec{\omega}$
for a divisor $D =
\sum\eta_i P_i$.  Then $f(P) = \theta(z+I(P)|B)$ either vanishes identically
or it has exactly $g$ zeros $P_1,\cdots,P_g$.  In the latter case there
exists a vector $\Delta \sim$ Riemann divisor
class, $\sim -K$ in Section 2, depending only on $P_0$
and the canonical homology basis, such that $z+I(\sum P_i) = \Delta$.
The set of $z\in J(\Sigma_g)=$ Jacobian variety where $\theta$ vanishes
is a subset of complex codimension one called the $\theta$ divisor.
Then $\theta(z|B) = 0$ if and only if there exist $g-1$ points $P_1,
\cdots,P_{g-1}$ in $\Sigma_g$ such that $z = \Delta - I(P_1+\cdots+
P_{g-1})$.  The bundle of holomorphic one forms is called ${\cal K}$ and this
is the canonical line bundle of degree $2g-2$; spinor bundles $L$ satisfy
$L^2 = {\cal K}$.  We note that to any
holomorphic line bundle $L$ on $\Sigma_g$,
with $s$ a meromorphic section, one associates a bundle
$L_s$ with the equivalence class $D_s$ of divisor $s$.  Similarly for any
divisor $D$ we can associate a line bundle (locally build transition
functions via divisors restricted to open sets).  The relation between
spin structures and theta functions is summarized well in recalling that
the Riemann class $\Delta$ satisfies $2\Delta = {\cal K}$.
Hence $\Delta$ is the
divisor class of a spin structure $S_0$ and the divisor classes of other
spin structures $S_{\alpha,\beta}$ are given by $D_{\alpha} = \Delta -
B\alpha -\beta,\,\,(\alpha,\beta)\in (\frac{1}{2}{\bf Z}/{\bf Z})^{2g}$
to which one associates a theta function $\theta[\alpha,\beta](z|B)$.
For such theta functions one has $\theta[\alpha,\beta](-z|B) = (-1)^
{4\alpha\cdot\beta}\theta[\alpha,\beta](z|B)$ leading to a classification
of spin structures as even or odd depending on whether the corresponding
theta function is even or odd.  For odd characteristics $\alpha,\beta$
we have $\theta[\alpha,\beta](0|B)=0$ so by Riemann's vanishing theorem
there are $g-1$ points $P_i$ such that $D_{\alpha,\beta} = [\sum P_i]$
($[D]$ denotes divisor class).  This suggests that $S_{\alpha,\beta}$ has a
holomorphic section $h_{\alpha,\beta}(z)$ having $g-1$ zeros for $s = P_i$
and indeed this section can be written as follows.  Define
\be
f_{\alpha,\beta}(z,w) = \theta[\alpha,\beta](\int^z_w\omega|B)
\label{HF}
\ee
Then by the vanishing theorem $f_{\alpha,\beta}$ has single zeros for
$z=w,\,z=P_i,$ or $w=P_i$ (obvious with a moment's thought) so for both
$z$ and $w$ in the neighborhood of one of the $P_i,\,\,f_{\alpha,\beta}
(z,w)$ looks like $constant\cdot(z-w)(z-P_i)(w-P_i)$.  Differentiating this
with respect to $w$ at $w=z$ one finds that the one form $g_{\alpha,\beta}
(z) = \sum\partial_i\theta[\alpha,\beta](0|B)\omega_i(z)$ has $g-1$ double
zeros for $z=P_i$.  Note that since the divisor of $g_{\alpha,\beta}$ is
in ${\cal K}$ we have indeed
$2D_{\alpha,\beta} = {\cal K}$ and there exists a holomorphic
$\frac{1}{2}$-differential $h_{\alpha,\beta}$ such that $g_{\alpha,\beta}
=h^2_{\alpha,\beta}$ which is the holomorphic section of $S_{\alpha,\beta}$
desired.  Now one can define a holomorphic differential form of weight
$(-\frac{1}{2},0)\times(-\frac{1}{2},0)$ via
\be
E(z,w) = \frac{f_{\alpha,\beta}(z,w)}{h_{\alpha,\beta}(z)h_{\alpha,\beta}
(w)};\,\,(\alpha,\beta)\,\,odd)
\label{HG}
\ee
Let us note also that one can construct arbitrary meromorphic functions
or differentials via $E$.  Thus given that $D=P_1 +\cdots + P_n-Q_1-
\cdots - Q_n$ is the divisor of a meromorphic function one can
express this function via (cf. \cite{ma})
\be
f(z) = \frac{\prod_1^nE(z,P_i)}{\prod_1^nE(z,Q_i)}
\label{HH}
\ee
On the other hand for example a differential of third kind with first
order poles at $P$ and $Q$ with residues $\pm 1$ respectively is given by
\be
\omega(z;P,Q) = \partial_z\,log\frac{E(z,P)}{E(z,Q)}dz
\label{HI}
\ee
For the prime form (\ref{GK}) we note also the following representation
from \cite{ma}, Vol. 1, p. 160 ($E$ is treated as a function here).
\be
E(x,y) \sim E_e(x,y) = \theta(e+\int_x^y\vec{\omega});\,\,\theta(e)=0
\label{MD}
\ee
Here $\vec{\omega}\sim(\omega_j)$ and we note that $\theta(e) = 0$ if
and only if there exist points $P_1,\cdots,P_{g-1}$ such that $e =
K-\sum_1^{g-1}\int_{P_0}^{P_i}\vec{\omega}$.

\subsection{Remarks on theta functions}
Let us now extract some information from \cite{fe} regarding various
kernel formulas and theta functions.  We recall the prime form as in
(\ref{HG}) rewritten here as ($(\alpha,\beta)\sim\delta$)
\be
E(x,y) = \frac{\theta[\delta](y-x)}{h_{\delta}(x)h_{\delta}(y)};\,\,
h^2_{\delta}(x) = \sum_1^g\frac{\partial\theta[\delta](0)}{\partial
z_i}\omega_i(x)
\label{LA}
\ee
\be
\theta(z) = \theta[0](z) = \sum e^{\frac{1}{2}mBm^T + mz^T};\,\,
\theta{x \brack y}(z) =
\label{LB}
\ee
$$\sum e^{\frac{1}{2}(m+\alpha)B(m+\alpha)^T + (z+2\pi i\beta)(m+\alpha)^T}
= e^{\frac{1}{2}\alpha B\alpha^T +(z+2\pi i\beta)\alpha^T}\theta(z+e)$$
Note that $\theta(-z) = \sum exp[(1/2)mBm^T-mz^T]=\theta(z)$ via
$m\to -m$.
Here $e = 2\pi i\beta + \alpha B = (\beta,\alpha){2\pi i \choose B}$
and the period matrix $B$ is symmetric with $\Re B<0$ (Siegel left half
plane).  $E(x,y)$ will have the following properties (cf. also Section 6.4
for a physics approach).  ({\bf A})  $E(x,y) = -E(y,x)$ (recorded earlier)
({\bf B})  If ${\cal A} = \sum_1^na_i$ and ${\cal B} = \sum_1^nb_i$ are
divisors then (cf. (\ref{HI}))
\be
d\,log\prod_1^n\frac{E(x,b_i)}{E(x,a_i)} = \omega_{{\cal B}-{\cal A}}(x)
\label{LC}
\ee
where $\omega_{{\cal B}-{\cal A}}$ is a differential of third kind having
poles at $b_i$ and $a_i$ with residues $\pm 1$ respectively.  One will
have also for $X,\,Y$ divisors of degree $n$
\be
\int_X^Y\omega_{{\cal B}-{\cal A}} = \int^{{\cal B}}_{{\cal A}}
\omega_{Y-X}
\label{LD}
\ee
({\bf C})  If $\Sigma_g$ is realized as a covering of ${\bf P}^1$ via
$z:\,\,\Sigma_g\to {\bf P}^1$ then
\be
E^2(x,y) = \frac{(z(y)-z(x))^2}{dz(x)\,dz(y)}exp\{\int^{-y+z^{-1}z(y)}_
{-x+z^{-1}z(x)}\omega_{y-x} + \sum_1^g\int^y_xm_i\omega_i\}
\label{LE}
\ee
where $m_j = (1/2\pi)\int_{a_j}d\,arg[(z-z(y))/z-z(x))]$ and paths of
integration are taken within $\Sigma_g$ cut along its homology basis
in a standard manner.  Note here also that from (\ref{LC})
\be
exp\{\int_p^q\omega_{y-x}\} = \frac{E(y,q)E(x,p)}{E(x,q)E(y,p)}
\label{LF}
\ee
({\bf D})  For any nonsingular $f\in (\theta)$ (i.e. $\theta(f)=0$ and
$(\partial\theta/\partial z_i)(f)\not= 0$ for some $i$) define
\be
H_f(x) = \sum_1^g\frac{\partial\theta}{\partial z_i}(f)\omega_i(x);\,\,
Q_f(x) = \sum_{i,j=1}^g\frac{\partial^2\theta}{\partial z_i\partial z_j}
(f)\omega_i(x)\omega_j(x)
\label{LG}
\ee
Then one has holomorphic Prym differentials with $g-1$ double zeros
\be
\left(\frac{\theta(y-x-f)}{E(x,y)}\right)^2 = H_f(x)H_f(y)exp(\int^x_y\frac
{Q_f}{H_f})
\label{LH}
\ee
It follows that for $f\in (\theta)$ nonsingular
\be
\frac{\theta(y-x-f)\theta(y-x+f)}{H_f(x)H_f(y)} = E(x,y)E(y,x) =
-E(x,y)^2
\label{LI}
\ee
({\bf E})  For $\Xi = d_xd_y\,log\,E(x,y)$ (cf. (\ref{GM}))
\be
\Xi(x,y) = d_xd_y\,log\,E(x,y)dxdy=d_xd_y\,log\theta(y-x-f)dxdy
\label{LJ}
\ee
is a well defined bilinear meromorphic differential independent of the
nonsingular $f\in (\theta)$.  ({\bf F})  ${\cal Z}_p(x) = (d/dx)log\,E(x,p)$
(cf. (\ref{HI})) has a pole of residue $1$ at $x=p$ and ${\cal Z}_b(x) -
{\cal Z}_a(x) = \omega_{b-a}(x) = \int^b_a\Xi(x,y)dy$.  For $f$ holomorphic
in a neighborhood $U$ of $p$ one has
\be
f(p) = \frac{1}{2\pi i}\int_{\partial U}f(x){\cal Z}_p(x)
\label{LK}
\ee
so $Z_p$ is a local Cauchy kernel.

\subsection{Expressions for $\omega$}
First note that (\ref{GO}) is the
same as (\ref{GD}) for $z = 1/k$ and the expression (\ref{GM}) is a
standard way of obtaining a differential with a second order pole
(cf. \cite{ag}).  Thus $\hat{\omega}_2(z,\zeta) = \partial_{\zeta}
\partial_z log E(z,\zeta)
= \partial_{\zeta}(E_z/E) = (E_{z\zeta}/E) - (E_z
E_{\zeta}/E^2)$ and via $E_z\not= 0,\,\,E_{\zeta}\not= 0$ at $z=\zeta$
the last term looks like $1/(z-\zeta)^2$ (note also $\hat{\omega}_2
dz$ has zero $a_i$ periods - cf. \cite{ma}).  From this one picks up
differentials of the second kind with poles of order $n+1$ at $z=\zeta$
via $\hat{\omega}_{n+1} = \partial^{n-1}_{\zeta}\hat{\omega}_2(z,\zeta)/
n!\,\,(n=2,\cdots)$ for example and we note that (cf. \cite{ag,sb})
\be
\oint_{b_i}\hat{\omega}_{n+1}(z,\zeta)dz = \frac{2\pi i}{n!}D_{\zeta}^
{n-1}f_i(\zeta)
\label{HJ}
\ee
where $\omega_i = f_id\zeta$.  Further (cf. (\ref{HI}))
\be
\oint_{\vec{b}}\,\,\omega(z,P,Q) = 2\pi i\int_Q^P\omega
\label{HK}
\ee
where $\vec{b}\sim (b_1,\cdots,b_g)$ and $\omega\sim (\omega_1,\cdots,
\omega_g)$ represents the standard holomorphic differentials.  Here the
relation (\ref{HK}) follows from standard bilinear identities as does
\be
\oint_{b_k}\hat{\omega}_2dz = 2\pi i\frac{\omega_k}{d\zeta}
\label{HL}
\ee
and (\ref{HJ}) follows from (\ref{HL}).
\\[3mm]\indent
We can make further comments now relative to (\ref{GM}) - (\ref{GO})
by writing ($z\sim 1/k$)
\be
\hat{\omega}_2(z,\zeta) = d_zd_{\zeta}\,log\,E(z,\zeta) = -\sum_1^{\infty}
\Omega_s(z)\zeta^{s-1}d\zeta;\,\,\hat{\omega}_{n+1}(z,\zeta) =
\label{HN}
\ee
$$=\partial_{\zeta}^{n-1}\frac{\hat{\omega}_2(z,\zeta)}{n!} = -[\frac
{(n-1)!}{n!}\Omega_n +\sum_{n+1}^{\infty}{s-1 \choose n}\Omega_s(z)
\zeta^{s-n}]d\zeta$$
(the latter for $n\geq 2$).  This leads to
\be
\Omega_n(z)d\zeta = -n\hat{\omega}_{n+1}(z,0) = -\frac{1}{(n-1)!}
\partial_{\zeta}^{n-1}\hat{\omega}_2(x,\zeta)|_{\zeta=0}\,\,\,(n\geq 1)
\label{HNN}
\ee
\be
2\pi i(U_s)_m\sim\oint_{b_m}\Omega_s\sim\oint_{b_m}\hat{\omega}_{s+1}dz|_
{\zeta=0} =
\label{HO}
\ee
$$ = \frac{1}{s!}D_{\zeta}^{s-1}\oint_{b_m}\hat{\omega}_2dz|_{\zeta=0} =
\frac{2\pi i}{s!}
D_{\zeta}^{s-1}f_m(\zeta)|_{\zeta=0}$$
This implies in particular that
$(1/s!)(U_s)_m\sim D_{\zeta}^{s-1}f_m(\zeta)|_
{\zeta=0}$.  We note also that differentiating (\ref{GN}) one obtains
\be
d_{\zeta}d_z (log\frac{E(\zeta,z)}{\zeta-z}) = d_{\zeta}d_z log E(\zeta,z)
-d_z(\frac{d\zeta}{\zeta-z}) =
\label{HP}
\ee
$$= d_{\zeta}d_z log\,E - \sum_1^{\infty}\frac{nz^{n-1}}{\zeta^{n+1}}dzd\zeta
=\sum_1^{\infty}\sum_1^{\infty}\zeta^{m-1}z^{n-1}Q_{mn}dzd\zeta$$
while (\ref{GM}) and (\ref{GO}) imply
\be
d_{\zeta}d_zlog\,E = -\sum_1^{\infty}z^{s-1}dz[d(\frac{1}{\zeta^s}) -
\sum_1^{\infty}Q_{sm}\zeta^{m-1}d\zeta] =
\label{HQ}
\ee
$$=[\sum_1^{\infty}sz^{s-1}\zeta^{-s-1} + \sum_1^{\infty}\sum_1^{\infty}
Q_{sm}z^{s-1}\zeta^{m-1}]dzd\zeta$$
One sees that these equations are consistent and the terms involving
$Q_{m0}$ in (\ref{GN})
play the role of integration ``constants" used
for normalization in \cite{ge} (they are absent in \cite{ag}).
\\[3mm]\indent  Now
consider from (\ref{GJ}) $\partial_x\omega(k,k',\vec{t}) = (1/2\pi i)
\psi(k,\vec{t})\psi^{\dagger}(k',\vec{t})$ (see below for relations
between $\psi^*$ and $\psi^{\dagger}$ and further discussion
relative to the definition of $\omega$).
Thus first let us go carefully over the construction of BA functions
etc. in Section 2.
In (\ref{AA}) for example we pick $\Omega^1\sim
\Omega_1 = dk + \cdots,\,\,\Omega^j\sim\Omega_j = d(k^j) + \cdots$ with
$\oint_{a_i}\Omega_j = 0$ (or sometimes $\Re\oint_{a_i}\Omega_j = 0 =
\Re\oint_{b_i}\Omega_j$ as in Section 3.2 for example).  Set $U_j =
\oint_{b_j}\Omega_1,\,\,V_j = \oint_{b_j}\Omega_2,\cdots$ up to factors
of $2\pi i$ and recall $dp\sim \Omega_1,\,\,dE\sim \Omega_2$
(or $\Omega_3$ for KdV).  The flow variables arise via $q(k)$ in
Section 2 but the Riemann surface contributes via the argument $xU+
yV+yW$ in the theta function to establish linear flows on the Jacobian
$J(\Sigma_g)$.  We recall a few additional facts about BA functions etc.
from \cite{dc,do}.  First from (\ref{AA}) we have
\be
\psi = exp(\int^P_{P_0}\Omega)\cdot\frac{\theta({\cal A}(P) + \vec{U}+z_0)}
{\theta({\cal A}(P)+z_0)}
\label{HW}
\ee
for $z_0 = -{\cal A}(D)-K\,\,(\Omega\sim x\Omega_1+\cdots,\,\,\vec{U}\sim
xU+yV+\cdots)$.  We recall also that $F(P) = \theta({\cal A}(P)-e)$ is
either identically zero or $F(P)$ has exactly $g$ zeros including
multiplicities on the cut surface $\tilde{\Sigma}_g$.  Note here
$e=(e_1,\cdots,e_g)$ and ${\cal A}(P) = (\int^P_{P_0}\omega_1,\cdots,
\int^P_{P_0}\omega_g)$.  Then if $F(P)\not\equiv 0$ and $P_1,\cdots,P_g$
are its zeros one has ${\cal A}(P_1,\cdots,P_g) = {\cal A}(P_1) +
\cdots + {\cal A}(P_g) \equiv e-K$ on $J(\Sigma_g)$ where
${\cal A}(P_1,\cdots,P_g)=
\sum_1^g(\int^{P_i}_{P_0}\omega_1,\cdots,
\int^{P_i}_{P_0}\omega_g) = (\sum_1^g\int^{P_i}_{P_0}\omega_1,\cdots,
\sum_1^g\int^{P_i}_{P_0}\omega_g)$.  Recall here that $K$ is the vector
of Riemann constants and $2K=-{\cal A}(K_{\Sigma})$ where $K_{\Sigma}\sim$
equivalence class of meromorphic differentials.  Hence the function
$\theta({\cal A}(P)-e)$ is identically zero if and only if $e$ can be
written as $e = \sum_1^g{\cal A}(Q_i) +K$ where $Q=\sum_1^g Q_i$ is a
special divisor (i.e. there exists a nonconstant meromorphic function
with poles only at the $Q_i$); then for $D=\sum P_i,\,\,{\cal A}(D)
=e-K\sim{\cal A}(D) = {\cal A}(\sum Q_i) = 0$ for any $D$.
A divisor in general position is not special.  The above leads to ($e=
\zeta +K,\,\,\zeta$ arbitrary) ${\cal A}(P_1,\cdots,P_g) = \zeta$ and
further the zeros $e$ of $\theta(e) = 0$ have a representation $e =
{\cal A}(P_1)+\cdots+{\cal A}(P_{g-1}) +K$ where $P_1,\cdots,P_{g-1}$
are arbitrary (cf. \cite{dc}).  In particular one sees that the poles
of the BA function (\ref{HW}) arising from zeros of the denominator
lie at the points $P_i$ (i.e. for $D$ nonspecial, $D=\sum P_i,\,\,
\theta({\cal A}(P) -{\cal A}(D)-K)$ has exactly $g$ zeros $P_1,\cdots,
P_g$).
\\[3mm]\indent
For completeness let us also include here some of the formulation
of \cite{ks} (cf. also \cite{ca}) but omitting any conformal field
theory.  One writes $\xi(\vec{t},z) = \sum_1^{\infty}t_nz^n$ with
$\Psi(\vec{t},z)\sim exp(-\xi)(1+\sum_1^{\infty}w_kz^{-k})\sim
exp(-\xi)\tau(\vec{t}+[z])/\tau(\vec{t}\,)$ where $\vec{t}+[z]\sim
(t_j+1/jz^j)$.  Similarly $\tilde{\Psi}(\vec{t},z)\sim exp(\xi)
(1+\sum_1^{\infty}\tilde{w}_kz^{-k})\sim exp(\xi)\tau(\vec{t}-[z])/
\tau(\vec{t})$ ($\bar{\Psi}$ is used in \cite{ks}).  Thus $\tilde
{\Psi}\sim$ our normal $\psi$ and $\Psi\sim$ our normal $\psi^*$ but
poles and zeros must be examined.
Then on a Riemann surface $\Sigma_g$ one takes holomorphic differentials
(Abelian differentials of the first kind) in the form $\omega_j = \omega_j
dz$ with $\oint_{a_i}\omega_j = \delta_{ij},\,\,\oint_{b_i}\omega_j =
B_{ij},\,\,\omega_j(z)dz = d(\sum_{n>0}I_n^j z^{-n}/n) = -\sum
I^j_nz^{-n-1}dz$ (we inserted a plus sign in $\omega_j$).  For
differentials of the second kind one writes ($Q\in \Sigma,\,\,z(Q) =
\infty$)
\be
\omega_Q^n=\omega_Q^ndz;\,\,\oint_{a_i}\omega_Q^n = 0;\,\,
\oint_{b_j}\omega^n_Q = 2\pi i I^j_n;
\label{ZB}
\ee
$$\omega_Q^n\sim d(z^n - \sum_{m>0}\frac{q_{nm}}{m}z^{-m})$$
If we write $z=1/\zeta$ this becomes $\omega^n_Q
\sim d(1/\zeta^n) - \sum q_{nm}\zeta^{m-1}d\zeta$
which agrees with $\Omega_n$ in (\ref{GO}) for $q_{nm} = Q_{nm}$.  Further
\be
d_zd_w\,log\,E(z,w) = [\frac{1}{(z-w)^2}+\sum_{m,n>0}q_{nm}
z^{-n-1}w^{-m-1}]dzdw
\label{ZC}
\ee
which agrees with (\ref{HQ}) (since $1/(z-w)^2 = (d/dz)(1/(w-z)
=(d/dz)\sum_0^{\infty}z^nw^{-n-1} = \sum_1^{\infty}nz^{n-1}w^{-n-1}$).
The Szeg\"o kernel is written now as
\be
S_c(z,w) = \frac{\theta(I(z)-I(w)+c|B)}{\theta(c|B)E(z,w)}
=\frac{1}{z-w}+\sum_{\mu,\nu>0}c_{\mu\nu}z^{-\mu-\frac{1}{2}}w^{-\nu-
\frac{1}{2}}
\label{ZD}
\ee
where $c\in {\bf C}^g$ with $\theta(c|B)\not= 0$
(note the similarity to (\ref{EEE}) which however is not equivalent).
We will not consider Szeg\"o kernels further (nor deal with Krichever-
Novikov kernels as in \cite{gd,kp}).
Here the Abel-
Jacobi map is written as
\be
I(z) = (\int_Q^z\omega_j) = (\sum_n \frac{I^j_n}{n}z^{-n})
\label{ZE}
\ee
Here we have inserted a factor of $1/n$ in (\ref{ZE}), apparently
forgotten in \cite{ks}.  Note our $\omega_j\sim z^{-2}dz$ whereas
in (\ref{FA}) or (\ref{FH}) for hyperelliptic situations $\omega_j
\sim \lambda^{-3/2}d\lambda\sim 2k^{-2}dk$ for $\lambda = k^2$ so the
growth condition matches.  Finally set $\phi^0(z) = 1$ with
\be
\phi^n(z) = z^n - \sum_1^{\infty}\frac{q_{nm}}{m}z^{-m} = \int^z\omega_Q^n\,
\,\,(\omega_Q^n\sim\Omega_n);
\label{ZF}
\ee
$$\phi^n(z+a_i) = \phi^n(z);\,\,\phi^n(z+b_j) = \phi^n(z) + 2\pi i I_n^j$$
We record now the tau function via
\be
\tau_c(\vec{t},\Sigma_g) = e^{\frac{1}{2}q(\vec{t}\,)}\theta(I(\vec{t}\,)
+c|B);
\label{ZG}
\ee
$$q(\vec{t}\,) = \sum_{m,n>0}q_{nm}t_nt_m;\,\,I(\vec{t}\,) = (I^j(\vec{t}\,) =
(\sum_1^{\infty}I^j_nt_n)$$
and the BA functions have the form indicated above, or more generally,
in the form required here
\be
\Psi\sim f(z)e^{-\sum_1^{\infty}t_n\phi^n(z)}\frac{\theta(I(\vec{t}\,)
+I(z)+c|B)}{\theta(I(\vec{t}\,)+c|B)};
\label{ZH}
\ee
$$\tilde{\Psi}\sim f(z)e^{\sum_1^{\infty}t_n\phi^n(z)}\frac
{\theta(I(\vec{t}\,)-I(z)+c|B)}{\theta(I(\vec{t}\,)+c|B)}$$
where $f(z)\sqrt{dz} = (\sqrt{dz}w/E(z,w))|_{w\to\infty} =
[1+o(z^{-1})]\sqrt{dz}$.  Poles and zeros of $\Psi,\,\tilde{\Psi}$
should however be examined.  For comparison to $\psi,\,\,\psi^*$
evidently $I(\vec{t})$ must correspond to $\vec{U} = xU+yV+\cdots$, and
$I(z)\sim -{\cal A}(P)$ (i.e. $z\sim P$ and perhaps we should restore
the minus sign here in $\omega_j$).  In this spirit (\ref{HW}) corresponds
to $\tilde{\Psi}$ via
\be
\psi = e^{\sum_1^{\infty}t_n\int^z_{\infty}\Omega_n}\cdot\frac
{\theta(\vec{U}-I(z)+z_0)}{\theta(z_0-I(z))}\sim\frac
{\tilde{\Psi}}{f}\cdot\frac{\theta(I(\vec{t}+c|B)}{\theta(z_0-I(z))}
\label{HWW}
\ee
and as $P\to P_0\,\,(z\to\infty$) one has $\psi\sim\tilde{\Psi}/f$ as
required.  We will not pursue this.
\\[3mm]\indent
Now look at $\omega$ in (\ref{GJ}).  One can take $\psi$ as
in (\ref{HW}) and for $\psi^*$ we know it is determined via a dual
divisor $D^*$ (cf. \cite{cm,de})
such that $D+D^*$ is the null divisor of
some meromorphic differential $\Omega = dk +\alpha(dk/k^2)+\cdots$.
Thus $D+D^*-2Q\sim K_{\Sigma}\,\,(Q\sim\infty,\,\,2K\sim -{\cal A}
(K_{\Sigma}$) so ${\cal A}(D^*) - {\cal A}(Q) +K = -[{\cal A}(D) -
{\cal A}(Q) +K]$.  The prescription of $\psi^{\dagger}$ in
({\bf A}) - ({\bf F}) seems to
require for example via (\ref{star}) ($P\sim k,\,\,P'\sim k'$)
\be
\psi^{\dagger}(k',\vec{t}\,)\sim\theta({\cal A}(P')+z_0)\theta({\cal A}(P')
+z_0^*)\psi^*(k',\vec{t}\,)dk'
\label{stdg}
\ee
(modulo some normalization and adjustment for path independence).
We recall now that by definition of $\psi^*,\,\, \psi(k',\vec{t}\,)
\psi^*(k',\vec{t}\,)\Omega(k') = \Phi(k',\vec{t}\,)$ is a meromorphic
differential in $k'$ having its only (double) pole at $\infty$.  Given
$\psi$ and $\psi^*$ as in (\ref{AA}) - (\ref{star}) this means we could
take $\Omega$ in the form $\theta({\cal A}(P')+z_0)\theta({\cal A}(P')
+z_0^*)\tilde{\Omega}$ where $\tilde{\Omega}$ is any meromorphic differential
with its only pole at $\infty$ of the type indicated.  But this is
exactly what we achieve via (\ref{stdg}) except for the pole at
$\infty$ and the question of global definition.
We recall from (\ref{AA}) and remarks after Theorem 6.2 the
changes induced by a path change
$\gamma = \sum_1^gn_ka_k +\sum_1^gm_jb_j$.  Thus (cf. \cite{ca}, p. 56)
one has $\int_{P_0}^P\Omega\to\int_{P_0}^P\Omega + <M,U>$ and
${\cal A}(P)\to {\cal A}(P) + 2\pi iN+BM$ with $\theta(z+
2\pi iN+BM)=\theta(z)exp(-(1/2)<BM,M>-<M,z>)$.  Thus introducing
theta functions to cancel poles etc. leads to problems of global
definition and we are better advised to use the prime form $E$.
If we take $E$ in the form $E(x,y) = \theta(e+\int_x^y\vec{\omega})$
for example (cf. (\ref{MD})) then one has, generically, for suitable
paths, upon cutting the Riemann surface in a standard manner, $\theta
(e+\int_{P_i}^P\vec{\omega})=\theta(e+{\cal A}(P)-{\cal A}(P_i))$ and
this is unchanged with a path $\gamma$ (which affects both ${\cal A}(P)$
and ${\cal A}(P_i)$).  However $E$ is multivalued as indicated after
(\ref{GK}).  Then instead of (\ref{stdg}), one can
think of
\be
\psi^{\dagger}(k',\vec{t}\,) = \psi^*(k',\vec{t}\,)\prod_1^gE(P',P_i)
\prod_1^gE(P',P^*_i)dk'
\label{psinow}
\ee
(recall $E$ is treated as a function here).
Thus we think of defining $\omega$ in (\ref{GJ}) as
\be
\omega(P,P',x,y,t,\cdots) = \frac{1}{2\pi i}\int_{\pm\infty}^x
\psi(k,x',y,t,\cdots)\psi^*(k',x',y,t,\cdots)\check{\Omega}(k')dx'
\label{kkernel}
\ee
(up to some factors for normalization and global definition indicated below)
where $\psi$ and $\psi^*$ are given in (\ref{AA}) - (\ref{star}) and
\be
\check{\Omega}(k') =\prod_1^gE(P',P_i)\prod_1^gE(P',P^*_i)dk'
\label{omega}
\ee
Then $\omega$ will coincide with (\ref{GJ}) when $\psi^{\dagger}$ is
given by (\ref{psinow}) and
properties ({\bf A}) - ({\bf F}) will hold.
Thus the kernel in (\ref{GJ})
can be written tentatively via
\be
\omega_x(P,P',\vec{t}\,) = \frac{1}{2\pi i}\psi(x',k,y,t,\cdots)
\psi^*(x',k',y,t,\cdots)\check{\Omega}(k') =
\label{ker}
\ee
$$\frac{e^{(\int_{P_0}^{P}-\int_{P_0}^{P'})
[x\Omega_1
+\cdots]}}{2\pi i}
\cdot\frac{\theta({\cal A}(P)+\vec{U}+z_0)\theta
({\cal A}(P')-\vec{U}+z_0^*)\prod_1^gE(P',P_i)E(P',P^*_i)}
{\theta({\cal A}(P)+z_0)\theta({\cal A}(P')+z_0^*)}dk'$$
Now to determine
global definition we are concerned only with separate path changes
for $P$ or $P'$ integrals.  Thus for $P$ the exponential term contributes
a multiplier $exp(<M,U>)$ and the theta functions give $exp(-<M,U>)$
which cancels.  For $P'$ we get $exp(-<M,U>)$ from the exponential and
$exp(<M,U>)$
from the theta functions which again cancels.  Hence
(\ref{ker}) is globally well defined (but multivalued) and we have tentatively
\\[3mm]\indent {\bf THEOREM 6.2}.$\,\,$  A path independent expression
for $\omega_x$ having the essential poles and singularities
stipulated in ({\bf A}) - ({\bf F}) can be
written as in (\ref{ker}).  $\bullet$
\\[3mm]
\indent
A definition of $\omega$ via $\psi\psi^*$ (instead of $\psi\psi^{\dagger}$)
seems more appropriate in dealing with dispersionless limits (as seen
in Section 6.9) but in order to have a Cauchy kernel the poles from
$\psi\psi^*$ should be eliminated (hence $\psi^{\dagger}$).
This leads to $\psi^{\dagger}$ as in (\ref{psinow}) and $\omega_x$ as
in (\ref{ker}).  The pole in $\omega$ at $k=k'$ will emerge from the
integration as indicated before
(cf. also Theorem 6.3) and we will have a discontinuous Cauchy
kernel analogue in the spirit of Section 6.5.
We have
to rethink (\ref{GH}), (\ref{GI}), and $\partial_n\psi^{\dagger} =
-B_n^*\psi^{\dagger}$ now but, since no $x,y,t$ dependence has been
introduced, it seems that $\partial_n\psi^{\dagger}
=-B_n^*\psi^{\dagger}$ will still hold, (\ref{GH}) should be OK,
and (\ref{GI}) should be preserved up to a multiplier (which we
don't compute here).

\subsection{Averaging for theta functions}
In terms of averaging one has ($\int_{P_0}^P\Omega_1\sim p$)
\be
\partial_x\,log\omega_x = \frac{\omega_{xx}}{\omega_x} = [p(k)-p(k')]
+\partial_x\,log\theta({\cal A}(P)+\vec{U}+z_0) +
\label{aver}
\ee
$$+\partial_x\,log\theta({\cal A}(P')-\vec{U}+z_0^*)$$
The averaging process
should now evidently revert to
integrals $(1/2\pi)^{2g}\int\cdots\int\partial_x
log\theta
\linebreak
\sim (1/2\pi)^{2g}\int\cdots\int\sum U_i(\partial/\partial\theta_i)
log\theta d^{2g}\theta_i$ and we recall that $\theta(\vec{z}+2\pi ie_k) =
\theta(\vec{z})$ and $\theta(\vec{z}+Be_k) = exp(-(1/2)B_{kk}-z_k)
\theta(\vec{z})$ (cf. \cite{dc}).  This would imply
for the fourth term in (\ref{aver}) (note ($\theta_i,
\theta_{g+i})\sim z_i$), running $\theta_j$ between $0$ and
$2\pi$,
\be
\int\cdots\int\partial_x\,log\theta = -\sum_1^g U_i(-\frac{1}{2}
B_{ii})
\label{KA}
\ee
(here $\theta_j\sim {\cal A}(P)+\vec{U}_j+(z_0)_j$ and running
this in $(0,2\pi)$
eliminates the $z_j$ terms from $-(1/2)B_{jj}-z_j$ above).
But from the fifth term in (\ref{aver}) we would obtain correspondingly
$-\sum_1^gU_i(-(1/2)B_{ii})$ yielding formally
\be
(4)+(5) = -\sum_1^g U_i(-\frac{1}{2}B_{ii}) + \sum_1^gU_i(-\frac
{1}{2}B_{ii})] =0
\label{KB}
\ee
Note here $\vec{U}_i\not= U_i$ since $\vec{U} = xU+yV+\cdots$ and
$\vec{U}_i = xU_i+yV_i+\cdots$ (althought this is not used here).
The quantities $B_{ii},\,U_i$ are then
thought of as functions of slow variables $X,Y,T,\cdots$.  This is
perhaps somewhat cavalier but starting a period at some $z_i\not= 0$ will
leave terms in (\ref{KB}) depending on $x,y,t,\cdots$;
recall also that (up to possible factors of $2\pi i$)
\be
U_i=\oint_{b_i}\Omega_1;\,\,U'_i =\oint_{b_i}\Omega'_1;\,\,
B_{ii} = \oint_{b_i}\omega_i;\,\,B'_{ii} = \oint_{b_i}\omega'_i;
\label{KC}
\ee
Consequently one obtains
\be
<\partial_x\,log\omega_x> = p(k)-p(k')
\label{log}
\ee
which agrees with (\ref{FE}) and (\ref{FL}) for example in the
hyperelliptic case where $k\sim i\sqrt{\lambda}$.

\subsection{Connections to differential Fay identity}
We recall from \cite{ab} that the differential Fay identity leads to
\be
\psi^*(\vec{t},\lambda)\psi(\vec{t},\mu) =\frac{1}{\mu-\lambda}
\partial\{\frac{X(\vec{t},\lambda,\mu)\tau(\vec{t}\,)}{\tau(\vec{t}\,)}\}=
\label{KH}
\ee
$$= \frac{1}{\mu-\lambda}\partial\{e^{\sum t_j(\mu^j-\lambda^j)}
\frac{\tau(\vec{t}+[\lambda^{-1}]-[\mu^{-1}])}{\tau(\vec{t}\,)}\}$$
Given the equivalence of the dispersionless differential Fay identity
with the kernel expansion (\ref{GA}) (cf. \cite{ch}) one expects
(\ref{KH}) to be at least formally useful in studying $\omega$ in (\ref{GJ})
since (\ref{KH}) implies (modulo adjustment for $\psi^{\dagger}$)
\be
\omega\sim\frac{1}{2\pi i}\partial_{x'}^{-1}[\psi^*(x',t_n,\lambda)\psi
(x',t_n,\mu)]\sim\frac{1}{2\pi i}\frac{1}{\mu-\lambda}\frac
{X(\vec{t},\lambda,\mu)\tau(\vec{t}\,)}{\tau(\vec{t}\,)}
\label{KI}
\ee
modulo integration or normalization factors (note (\ref{KI}) implies
$\omega\sim 1/2\pi i(\mu-\lambda)$ as $\mu\to \lambda$).   Hence one
should be able to determine easily a dispersionless limit for $\omega$.  Thus
as in \cite{ch,ta} express $\tau$ via $\tau\sim exp(F(T)/\epsilon^2)$ and
write
\be
\omega_{\epsilon}\sim\frac{1}{2\pi i}\frac{1}{\mu-\lambda}e^{\frac{1}
{\epsilon}\sum T_i(\mu^i-\lambda^i)}\cdot\frac{\tau(\vec{T}+
\epsilon[\lambda^{-1}]-\epsilon[\mu^{-1}])}{\tau(\vec{T}\,)}
\label{KJ}
\ee
Take logarithms to obtain then
\be
log\omega_{\epsilon}\sim log\frac{1}{2\pi i}- log(\mu-\lambda) +
\frac{1}{\epsilon}\sum T_i(\mu^i-\lambda^i) +
\label{KK}
\ee
$$+ \frac{1}{\epsilon^2}\{F(\vec{T}+\epsilon [\lambda^{-1}] -\epsilon
[\mu^{-1}])-F(\vec{T}\,)\}$$
The last term can be written as ($\chi_n\sim$ elementary Schur functions)
\be
\frac{1}{\epsilon^2}\{e^{\sum\lambda^{-i}\frac{\epsilon\partial_i}
{i}}\cdot e^{-\sum\mu^{-1}\frac{\epsilon\partial_i}{i}}F - F\}=
\label{KL}
\ee
$$=\frac{1}{\epsilon^2}\{\sum_0^{\infty}\chi_n(\epsilon\tilde{\partial})
\lambda^{-n}\cdot\sum_0^{\infty}\chi_m(-\epsilon\tilde{\partial})\mu^{-m}F
-F\} =$$
$$=\frac{1}{\epsilon^2}\{\sum_1^{\infty}(\,\,\,)_n + \sum_1^{\infty}(\,\,\,)_m
+ \sum_1^{\infty}\sum_1^{\infty}(\,\,\,)_n(\,\,\,)_m\}$$
Now we have (cf. (cf. \cite{ch,ta})
\be
\frac{1}{\epsilon^2}\sum_1^{\infty}\sum_1^{\infty}(\,\,\,)_n(\,\,\,)_m\to
-\sum_1^{\infty}\sum_1^{\infty}\frac{F_{nm}}{nm}\lambda^{-n}\mu^{-m}
\label{KM}
\ee
and we
recall here from \cite{ch}
\be
\sum_1^{\infty}\sum_1^{\infty}\frac{F_{nm}}{nm}\lambda^{-n}\mu^{-m} =
-log(1-\frac{\mu}{\lambda})-\sum_1^{\infty}\frac{Q_n(\mu)}{\lambda^n}=
\label{KN}
\ee
$$=-log(\lambda-\mu)+log\lambda - \sum_1^{\infty}Q_n(\mu)\lambda^{-n}
= log[\frac{P(\lambda)-P(\mu)}{\lambda-\mu}]$$
(cf. also Section 4).
Now try an expression $\omega_{\epsilon} =
f(\lambda,\mu)exp(R/\epsilon)$ which will
entrain
\be
\frac{R}{\epsilon}+log\,f\sim log(\frac{1}{2\pi i})-
log(\mu-\lambda) +
\frac{1}{\epsilon}\sum T_i(\mu^i-\lambda^i) +
\label{KO}
\ee
$$+ \frac{1}{\epsilon^2}\{\sum_1^{\infty}(\,\,\,)_n + \sum_1^{\infty}(\,\,\,)_m
-\sum_1^{\infty}\sum_1^{\infty}(\,\,\,)_n(\,\,\,)_m\}$$
Multiply by $\epsilon$ and let $\epsilon\to 0$ to obtain
\be
R\sim\sum T_i(\mu^i-\lambda^i) + \sum_1^{\infty}\frac{F_n}{n}\lambda^{-n}
-\sum_1^{\infty}\frac{F_m}{m}\mu^{-m}
\label{KP}
\ee
which leads to
\be
log\,f\sim log(\frac{1}{2\pi i})-log(\mu-\lambda)-log(\frac
{P(\mu)-P(\lambda)}{\mu-\lambda})
\label{KPP}
\ee
This implies via \cite{ch} and Section 4
\be
\partial_XR = (\mu-\lambda) + \sum_1^{\infty}\frac{F_{1n}}{n}(\lambda^{-n}
-\mu^{-n}) = P(\mu)-P(\lambda)
\label{KQ}
\ee
and
\be
log\,f=-log[P(\mu)-P(\lambda)]+log(\frac{1}{2\pi i})
\label{KQQ}
\ee
Thus $\omega_{\epsilon}\sim (1/2\pi i)[1/(P(\mu)-P(\lambda))]exp\{1/
\epsilon)[S(\mu)-S(\lambda)]\}$
and $\partial_x\,log\omega_{\epsilon}\sim P(\mu) -
P(\lambda)\,\,(\partial_x = \epsilon\partial_X)$.
Note $\partial_XS = P$ in the notation of \cite{ch}
implies
\be
\partial_X R =  \partial_X[S(\mu)-S(\lambda)]
\Rightarrow R \sim S(\mu)-S(\lambda)
\label{KR}
\ee
One can therefore state
\\[3mm]\indent {\bf THEOREM 6.3}.$\,\,$
A dispersionless kernel analogous to $\omega$ can be modeled on
(\ref{KI}) to be extracted from
\be
2\pi i\omega_{\epsilon}\sim\frac{1}{P(\mu)-P(\lambda)}e^{\frac{1}{\epsilon}
[S(\mu)-S(\lambda)]}
\label{KS}
\ee
(recall also $\psi\sim exp(S/\epsilon)$ and $\psi^*\sim
exp(-1/\epsilon)$ in the dispersionless theory).
We see here how ${\cal K}\sim (1/[P(\mu)-P(\lambda)])$ arises in looking
for a dispersionless form of $\omega$
in the genus zero situation.
$\bullet$
\\[3mm]\indent
We note that $2\pi i\omega_x\sim\psi\psi^*$
corresponds to
\be
\partial_x(log\,W_x)_{\epsilon}\sim\partial_x\frac{1}{\epsilon}
[S(\mu)-S(\lambda)]\sim P(\mu)-P(\lambda)
\label{KT}
\ee
so results like
(\ref{FE}) seem natural - i.e. the dispersionless limit agrees with
averaging upon identification of the differently defined quantities
$p(\lambda)$.  We do not bother here with the factor of $i$ which arises
from a particular normalization (our $\lambda,\mu$ above correspond to
$k',k$).

\section{FURTHER PERSPECTIVES AND CONCLUSIONS}
\renewcommand{\theequation}{7.\arabic{equation}}\setcounter{equation}{0}

We begin with some remarks related to results in \cite{ch,cl}.

\subsection{Connections to inverse scattering}
Now referring to KdV situations and we recall the
classical KdV picture \`a la \cite{ca,ce,ch,ds,za}.  Thus for $u_t =
u'''-6uu'$ with $\psi''-u\psi = -k^2\psi$ and $\psi_{\pm}\sim
exp(\pm ikx)$ as $x\to \pm\infty$ one has
$T(k)\psi_{-}(k,x) = R(k)\psi_{+}(k,x) + \psi_{+}(-k,x)$
where $R,\,T$ represent reflection and transmission coefficients
respectively.  For $\psi_{-}=exp(-ikx+\phi(k,x))$ this leads to
\be v = -
\partial \log (\psi_-) =ik - \phi'; \, \,  \phi'' - 2ik\phi' + \phi'^2 = u
\label{MU}
\ee
\be
\phi'= \sum_1^{\infty}{\phi_n \over (ik)^n};\,\,\, v= ik +
\sum_1^{\infty}{v_n \over (ik)^n};\,\,
 -\log(T) =
\sum_1^{\infty}\frac{1}{(ik)^n}\int_{-\infty}^{\infty} \phi_n dx
\label{MV}
\ee
involving $\phi_n=-v_n$, where these $\phi_n$ are \`a priori
unrelated to the $\phi_n$ of (\ref{FN}).
In the dKdV context with $\psi_{+}\sim\psi
=exp(S/\epsilon)$ and $\psi_{-}\sim\psi^* = exp(-S/\epsilon)$ one obtains
(cf. \cite{ch})
\be
{1 \over T} = {P \over ik} = 1 - \sum_1^{\infty} {\tilde{P}_{n+1} \over
(ik)^{n+1}}
\label{MX}
\ee
(we have written $\tilde{P}_n\sim \tilde{P}_n(X,\cdots)$  here to
distinguish these coefficients \`a priori
from the $P_n$ in (\ref{TMT})).
One also obtains
$P = {\partial S \over \partial X} = ik - \phi' = v; \,\,\, P^2 - U = -k^2$
where $\psi_{-}\sim exp(-S/\epsilon),\,\,\phi'(x,t)\sim \tilde{\phi}'
(X,T)+O(\epsilon),\,\,P = \partial_XS,$ etc.  These relations are quite
interesting as pointed out in \cite{ch}.  The scattering data $T$ determined
by asymptotics, has meaning in some sense related to Whitham averaging on a
degenerate Riemann surface, or to dispersionless limits,
and e.g. the classical action variable
$log|T|$ can be thought of as depending on the slow variables $X$, etc.
Further, formally, from (\ref{MV}) and (\ref{MX}) we get
\be
P=ik-\tilde{\phi}'\sim ik-\sum_1^{\infty}\frac{\tilde{\phi}_n}{(ik)^n}=
\label{MYY}
\ee
$$= ik-\sum_1^{\infty}\frac{\tilde{P}_{n+1}(X,\cdots)}{(ik)^n}\Rightarrow
\tilde{\phi}_n = \tilde{P}_{n+1}$$
which indicates how the classical $\phi_n$ are related to slow variables
via $\tilde{P}_{n+1}$
Also from (\ref{MX})
\be
-log(T) = log\left(1-\sum_1^{\infty}\frac{\tilde{P}_{n+1}}{(ik)^{n+1}}\right)
\label{MYM}
\ee
Because of the first order term in (\ref{MV}) for this to be correct
we must have $\int_{-\infty}^{\infty}\phi_1dx = 0$ in (\ref{MV}), which
corresponds to $H_0 =0$ or $\int_{-\infty}^{\infty}udx=0$ (cf. \cite{ds}).
With this assumption we can expand the logarithm in (\ref{MYM}) to obtain
via $log(1-x) = x-(1/2)x^2+(1/3)x^3+\cdots$
\be
-log(T) = \sum_1^{\infty}\frac{\tilde{P}_{n+1}}{(ik)^{n+1}} -
\frac{1}{2}\left(\sum_1^{\infty}\frac{\tilde{P}_{n+1}}{(ik)^{n+1}}\right)^2
+ \cdots =
\label{ND}
\ee
$$= \frac{\tilde{P}_2}{(ik)^2} +\frac{\tilde{P}_3}{(ik)^3} +\cdots
-\frac{1}{2}\left(\frac{\tilde{P}^2_2}{(ik)^4} + \frac{2\tilde{P}_2
\tilde{P}_3}{(ik)^5} + \cdots\right) +\cdots$$
Hence from (\ref{MV}) one obtains
\be
0=\int_{-\infty}^{\infty}\phi_2dx\sim\tilde{P}_2;\,\,\int_{-\infty}^{\infty}
\phi_3dx\sim\int_{-\infty}^{\infty}P_1dx\sim\tilde{P}_3;
\label{NE}
\ee
$$0=\int_{-\infty}^{\infty}\phi_4dx\sim\tilde{P}_4-\frac{1}{2}\tilde{P}_2^2;
\,\,\cdots$$
This gives connections of the universal coordinates $\tilde{P}_s$ from
dKdV with classical scattering data.
In \cite{ce} one uses $u_t-uu_x+u_{xxx}=0$ with $\psi_{+}\sim
exp(ikx+\hat{\phi}(k,x))$ to obtain (modulo $\pm$ signs)
\be
\hat{\phi}'\sim\sum_1^{\infty}\frac{\hat{\phi}_n}{(ik)^n};\,\,H_n\sim
2^n\cdot18\int_{-\infty}^{\infty}\hat{\phi}_{2n+1}dx;
\label{MY}
\ee
$$E\frac{\delta H_n}{\delta u}=\partial\frac{\delta H_{n+1}}{\delta u};\,\,
E = \partial^3+\frac{1}{3}(\partial u+u\partial);\,\,
H_1\sim\frac{1}{2}\int_{-\infty}^{\infty}u^2dx;\,\,H_2\sim\int_{-\infty}^
{\infty}(\frac{u^2}{6}-\frac{u_x^2}{2})dx$$
We note a few obvious comparisons.  First one is dealing
with $L\psi = \lambda\psi$ in Remark 6.2, where $L=-\partial^2+u$ say,
so that $\lambda\sim k^2$ with $\sqrt{\lambda}\sim k$.  Hence (\ref{FL})
corresponds to $-i(log\psi)_x\sim k +\sum_1^{\infty}(P_s/(2k)^{2s+1}$.
Compare this to (\ref{MU}) for example to get (for $\psi\sim\psi_{-}$)
\be
-i[log(\psi_{-})]_x = k +\phi';\,\,\phi\sim\sum_1^{\infty}\frac
{\int_{-\infty}^{\infty}\phi_n}{(ik)^n}=-i\sum_0^{\infty}(-1)^m\frac
{\int_{-\infty}^{\infty}\phi_{2m+1}dx}{k^{2m+1}}
\label{MZ}
\ee
since $\int_{-\infty}^{\infty}\phi_{2m}dx=0$ for KdV
(cf. \cite{ds}).  Hence one thinks
of $\phi\sim\sum_0^{\infty}(\int_{-\infty}^{\infty}P_sdx/(2k)^{2s+1})$
yielding $(1/2^{2s+1})\int P_sdx\sim -i(-1)^s\int\phi_{2s+1}dx$
or
\be
P_s\sim -i(-1)^s2^{2s+1}\phi_{2s+1} =\alpha_s\phi_{2s+1}\Rightarrow
H_s\sim\int P_sdx\sim\alpha_s\int\phi_{2s+1}
\label{NA}
\ee
(the latter from \cite{ce,ds}).  We also see from (\ref{MV}) that
in some sense
$-log(T)=-i\sum_0^{\infty}(-1)^s[\int_{-\infty}^{\infty}\phi_{2s+1}dx/
k^{2s+1})$ so that
\be
-log(T)\sim\sum_0^{\infty}\frac{\int_{-\infty}^{\infty}P_sdx}{(2k)^{2s+1}}
\label{NB}
\ee
which it is tempting to compare with (\ref{FL}) for $\sqrt{\lambda} = k$,
i.e. $p(\lambda)\sim \tilde{p}(k)\sim k+\sum_0^{\infty}(I_s/
(2k)^{2s+1})$.  Such a comparison would involve
\be
k-log(T)\sim \tilde{p}(k);\,\,I_s=<P_s>_x\sim\int_{-\infty}^{\infty}
P_sdx
\label{NC}
\ee
This suggests an heuristic theorem of the form
\\[3mm]\indent{\bf THEOREM 7.1}.$\,\,$  Inverse scattering data $k-
log(T)$ for classical KdV on {\bf R} corresponds to $p(\lambda)\sim
\tilde{p}(k)$ in the finite zone situation where the quasi-momentum
$p(\lambda)\sim\int^P_{\infty}\Omega^1$ as in (\ref{FF}) (with general
$\Omega_1\sim\Omega^1$ as in (\ref{GD}), (\ref{HNN}), etc.).  Relations
to averaging of square eigenfunctions arise via formulas of the form
(\ref{AP}) (where
a different normalization is used) or simply via (\ref{AN}) and (\ref{NA}),
i.e. one relates the ${\cal T}_j,\,\,{\cal X}_j$ arising from square
eigenfunction expansions to the $H_s$ and the $H_s$ can be defined via
inverse scattering or via differentials.  Further, a direct connection
of classical scattering ideas to slow variables is expressed via
(\ref{MX}) and $\tilde{\phi}_n\sim\tilde{P}_{n+1}$ in (\ref{MYY}) when
$\phi'(x)\sim\tilde{\phi}'(X) +O(\epsilon)$.  Other formulas hold as
indicated above.

\subsection{Relations between differentials and averaged quantities}
We recall that averaging in Section 3.5 involved $\gamma=\psi\psi^*$ directly,
whereas in Section 3.6 expansions were used involving terms like
$<\psi^*L^1\psi>,\,\,<\psi^*A^1\psi>$, etc.  In the dispersionless limit
situation one could also conceivably adapt
Section 6.9 to the latter developments
but this is not immediate.  Let us write
from Section 3.6 ($\Upsilon\sim px+Ey+\Omega t+
\sigma\cdot\zeta$ and $\Xi=xU+yV+tW+\zeta$)
$$
<\psi^*L^1\psi> = <\psi^*(-2\partial)\psi> = -2<e^{-\Upsilon}\phi^*
[pe^{\Upsilon}\phi +Ue^{\Upsilon}\phi_{\theta}]>=$$
\be
-2[p<\phi^*\phi> + U<\phi^*\phi_{\theta}>]
\label{PA}
\ee
Similarly from $A^1 = -3\partial^2+(3/2)u_0$ we have
\be
<\psi^*A^1\psi> = -3<\psi^*[p^2e^{\Upsilon}\phi+2pUe^{\Upsilon}\phi_
{\theta}+U^2e^{\Upsilon}\phi_{\theta\theta}]>+
\label{PB}
\ee
$$+\frac{3}{2}<\psi^*u_0\psi>=-3[p^2<\phi^*\phi>+2pU<\phi^*\phi_{\theta}>
+U^2<\phi^*\phi_{\theta\theta}>]+\frac{3}{2}<\phi^*u_0\phi>$$
We note also a natural comparison with (\ref{NZ}) for the formula
\be
<\psi\psi^*> = \sum_1^{\infty}\frac{<s_n>}{\lambda^n}
\label{PC}
\ee
We remark here that such a series is natural from asymptotic expansions
but when $\psi,\,\,\psi^*$ are written in terms of theta functions it
requires expansion of the theta functions in $1/\lambda$ (such
expansions are documented in \cite{dc}, p. 49 for example).  It is now
natural to ask whether one can express $dp,\,\,dE,\,\,d\Omega$ from
(\ref{BK}) in more detail.  Thus note first from (\ref{BK})
\be
dE<\phi^*\phi> = 2dp[p<\phi^*\phi> +U<\phi^*\phi_{\theta}>];\,\,
d\Omega<\phi^*\phi> =
\label{PD}
\ee
$$3dp[p^2<\phi^*\phi> + 2pU<\phi^*\phi_{\theta}> + U^2<\phi^*\phi_{\theta
\theta}>]-\frac{3}{2}dp<\phi^*u_0\phi>$$
which says e.g.
\be
dE = 2pdp + \frac{2U<\phi^*\phi_{\theta}>}{<\phi^*\phi>}dp;\,\,
d\Omega =
\label{PE}
\ee
$$= 3p^2dp +\left[6pU\frac{<\phi^*\phi_{\theta}>}{<\phi^*\phi>} +
3U^2\frac{<\phi^*\phi_{\theta\theta}>}{<\phi^*\phi>} -\frac{3}{2}
\frac{<\phi^*u_0\phi>}{<\phi^*\phi>}\right]dp$$
Given $p\sim\lambda+\cdots,\,\,E\sim\lambda^2+\cdots,\,\,
\Omega\sim\lambda^3+\cdots$ this is reasonable.  One would also like
to compare (\ref{NS}) and the equation $\partial_Tdp +\partial_Xd\Omega
+\partial_YdE=0$ (recall $\hat{\cal L}=(1/4)\partial^2\gamma+3\partial^{-1}
(u_0\partial\gamma)$ and $\hat{\cal G} = (3/4)\partial_y\partial^{-1}\gamma$).
One expects $dp = d\lambda+\cdots,\,\,dE = 2\lambda d\lambda+\cdots,$
and $d\Omega=3\lambda^2 d\lambda+\cdots$ so a formula like (\ref{PC})
has an order of magnitude compatibility with a possible identification
$dp\sim<\gamma>d\lambda$.
Note here the analogy with (\ref{NZ}) etc. where $<\Psi>\sim <{\cal T}>$
with $\hat{\Omega}_1\sim -<{\cal T}>(d\mu/2\sqrt{\mu})$ (the $1/\sqrt{\mu}$
weight factor arises from KdV notation here, linked to a
hyperelliptic Riemann surface); recall also $<{\cal X}>$ is
similarly connected to $\hat{\Omega}_2$.  Let us assume the ``ansatz"
$(\diamondsuit)$:  For some choice of standard homology basis and
local coordinate $dp\sim <\psi\psi^*>d\lambda$.  We have seen that a
variation on this is valid for KdV situations and
via changes in homology and local
coordinate is reasonable in general.  In fact it generally true via
\\[3mm]\indent {\bf LEMMA 7.2}.$\,\,$  The ansatz $(\diamondsuit)$ is
valid for KP situations.
\\[2mm]\indent {\it Proof}:$\,\,$  We refer here to \cite{cd} for
background and notation (cf. also \cite{ci} for dispersionless
genus zero situations).  We write $\partial
= L + \sum_1^{\infty}\sigma_j^1 L^{-j}$ which implies
$\frac{\partial\psi}{\psi}=\lambda + \sum_1^{\infty}
\sigma_j^1\lambda^{-j}$. (this is motivated by the proof
of Lemma 2.1 in \cite{ba}).  Then
$(log\psi)_x=\lambda +\sum_1^{\infty}\sigma_j^1\lambda^{-j}$ and
$\overline{(log\psi)_x} = p
= \lambda + \sum_1^{\infty} <\sigma_j^1>\lambda^{-j}$.  But $s_{n+1} =
-n\sigma_n^1 - \sum_1^{n-1}\partial_j\sigma_{n-j}^1$ and one
obtains $<s_{n+1}> =
-n<\sigma_n^1> -\sum_1^{n-1}<\partial_j\sigma_{n-j}^1> = -n<\sigma_n^1>$. Thus
$dp = d\lambda - \sum_1^{\infty}j<\sigma_j^1>\lambda^{-j-1}d\lambda = d\lambda
+\sum_1^{\infty}<s_{j+1}>\lambda^{-j-1}d\lambda=<\psi\psi^*>$ since
$\psi\psi^*=\sum_0^{\infty}s_n\lambda^{-n}$. {\bf QED}
\\[3mm]\indent
Consequently from
(\ref{BK}) we obtain
\be
d\Omega\sim -<\psi^*A^1\psi>d\lambda;\,\,dE\sim -<\psi^*L^1\psi>d\lambda
\label{PF}
\ee
This would require $<\psi^*L^1\psi> = O(2\lambda)$ which by (\ref{PA})
is correct while $<\psi^*A^1\psi>$ should be $O(3\lambda^2)$ which is
also correct.  Therefore one has
\\[3mm]\indent {\bf THEOREM 7.3}.$\,\,$  With
the notations as above $dp = <\psi\psi^*>d\lambda,\,\,
dE\sim -<\psi^*L^1\psi>$, and
$d\Omega \sim -<\psi^*A^1\psi>d\lambda$.
\\[3mm]\indent
It follows then from (\ref{NS}) (holding $Y$ or $T$ constant) that
\\[3mm]\indent {\bf COROLLARY 7.4}.$\,\,$ From Theorem 7.3 results
$<\hat{{\cal L}}>d\lambda\sim d\Omega$ and $<\hat{{\cal G}}>
d\lambda\sim dE$; consequently $-<\psi^*L^1\psi>\sim <\hat{{\cal G}}>$
with $-<\psi^*A^1\psi>\sim<\hat{{\cal L}}>$.
\\[3mm]\indent
In summary we can also state
\\[3mm]\indent {\bf THEOREM 7.5}.$\,\,$
The quantity $\psi\psi^*$ is seen to determine the Whitham hierarchy,
the differentials $dE,\,\,d\Omega$, etc., and via $\omega\sim {\cal K}$
as in Theorem 6.4
the theory is connected to the dispersionless Whitham theory.
\\[3mm]\indent {\bf REMARK 7.6}.$\,\,$  Since $D+D^*-2\infty\sim K_{\Sigma},
\,\,\psi\psi^*$ is determined by a section of $K_{\Sigma}$ (global
point of view) but it is relations based on $<\psi^*L^1\psi>,\,\,
<\psi^*A^1\psi>,$ etc. (based on the Krichever averaging process)
which reveal the ``guts" of $\psi\psi^*$ needed for averaging.

\subsection{The moduli space $M_{gn}$}
One can formulate the LG theory modeled on $A_{n-1}$ (or $A_n$) in a
genus $g$ situation as follows (cf. \cite{dj,dk}).  One considers
(*)  $\partial_{\alpha}L = [L,L^{\alpha/n}_{+}],\,\,\alpha\not=
kn,$ for $L = \partial^n + q_{n-2}\partial^{n-2}+\cdots+q_0$
(nKdV).  Then one averages (*) over the family of $g$ gap solutions,
with averaged Poisson brackets, Hamiltonian structures, etc.  The
space $M=M_{gn}\sim \{parameters\,\, of\,\, g\,\, gap\,\, solutions\}$
is the moduli space of Riemann surfaces $\Sigma_g$ with genus $g$,
a marked point $Q\sim\infty$, and $\lambda(P)$ a function with a pole
at $\infty$ of order $n$ (note this is not the same as
${\cal M}_{gn}$ of Section 3.7).
The dimension of $M_{gn}$ is $N=2g+n-1$ as in
Section 5 ($2g+n_0+\cdots+n_m+2m = 2g+n-1,\,\,n_0=n-1,\,\,m=0$).  One
takes $u^j=\lambda(P_j),\,\,j=1,\cdots,N$ where $d\lambda|_{P_j}=0$.
Evidently the number of critical points is greater than $n-1$ in general
so $\lambda$ cannot simply be a polynomial of degree $n$ as in the
$A_{n-1}$ situation.  The exact form of $\lambda$ can vary (note
e.g. even a function like $p^n+q_{n-2}p^{n-2}+\cdots+q_0+(\alpha/p)
+\cdots+(\beta/p^s)$ could generate more than $n-1$ critical points).
In fact $\lambda$ will have various natural forms illustrated by
the case $M_{11}$ where $\lambda(p) = {\cal P}(2\omega p,\omega,\omega')$
where ${\cal P}$ is the Weierstrass function (cf. \cite{di}).  One takes
now a covering $\hat{M}_{gn}$ of $M_{gn}$ by fixing a homology basis
$(a_i,b_i)$ and choosing a local coordinate $k^{-1}$ near $\infty$
such that $k^n=\lambda$.  Then one fixes an abelian differential $dp$ such
that $dp=d(k+O(1))$ as $\lambda\to\infty$ with $\oint_{a_i}dp=0$.
The flat Egoroff metric on $M$ in local coordinates $u^j$ has the form
\be
ds^2=\sum_1^Ng_{ii}(u)(du^i)^2;\,\,g_{ii}(u) = -\frac{1}{n}Res_{P_i}
\frac{(dp)^2}{d\lambda}
\label{QA}
\ee
and the flat coordinates $T^1,\cdots,T^N$ for $ds^2$ are
\be
T^{\alpha} = -nRes\frac{\lambda^{\frac{n-\alpha}{n}}}{n-\alpha}dp\,\,
(\alpha=1,\cdots,n-1);
\label{QB}
\ee
$$T^{n-1+\alpha}=\frac{1}{2\pi i}\oint_{a_{\alpha}}pd\lambda\,\,
(\alpha=1,\cdots,g);\,\,T^{g+n-1+\alpha}=\oint_{b_{\alpha}}dp\,\,
(\alpha=1,\cdots,g)$$
Note that this is a global definition of the flat coordinates so $M$ is
unramified over some domain in ${\bf C}^N$.  Further the metric
$ds^2$ in the flat coordinates has the form
\be
<dT^{\alpha},dT^{\beta}>=\delta^{\alpha+\beta,n}\,\,(1\leq \alpha,\beta
\leq n-1);
\label{QC}
\ee
$$<dT^{n-1+\alpha},dT^{g+n-1+\beta}>=\delta^{\alpha,\beta}\,\,(1\leq
\alpha,\beta\leq g)$$
and is otherwise zero.
\\[3mm]\indent
Let us continue this example following \cite{dj,dk}.  Thus $p =
\int_{P_0}^Pdp$ is a multivalued function on $\Sigma_g$ with $P_0$
chosen so that $\lambda(P_0)=0$ in some domain on $M_{gn}$.
Then the multivalued differential $pd\lambda$ has the form
(as $k\sim\lambda^{1/n}\to\infty$)
\be
pd\lambda = kd\lambda-\left(\sum_1^{n-1}T^{\alpha}k^{\alpha-1}
+O(k^{-1})\right)dk;
\label{QD}
\ee
$$\oint_{a_s}pd\lambda=2\pi iT^{n+s-1};\,\,\Delta_{a_s}(pd\lambda)=0;
\,\,\Delta_{b_s}(pd\lambda)=T^{g+n-1+s}d\lambda\,\,(s=1,\cdots,g)$$
where $\Delta_{a_s}(f(P))=f(P+a_s)-f(P),\,\,\Delta_{b_s}(f(P)) =
f(P+b_s)-f(P)$
for any function or differential.  The primary differentials
(perhaps multivalued) are defined via
\be
\phi_{\alpha}=\partial_{\alpha}(pd\lambda)|_{\lambda=c}=
-\partial_{\alpha}(\lambda dp)|_{p=\hat{c}}\,\,(\alpha=1,\cdots,N)
\label{QE}
\ee
so that $\lambda=\lambda(p)$ is the LG potential of the model.  Explicitly
(for $k\to\infty$)
\be
\phi_{\alpha}=dp^{\alpha}=(-k^{\alpha}+O(k^{-2})dk;\,\,\oint_{a_s}
dp^{\alpha}=0\,\,(s=1,\cdots,g)
\label{QF}
\ee
are normalized differentials of the second kind with $\phi_1=dp^1=-dp$.
Next
\be
\phi_{n-1+\alpha}\equiv \omega_{\alpha}\,\,(\alpha=1,\cdots,g)
\label{QG}
\ee
are the normalized holomorphic differentials ($\oint_{a_s}\omega_{\alpha}
=2\pi i\delta_{\alpha,s}$) and
\be
\phi_{g+n-1+\alpha}\equiv \sigma_{\alpha}\,\,(\alpha=1,\cdots,g)
\label{QH}
\ee
are holomorphic (modulo $d\lambda$) multivalued differential with
increments
\be
\Delta_{a_{\beta}}\sigma_{\alpha}=0;\,\,\Delta_{b_{\beta}}=
\delta_{\alpha,\beta}d\lambda
\label{QHH}
\ee
\indent
Then the primary part of the associated hierarchy for
the solution where $T^{\alpha,0}\equiv
T^{\alpha}$ and $\partial_Xu^j=1\,\,(j=1,c\cdots,N)$ has the FFM form
(from \cite{fa})
\be
\partial_{T^{\alpha}}dp=-\partial_X\phi_{\alpha}
\label{QI}
\ee
or equivalently in the diagonal variables $u^1,\cdots,u^N$
\be
\partial_{T^{\alpha}}u^j=-(\frac{\phi_{\alpha}}{dp})|_{P_j}\partial_Xu^j
=-(\frac{\phi_{\alpha}}{dp})|_{P_j}\,\,(j=1,\cdots,N)
\label{QJ}
\ee
(we omit discussion of the theory involving $T^{\alpha,0}$ - cf. \cite{dj}).
Equations (\ref{QI}) for $\alpha=1,\cdots,n-1$ correspond to the
averaged Gelfand-Dickey hierarchy over $g$ gap solutions.  The extension
for $\alpha=n,\cdots,N$ is necessary to construct a closed primary
operator algebra.  One has
\be
\eta_{\alpha,\beta}=\sum_1^NRes_{P_i}\frac{\phi_{\alpha}\phi_{\beta}}
{d\lambda};\,\,c_{\alpha\beta\gamma}(T)=-\sum_1^NRes_{P_i}\frac
{\phi_{\alpha}\phi_{\beta}\phi_{\gamma}}{d\lambda dp}
\label{QK}
\ee
and this primary operator algebra can be represented via relations
\be
\phi_{\alpha}\phi_{\beta}=c^{\gamma}_{\alpha\beta}\phi_{\gamma} dp\,\,
(modulo\,\,d\lambda)
\label{QL}
\ee
This is the $M_{gn}$ model.  There are evident connections of this
approach to developments in \cite{kj,kl}, some of which we indicated
earlier.  The relation to Hurwitz spaces as in Section 5 is clear and one
can easily fill in the correspondences.

\subsection{Whitham and Toda}

We would be remiss not to mention some recent work (\cite{ea,gc,hc,mf,mg,me,nc,
sc} for example) on Whitham and Toda related to $N=2$ supersymmetric
gauge theories.  We sketch the development in \cite{nc} dealing
with $N=2$ supersymmetric $SU(N)$ Yang-Mills theory but will not
discuss the physics here and will omit proofs.
Of concern is
the differential
\be
dS = \frac{x\frac{dP(x)}{dx}}{y}dx
\label{WTA}
\ee
on the family of hyperelliptic curves
\be
y^2=P(x)^2-\Lambda^{2N};\,\,P(x)=x^N+\sum_0^{N-2}u_{N-k}x^k
\label{WTB}
\ee
Here $u=(u_2,\cdots,u_N)$ are parameters for flat moduli and $\Lambda$
is fixed.  The spectrum of excitations in the theory is measured by the
units
\be
a_i=\oint_{\alpha_i}dS;\,\,a_{D,i}=\oint_{\beta_i}dS\,\,\,(1\leq i\leq N-1)
\label{WTC}
\ee
where $(\alpha_i,\beta_i),\,\,1\leq i\leq N-1,$ is a standard homology
basis and the $\alpha_i$ circles counterclockwise the cut between
two neighboring branch points in the $x$ plane.  Let $p_{\infty},\,\,
\tilde{p}_{\infty}$ be the two points at $\infty$ (i.e. $x(p_{\infty})
=x(\tilde{p}_{\infty}) = \infty$).  One can write
\be
dS=dX_{\infty,1}+\sum_1^{N-1}a_i\omega_i
\label{WTD}
\ee
where $\omega_i\,\,(1\leq i\leq N-1)$ correspond to standard normalized
holomorphic differentials ($\oint_{\alpha_i}\omega_j=\delta_{ij}$).
Here $dX_{\infty,1}$ is a normalized meromorphic differential of second
kind with second order poles at $p_{\infty}$ and $\tilde{p}_{\infty}$
and $\oint_{\alpha_i}dX_{\infty,1}=0$.
\\[3mm]
\indent
Now define
$h=y+P(x)$ and $\bar{h}=-y+P(x)$ and consider infinitesimal deformations
of the moduli parameters $u$ with $h$ or $\bar{h}$ being fixed so e.g.
$(\partial/\partial u_{N-k})dS|_{h=c}=-(x^k/y)dx$.  Changing the
$u\to a=(a_1,\cdots,a_{N-1})$ this becomes $(\partial/\partial a_i)dS|_
{h=c}=\omega_i$ which implies
\be
\frac{\partial}{\partial a_i}\omega_j=\frac{\partial}{\partial a_j}
\omega_i\,\,(1\leq i,j\leq N-1)
\label{WTF}
\ee
where $(\partial/\partial a_i)$ fixes $a_j\,\,(j\not= i)$ and $h$.
These correspond to a subset of classical Whitham equations
\be
\frac{\partial}{\partial a_i}\omega_j=\frac{\partial}{\partial a_j}\omega_i;
\,\,\frac{\partial}{\partial T_A}\omega_i=\frac{\partial}{\partial a_i}
\Omega_A;\,\,\frac{\partial}{\partial T_A}\Omega_B=\frac{\partial}
{\partial T_B}\Omega_A
\label{WTG}
\ee
Then one specifies the $\Omega_A$ to be meromorphic differentials
$\Omega_{\infty,n}$ and $\tilde{\Omega}_{\infty,n}$ of second kind
($n\geq 1$) where $\Omega_{\infty,n}$ (resp. $\tilde{\Omega}_{\infty,n}$)
has a pole of order $n+1$ at $p_{\infty}$ (resp. $\tilde{p}_{\infty}$) and
is holomorphic elsewhere.  Let also $\Omega_{\infty,0}\,\,(\equiv\tilde
{\Omega}_{\infty,0}$) be a differential of third kind with
simple poles at $p_{\infty}$ and
$\tilde{p}_{\infty}$ of residues $\pm 1$ respectively.  All differentials
are normalized via $\oint_{\alpha_i}\Omega_{\infty,n}=\oint_{\alpha_i}
\tilde{\Omega}_{\infty,n}=0\,\,(n\geq 0)$ and they are determined by
$h\,\,(\bar{h})$ as follows.  Define local coordinates $z_{\infty}\,\,
(\bar{z}_{\infty})$ via $z^N_{\infty}=h^{-1}\,\,(\bar{z}^N_{\infty}=
\bar{h}^{-1})$ and near $p_{\infty}$ for example
\be
\Omega_{\infty,n}=[-nz_{\infty}^{-n-1}-\sum_{m\geq 1}q_{mn}z_{\infty}^
{m-1}]dz_{\infty}\,\,(n\geq 1);
\label{WTH}
\ee
$$\tilde{\Omega}_{\infty,n}=[\delta_{n,0}z_{\infty}^{-1}-\sum_{m\geq 1}
r_{mn}z_{\infty}^{m-1}dz_{\infty}]\,\,(n\geq 0)$$
with analogous expressions near $\tilde{p}_{\infty}$.  The integrability
condition (\ref{WTG}) implies there exists $dS$ such that
\be
\frac{\partial}{\partial a_i}dS=\omega_i;\,\,\frac{\partial}
{\partial T_n}dS = \Omega_{\infty,n};\,\,\frac{\partial}{\partial
\bar{T}_n}dS=\tilde{\Omega}_{\infty,n};\,\,\frac{\partial}{\partial T_0}dS
=\Omega_{\infty,0}
\label{WTJ}
\ee
where $1\leq i\leq N-1$ and $n\geq 1$.  One then constructs a function
$F(a,T,\bar{T})$ from dS via
\be
\frac{\partial F}{\partial a_i}=\frac{1}{2\pi i}\oint_{\beta_i}dS\,\,
(\equiv\frac{a_{D,i}}{2\pi i});
\label{WTK}
\ee
$$\frac{\partial F}{\partial T_n}=-Res_{p_{\infty}}z^{-n}_{\infty}dS;\,\,
\frac{\partial F}{\partial\bar{T}_n}=-Res_{\tilde{p}_{\infty}}
\bar{z}^{-n}_{\infty}dS;$$
$$\frac{\partial F}{\partial T_0}=-Res_{p_{\infty}}log\,z_{\infty} dS
+Res_{\tilde{p}_{\infty}}log\,\tilde{z}_{\infty}dS$$
Consistency of this definition follows from (\ref{WTG}) and the Riemann
bilinear relations.  The local behavior of $dS$ can be described by
$F$ as
\be
dS=\left[-\sum_{n\geq 1}nT_nz_{\infty}^{-n-1} + T_0z_{\infty}^{-1}
-\sum_{n\geq 1}\frac{\partial F}{\partial T_n}z_{\infty}^{n-1}\right]
dz_{\infty}\,\,\,(near\,\,p_{\infty});
\label{WTL}
\ee
$$dS = \left[-\sum_{n\geq 1}n\bar{T}_n\tilde{z}_{\infty}^{-1} -
T_0\tilde{z}_{\infty}^{-1} -
\sum_{n\geq 1}\frac{\partial F}{\partial \bar{T}_n}\tilde{z}_{\infty}^
{n-1}\right]d\tilde{z}_{\infty}\,\,\,(near\,\,\tilde{p}_{\infty})$$
\indent
An interesting class of solutions of the Whitham hierarchy are solutions
such that $\sum_1^{N-1}a_i(\partial F/\partial a_i) + \sum_{n\geq 0}
T_n(\partial F/\partial T_n) + \sum_{n\geq 1}\bar{T}_n(\partial F/\partial
\bar{T}_n) = 2F$ in which case the $dS$ of (\ref{WTJ}) satisfies
\be
dS = \sum_1^{N-1}a_i\omega_i+\sum_{n\geq 0}T_n\Omega_{\infty,n} +
\sum_{n\geq 1}\bar{T}_n\tilde{\Omega}_{\infty,n}
\label{WTM}
\ee
which is consistent with (\ref{WTL}).  Therefore one can reproduce
the $dS$ of (\ref{WTA}) by setting $T_1=-\bar{T}_1=1$ with the
other $T_A=0$.  The pre-potential ${\cal F}$ of $N=2$ supersymmetric
$SU(N)$ Yang-Mills theory is then given by ${\cal F}=2\pi iF$.  For this
$F$ one can write after some calculation ($\omega_i=-\sum_{m\geq 1}
\sigma_{im}z_{\infty}^{m-1}dz_{\infty}$ near $p_{\infty}$, with
$\bar{\sigma}_{im}$ defined analogously near $\tilde{p}_{\infty}$)
\be
F=\frac{1}{4\pi i}\sum_{i,j=1}^{N-1}\tau_{ij}a_ia_j +\sum_1^{N-1}
a_i\left[\sum_{k\geq 1}(\sigma_{ik}T_k+\bar{\sigma}_{ik}\bar{T}_k)
+\bar{\sigma}_{i,0}T_0\right]
\label{WTO}
\ee
$$+\frac{1}{2}\sum_{k,l\geq 1}q_{kl}T_kT_l +\sum_{k,l\geq 1}r_{kl}
T_k\bar{T}_l +\frac{1}{2}\sum_{k,l\geq 1}\bar{q}_{kl}\bar{T}_k\bar{T}_l$$
$$+\frac{1}{2}\bar{r}_{0,0}T_0^2 + T_0\sum_{k\geq 0}(r_{k,0}T_k
+\bar{r}_{0,k}\bar{T}_k)$$
\be
\tau_{ij}=\oint_{\beta_i}\omega_j;\,\,\bar{\sigma}_{i,0}=\frac{1}
{2\pi i}\oint_{\beta_i}\Omega_{\infty,0};
\label{WTP}
\ee
$$\bar{r}_{0,0}=-Res_{p_{\infty}}log\,z_{\infty}\Omega_{\infty,0} +
Res_{\tilde{p}_{\infty}}log\,\tilde{z}_{\infty}\Omega_{\infty,0}$$
\indent
To connect this with Toda let $t=(t_1,\cdots),\,\,\bar{t}=(\bar{t}_1,\cdots)$,
and $n$
be Toda lattice times where $n\in {\bf Z}$.  One knows from
\cite{zg} that the Toda lattice hierarchy has a quasi periodic solution
corresponding to a hyperelliptic curve as above, giving an $N$ periodic
solution of the Toda chain.  The moduli parameters are invariants
(integrals of motion) of such solutions.  One introduces new time
variables $\theta=(\theta_1,\cdots,\theta_{N-1})$ into the
associated BA function as
$$
\Psi(p,t,\bar{t},n,\theta)=e^{[-n\int^p\Omega_{\infty,0}+\sum_{n\geq 1}t_n
\int^p\Omega_{\infty,n}+\sum_{n\geq 1}\bar{t}_n\int^p\tilde{\Omega}_
{\infty,n}+i\sum_1^{N-1}\theta_i\int^p\omega_i]}\cdot$$
$$\cdot\frac{\theta(z(p)-z(D)+K-n(z(p_{\infty})-z(\tilde{p}_{\infty}))
+\sum_{n\geq 1}t_n\sigma_n+\sum_{n\geq 1}\bar{t}_n\bar{\sigma}_n
+\frac{1}{2\pi}\sum_1^{N-1}\theta_i\tau_i)}{\theta(z(p_{\infty}-z(D)+K
-n(z(p_{\infty}) -z(\tilde{p}_{\infty}))+\sum_{n\geq 1}t_n\sigma_n+
\sum_{n\geq 1}\bar{t}_n\bar{\sigma}_n +\frac{1}{2}\sum_1^{N-1}\theta_i
\tau_i)}$$
\be
\cdot\frac{\theta(z(p_{\infty})-z(D)+K)}{\theta(z(p)-z(D)+K)}
\label{WTQ}
\ee
Here $z$ is the Abel map ($\sim{\cal A}$), $D$ is a positive divisor
of degree $N-1$, $K$ is the Riemann constant, and $\sigma_i,\,\,\bar{\sigma}_i
\,\,(i\geq 1)$ and $\tau_i\,\,(1\leq i\leq N-1)$ are $N-1$ dimensional
vectors (e.g. $\sigma_i = {}^T(\sigma_{i1},\cdots,\sigma_{i,N-1})$).
As $\theta\to 0,\,\,\Psi\to$ BA function of the ordinary Toda
lattice.  A tau function can be now introduced near $p_{\infty}$ via
\be
\Psi(p,t,\bar{t},n,\theta)=z_{\infty}^{-n}e^{\sum_{k\geq 1}t_kz_{\infty}^{-k}}
\cdot\frac{\tau(t-[z_{\infty}],\bar{t},n,\theta)}{\tau(t,\bar{t},n,\theta)}
\label{WTR}
\ee
and near $\tilde{p}_{\infty}$ by
$$\Psi(p,t,\bar{t},n,\theta)=\tilde{z}_{\infty}^ne^{\sum_{k\geq 1}
\bar{t}_k\tilde{z}_{\infty}^{-k}}\cdot\frac{\tau(t,\bar{t}-[\tilde{z}_
{\infty}],n+1,\theta)}{\tau(t,\bar{t},n,\theta)}$$
(we recall that e.g. $[z]\sim(z_1,(1/2)z^2,(1/3)z^3,\cdots)$).
By matching this with (\ref{WTQ}) one obtains
\be
\tau(t,\bar{t},n,\theta) = e^{\hat{F}(t,\bar{t},n,\theta)}\cdot
\label{WTS}
\ee
$$\theta(-(n-1)z(p_{\infty})+nz(\tilde{p}_{\infty})-z(D)+K
+\sum_{n\geq 1}t_n\sigma_n+\sum_{n\geq 1}\bar{t}_n\bar{\sigma}_n
+\frac{1}{2\pi}\sum_1^{N-1}\theta_i\tau_i)$$
where $\hat{F}$ is a polyomial given by
\be
\hat{F}(t,\bar{t},n,\theta) = \frac{1}{2}\sum_{k,l\geq 1}q_{kl}t_kt_l
+\sum_{k,l\geq 1}r_{kl}t_k\bar{t}_l +\frac{1}{2}\sum_{k,l\geq 1}\bar{q}_
{kl}\bar{t}_k\bar{t}_l
\label{WTT}
\ee
$$-\frac{1}{4\pi i}\sum_1^{N-1}\tau_{ij}\theta_i\theta_j +\frac
{n(n-1)}{2}\bar{r}_{0,0} +i\sum_1^{N-1}\theta_i\left[
\sum_{k\geq 1}(\sigma_{ik}t_k+\bar{\sigma}_{ik}\bar{t}_k)-
n\bar{\sigma}_{i,0}\right]$$
$$-n\sum_{k\geq 1}r_{k,0}t_k-(n-1)\bar{r}_{0,k}\bar{t}_k
+\sum_{k\geq 1}d_kt_k + \sum_{k\geq 1}\bar{d}_k\bar{t}_k +\bar{d}_0n$$
Then following \cite{ba} one introduces slow time variables via
$T_i=\epsilon t_i,\,\,\bar{T}_i=\epsilon \bar{t}_i,\,\,T_0=-\epsilon n,$
and $a_i=i\epsilon\theta_i$.  Then from the asymptotics of (\ref{WTQ})
and (\ref{WTS}) one obtains
\be
d\,log\Psi(p,\frac{T}{\epsilon},\frac{\bar{T}}{\epsilon},-\frac{T_0}
{\epsilon},\frac{a}{i\epsilon})=\epsilon^{-1}\sum_{n\geq 0}\epsilon^n
dS^{(n)}(p,T,\bar{T},T_0,a)
\label{WTU}
\ee
$$log\tau(\frac{T}{\epsilon},\frac{\bar{T}}{\epsilon},-\frac{T_0}
{\epsilon},\frac{a}{i\epsilon})=\epsilon^{-2}\sum_{n\geq 0}
\epsilon^nF^{(n)}(T,\bar{T},T_0,a)$$
The leading order terms $dS^{(0)}$ and $F^{(0)}$ are the same as $dS$ and
$F$ in (\ref{WTO}) and one thinks now of the moduli parameters $u$ as
functions of the slow variables ($u_k=u_k(T,\bar{T},T_0,a)$.  One
obtains a system of modulation equations for the moduli parameters and these
represent the Whitham hierarchy (\ref{WTG}).

\end{document}